\documentclass[iop,twocolappendix]{emulateapj}

\usepackage{amsmath}
\usepackage{multirow}
\usepackage{nicefrac}
\usepackage{enumitem}
\usepackage{pbox}
\usepackage{soul}
\usepackage{float}
\usepackage{lineno}

\shorttitle{Habitability of exoplanet waterworlds}
\shortauthors{Kite \& Ford}

\begin{document}


\title{Habitability of exoplanet waterworlds}%
% \\ or The deep circulation of short-period rocky exoplanets} %\\ (Dynamics and outgassing of magma pools on
%short-period rocky extrasolar planets)}


\author{Edwin S. Kite}%\altaffilmark{1}}
\affil{Department of Geophysical Sciences, University of Chicago, Chicago, IL 60637, USA \\ kite@uchicago.edu}

\author{Eric B. Ford}\affiliation{Department of Astronomy and Astrophysics, The Pennsylvania State University, University Park, PA 16802, USA}
\affiliation{Center for Exoplanets and Habitable Worlds, The Pennsylvania State University, University Park, PA 16802, USA}
\affiliation{Institute for CyberScience, The Pennsylvania State University, University Park, PA 16802, USA}
\affiliation{Pennsylvania State Astrobiology Research Center, The Pennsylvania State University, University Park, PA 16802, USA}

%Center for Exoplanets \& Habitable Worlds, Department of Astronomy \& Astrophysics; \\
%Center for Astrostatistics; \& Institute of CyberScience,

%\email{kite@uchicago.edu}

%\textbf{switch notation throughout to Carbon kg/m2 not CO2 kg/m2; Fiegelson cites on 26-al]} 

\begin{abstract}
Many habitable zone (HZ) exoplanets are expected to form with water mass fractions higher than that of the Earth. For rocky exoplanets with 10-1000$\times$ Earth's H$_2$O but without H$_2$, we model the multi-Gyr evolution of ocean temperature and chemistry, taking into account C partitioning, high-pressure ice phases, and atmosphere-lithosphere exchange. Within our model, for Sun-like stars, we find that: (1)~the duration of habitable surface water is \mbox{strongly affected} by ocean chemistry; (2)~possible ocean pH spans a wide range; (3)~surprisingly, many waterworlds retain habitable surface water for $>$1 Gyr, and (contrary to previous claims) this longevity does not necessarily involve geochemical cycling. The key to this cycle-independent planetary habitability is that C exchange between the convecting mantle and the water ocean is curtailed by seafloor pressure on waterworlds, so the planet is stuck with the ocean mass and ocean cations that it acquires during the first 1\% of its history. In our model, the sum of positive
charges leached from the planetary crust by early water-rock interactions is -- coincidentally -- often within an order of magnitude of the early-acquired atmosphere+ocean inorganic C inventory overlaps. As a result, $p_{\mathrm{CO2}}$ is frequently in the ``sweet spot'' (0.2-20 bar) for which the range of semimajor axis that permits surface liquid water is about as wide as it can be. Because the width of the HZ in semimajor axis defines (for Sun-like stars) the maximum possible time span of surface habitability, this effect allows for Gyr of habitability as the star brightens. We illustrate our findings by using the output of an ensemble of \emph{N}-body simulations as input to our waterworld evolution code. Thus (for the first time in an end-to-end calculation) we show that chance variation of initial conditions, with no need for geochemical cycling, can yield multi-Gyr surface habitability on waterworlds.

\end{abstract}
\vspace{0.1in}


%Proxima Cen b, 
%Ross 128b, 
%TRAPPIST-1, 
%GJ 163c, 
%GJ 832c, 
%GJ 667C, 
%$\tau$ Ceti e, 
%KIC-5522786b, 
%Wolf 1061c
%LHS-1140 b, 
%K2-9b, 
%K2-18b, 
%K2-72e, 
%Kepler-22b, 
%Kepler-61b, 
%Kepler-62f,
%Kepler-174d, 
%Kepler-186f,
%Kepler-296e, 
%Kepler-296f, 
%Kepler-298b, 
%Kepler-440b,
%Kepler-442b, 
%Kepler-443b,
%Kepler-452b, 
%Kepler-705b, 
%Kepler-1090b, 
%Kepler-1229b, 
%Kepler-1410b, 
%Kepler-1552b, 
%Kepler-1540b,
%Kepler-1544b, 
%Kepler-1606b, 
%Kepler-1638b).}
\section{Introduction.}
\noindent

\noindent

Habitable zone (HZ) small-radius exoplanets are common \citep{Burke2015,DressingCharbonneau2015}. What fraction of them are habitable? For the purposes of this paper, a useful definition of a potentially habitable exoplanet is that it maintains $T$ $<$ 450K liquid water on its surface continuously for timescales that are relevant for biological macroevolution, $\gg$10$^7$ yr \citep{Vermeij2006, Carter1983, Bains2015}.\footnote{Sub-ice oceans in extrasolar planetary systems may be habitable, but this cannot be confirmed from Earth by remote sensing.} With this definition, two pathways allow long-term planetary habitability:
\vspace{-0.015in}
\begin{itemize}[leftmargin=0em]
\item[] \textbf{Habitability sustained by geochemical cycles:} Planets that stay habitable, due to a negative feedback, when climate is perturbed (by tectonics, stellar evolution, etc.). A proposed mechanism for the negative feedback is a geochemical cycle that balances volcanic outgassing and the fixation of atmospheric gases into rocks (carbonate-silicate weathering feedback; e.g.~\citealt{Walker1981,Kasting1993}). 
\end{itemize}
\vspace{-0.015in}

\noindent Climate-stabilizing geochemical cycles are a mainstay of textbooks and review papers (\citealt{CatlingKasting2017,Knoll2012,Kaltenegger2017}, and references therein). However, such feedbacks probably do not work on HZ rocky planets with water mass fractions $10 \times$ - $10^3 \times$ that of the Earth -- ``waterworlds'' \citep{Foley2015,Abbot2012}. Yet waterworlds should be common in the Galaxy (e.g.~\citealt{Mulders2015}). Are waterworlds doomed? Not necessarily, because there is another (less-studied) road to long-term planetary habitability.

\vspace{-0.015in}
\begin{itemize}[leftmargin=0em]
\item[] \textbf{Cycle-independent planetary habitability:} Planets where $p_{\mathrm{CO2}}$, ocean depth, and surface temperature $T_{surf}$ remain within the habitable range for~$\gg$10$^7$ yr without geochemical cycling.
\end{itemize}
\vspace{-0.015in}

\noindent In this paper, we first show that the long-term climate evolution of waterworlds can be modeled independently of long-term geochemical cycling between the atmosphere and the convecting silicate-rock mantle (Section 2). Next, we set up (Section 3) and run (Section 4) such a waterworld evolution model. Using the model, we estimate what combinations of water abundance, initial carbon abundance, and geologic processes allow habitable surface water to persist for~$>$1~Gyr. We focus on CO$_2$+H$_2$O($\pm$N$_2$) atmospheres (e.g. \citealt{WordsworthPierrehumbert2013}), for which the key climate-regulating greenhouse gas is CO$_2$. Our central result is that the partial~pressure of atmospheric~CO$_2$ ($p_{\mathrm{CO2}}$) in our waterworld model is frequently in the \mbox{$\sim$0.2-20 bar} range that enables $>$1 Gyr of surface liquid water. We ignore H$_2$ warming \citep{Stevenson1999}. We emphasize long-term climate evolution for planets around Sun-like stars, and defer discussion of nutrients to Section 6. In Section 5, we use an \emph{N}-body model of planet assembly to demonstrate how waterworlds form and migrate to the HZ. Readers interested only in astrophysics may skip Sections 3 and 4; readers interested only in geoscience may skip  Section 5. We conclude in Section 7. Our main results are shown in Figures~\ref{fig:mcuninformed}--\ref{fig:piecharts}.

 \begin{figure}[t]
%\epsscale{1.2}
\centering
\includegraphics[angle=-90,origin=c,width=1.0\columnwidth,trim={0mm 5mm 40mm 5mm}]{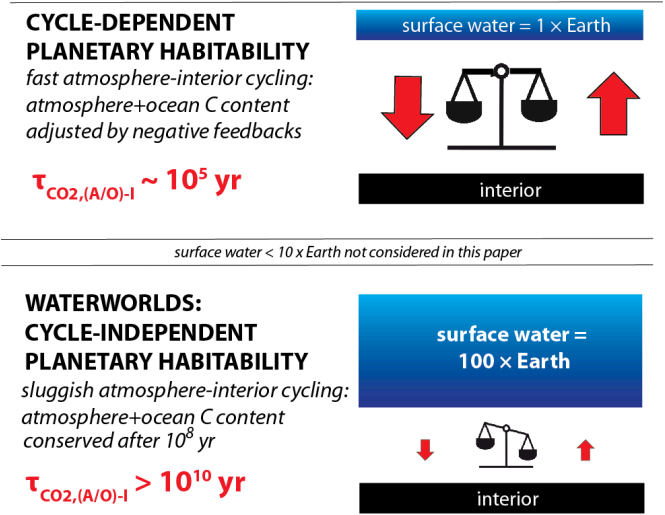}
\caption{Cycle-independent planetary habitability on exoplanet waterworlds. On shallow-ocean planets such as the Earth (\emph{top~row}), the timescale for (atmosphere/ocean)$\leftrightarrow$interior exchange of CO$_2$ ($\tau_{\mathrm{CO2,(A/O)\leftrightarrow I}}$) can be much shorter than the host star's main-sequence lifetime. With such fast cycling, small mismatches in the~C~cycle can quickly build up to a big change in greenhouse forcing, and big change in surface temperature - a threat to surface habitability (Figure \ref{fig:toymodels2}). Therefore, geochemical cycles are thought to be central to sustained surface habitability on shallow-ocean planets. By contrast, on deep-water planets (\emph{bottom~row}), atmosphere-interior cycling is curtailed by high seafloor pressure; $\tau_{\mathrm{CO2,(A/O)\leftrightarrow I}}$ can be $>$10~Gyr. Under such circumstances, long-term geochemical cycling between the atmosphere and the convecting silicate-rock mantle does not matter for determining whether or not the planet has habitable surface water for Gyr. Instead, the duration of habitable surface water is set by (i) atmosphere-ocean partitioning of C, and (ii) the initial cation content of the water ocean as it emerges from the first 100 Myr of giant impacts, late-stage delivery of C, and early crust formation.}  
\label{fig:toymodels}
\end{figure}

\begin{figure}
\begin{centering}
\epsscale{1.2}
\includegraphics[angle=-90,origin=c,width=1.00\columnwidth,clip=true,trim={0mm 20mm 0mm 20mm}]{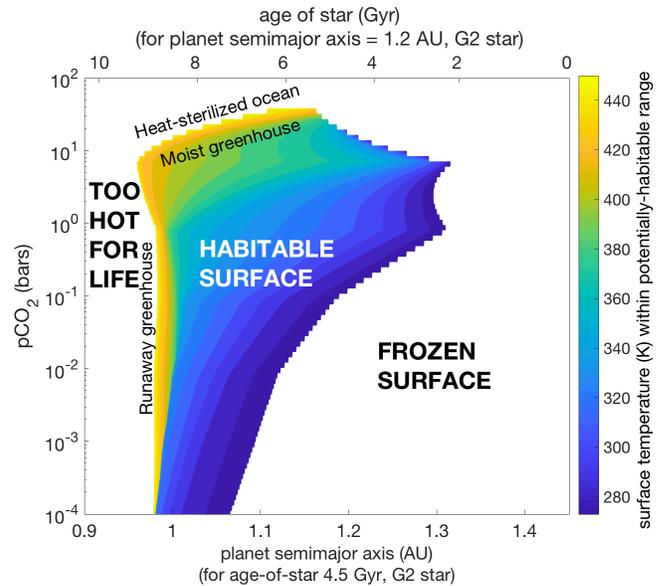} %\\
\caption{To show the effect of partial pressure of atmospheric CO$_2$ and of insolation $L_*$ on habitability. For G-type stars, insolation increases with age of the star on the main sequence (top axis) and decreases with distance from the star (bottom axis). The colored region corresponds to the habitable zone (HZ). The width of the HZ in semimajor axis defines the maximum possible time span of surface habitability. The planet fates define a lozenge of habitability that is broadest for 0.2-20 bars $p_{\mathrm{CO2}}$. In other words, for a given semimajor axis, the duration of surface habitability is maximized for a ``sweet spot'' of $p_{\mathrm{CO2}}$ in the range 0.2 - 20 bar. The color ramp corresponds to $T_{surf}$, and extends from 273 to 450K. Note that increasing $p_{\mathrm{CO2}}$ does not, by itself, cause a runaway greenhouse \citep{KastingAckerman1986,Ramirez2014}. This plot is based on \citet{WordsworthPierrehumbert2013}. (See Figure~\ref{fig:ramirezlozenge} below for a sensitivity test based on \citet{Ramirez2014}). 
} 
\label{fig:co2matters}
\end{centering}
\end{figure}
%\emph{Lower panel:} Cartoon showing the $p_{\mathrm{CO2}}$ needed to prevent development of a \mbox{HP-ice} layer, for 1 AU, as a function of planet mass $M_{pl}$ and planet water mass fraction . $>$0.5~bars $p_{\mathrm{CO2}}$ maximizes the chances of avoiding development of a \mbox{HP-ice} layer. (The dependence on planet mass in this rough, because we have not corrected the greenhouse effect for gravity).

\subsection{This paper in context}
Simulations of rocky-planet formation yield many rocky small-radius ($R_{pl}$ $<$1.6 $R_\earth$) planets that have planet water mass fractions $f_W$~=~10-1000 $\times~f_{W,\earth}$ \citep{Raymond2004,Raymond2007,Zain2018,Ciesla2015,Mulders2015}. Earth's $f_W$~($\sim10^{-4}$) appears to be the result of chance (e.g., \citealt{Raymond2007,Schonbachler2010,Morbidelli2016,Lichtenberg2016}) (Section 5.2). Indeed, simulations suggest  $f_W$~$\approx$~$f_{W,\earth}$ may be uncommon on planets in general (e.g.~\citealt{TianIda2015}). 

\begin{figure}
%\epsscale{1.2}
\begin{centering}
\includegraphics[width=0.875\columnwidth,clip=true,trim={13mm 60mm 13mm 50mm}]{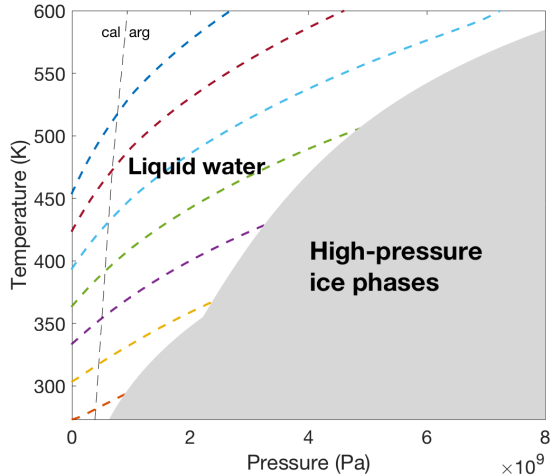}
\caption{Maximum ocean pressure to avoid a frozen seafloor (high-pressure ice, \mbox{HP-ice}). The thick dashed lines correspond to liquid water adiabats (spaced at 30K intervals in sea-surface temperature). For $T_{surf}$ $\gtrsim$ 375K, liquid water becomes a supercritical fluid with increasing $P$, avoiding  \mbox{HP-ice}. As the star brightens, sea-surface temperature increases, and it is possible for seafloor \mbox{HP-ice} to melt. The thin dashed line corresponds to the predominance boundary between the calcite (cal) and aragonite (arg) forms of CaCO$_3$ \citep{Redfern1989}. For details, see Appendix~B.
} 
\label{fig:H2Oadiabats}
\end{centering}
\end{figure}

Among previous studies of waterworlds (e.g., \citealt{Kuchner2003,Leger2004,Selsis2007,Fu2010,Abbot2012,Levi2017,Unterborn2018}), the closest in intent to our own are those of \cite{Kitzmann2015} and \cite{Noack2016}. \citet{Kitzmann2015} consider carbonate-system equilibria and find a ``sweet-spot'' in total planet C similar to our own, but they do not consider geological processes. \citet{Noack2016} emphasize the development of high-pressure H$_2$O-ice (\mbox{HP-ice}) layers that may isolate the liquid ocean from the nutrients supplied by silicate-rock leaching. \citet{Noack2016} also confirm the result of \citet{Kite2009} that deep oceans suppress C exchange between the convecting mantle and the water ocean. We conservatively do not count planets with \mbox{HP-ice} as examples of habitability (see Section 6.4 for a discussion). This can set a~$T_{surf}$-dependent~upper~limit on the depth of a habitable ocean (provided $T_{surf}$ $\lesssim$ 375K; Figure~\ref{fig:H2Oadiabats}), and we consider only $<$8~GPa oceans. We go beyond \citet{Noack2016} by considering the effect of C on climate, by using an \emph{N}-body code to calculate how volatile delivery and giant impacts ``load the dice'' by regulating the~fraction of planets that can have cycle-independent planetary habitability, and -- most importantly -- by tracking $T_{surf}$ and $p_{\mathrm{CO2}}$ for 10 Ga on the habitable worlds we model.

Most of the physical and chemical processes we discuss have been investigated previously in an Earth context. The novel aspect of our paper is that we apply these ideas to exoplanet waterworlds (Section 2).

%\textbf{Maybe add dots on this (or a copy of this) figure to show where planets emerging from the simulations plot.} 

\section{How to model habitability on waterworlds.}

\emph{Summary}. Waterworlds are buffered against H$_2$O loss (Section 2.1), but because CO$_2$ modulates surface temperature and thus surface habitability (Section 2.2), we need to consider the chemistry that sets ocean pH and thus CO$_2$ (Section 2.3). Fortunately, nature makes this task easier on waterworlds (Section 2.4), by chemically isolating the ocean+atmosphere from the convecting mantle after 10$^8$ yr. This isolation is a necessary condition for cycle-independent planetary habitability.

\subsection{Waterworlds are buffered against H$_2$O loss}

In a CO$_2$-H$_2$O atmosphere, high surface temperatures will cause high H$_2$O mixing ratios in the stratosphere: this is referred to as a moist greenhouse state. In the moist greenhouse, H$_2$O molecules are photolyzed by $\lambda$~$\lesssim$~240~nm photons, producing H$_2$. This H$_2$ is subsequently lost in a hydrodynamic outflow of the upper atmosphere, an outflow that is underpinned by absorption of $\lesssim$~100 nm photons \citep{Kasting1988}. Loss of H$_2$ from H$_2$O tends to dry out the planet, and so moist-greenhouse worlds are outside the ``conservative HZ'' limits of \citet{Kasting1993}. 

However, for an initial water endowment of $>$50 Earth oceans in the HZ of an FGK star, the moist greenhouse cannot cause complete ocean loss \citep{Kasting1988,Lammer2009,LugerBarnes2015,ZahnleCatling2017} (Figure~\ref{fig:toymodels}). For example, integrating the XUV-flux estimate of \citet{Lammer2009} for 1 AU and a G-type star gives a loss of no more than 20~Earth~oceans over 4.5~Ga. This is small compared to the water available on waterworlds. Moreover, the XUV-limited water loss is an upper bound, as other processes likely limit the water loss to a lower value (e.g. \citealt{WordsworthPierrehumbert2013,Kulikov2006,Tian2015,OwenAlvarez2016,Bourrier2017}). 

In principle, oceans can dry out if ocean water reacts with rocks to form hydrated minerals \citep{Mustard2018}. In practice, for waterworlds that have solid rock interiors, atmosphere-interior exchange of H$_2$O is not important for setting the surface water inventory. That is because solid Earth mantle rocks have an H-storage capacity of only 1-10~Earth~oceans (e.g.~\citealt{Hirschmann2016}); this is a small fraction of the total water available to a waterworld (\citealt{TikooElkinsTanton2017}; for a contrary view, see \citealt{Marty2012}). For such worlds, the ocean a planet has after cooling from the last giant impact is the ocean that planet will keep.

\subsection{Shallow-ocean worlds: vulnerable unless geochemical cycles maintain habitability.}

Planetary habitability is modulated by CO$_2$ (Figure~\ref{fig:co2matters}). Very high partial pressures of atmospheric CO$_2$ ($p_{\mathrm{CO2}}$) lead to temperatures too high for life. However, intermediate $p_{\mathrm{CO2}}$ can stave off global surface ice cover (up to a low-insolation limit where CO$_2$~must~condense) and thus extend habitability (Figure~\ref{fig:co2matters}) \citep{Kopparapu2013}. $p_{\mathrm{CO2}}$ would have varied widely if Earth had always lacked a negative feedback on $p_{\mathrm{CO2}}$ \citep{KastingCatling2003}. On Earth, $>$90\% of C is stored in rocks, and C cycles between the atmosphere+ocean+biosphere and rocks every~$\sim$300~Kyr in the modern era \citep{Knoll2012} -- this is geologically rapid. As a result of this rapid geochemical cycling, a small initial imbalance between the rate of C release from rocks (volcanic/metamorphic outgassing) and C uptake into rocks (weathering uptake) could lead to global surface ice cover (a snowball) or alternatively lead to a moist greenhouse. An initial moist greenhouse with XUV-limited ocean loss would lead to loss of Earth's ocean water (Figures ~\ref{fig:toymodels} and \ref{fig:toymodels2}). Shallow-ocean rocky planets are vulnerable, unless geochemical cycles provide a negative feedback (e.g. carbonate-silicate weathering feedback) that can maintain habitability.\footnote{Doom is not assured for a hypothetical shallow-ocean world without a negative feedback. Escape hatches include the following. (1) Neither the moist greenhouse transition nor the snowball transition is completely understood, and it is possible that they can self-arrest or reverse \citep{Abbot2011,Abe2011,Abbot2012, Hu2011,Kodama2015,Hoffman2017,Kodama2018}. (2) The desirable $p_{\mathrm{CO2}}$ for a planet near the inner edge of the HZ is small. For such a world, vigorous C uptake, feeble C outgassing, and an M-star host -- if combined -- could allow $\gg$10 Gyr of habitability.} 

For worlds with oceans a few times deeper than the Earth's ocean -- enough to drown the land, but not as deep as the oceans we model in this paper -- it is more difficult to equalize C uptake into rocks and C outgassing \citep{Foley2015,Abbot2012}. Even though C uptake by weathering of seafloor rocks might make up some of this imbalance \citep{Coogan2016,CooganGillis2018}, 
it is less likely that a small initial imbalance between C release and C uptake could be restored via a negative feedback on worlds with oceans a few times deeper than the Earth's.

\subsection{Calculating waterworld $p_{\mathrm{CO2}}$\rm{(}$t$): \emph{habitability is \\ strongly affected by ocean chemistry}}

Although waterworlds are buffered against water loss (Section 2.1), the surface temperature of that water -- icy, heat-sterilized, or somewhere in between? -- depends on $p_{\mathrm{CO2}}$ (Figure~\ref{fig:co2matters}). $p_{\mathrm{CO2}}$ on waterworlds is set by atmosphere-ocean partitioning of C. The ocean is a major reservoir of C that equilibrates with the atmosphere on $<$10$^8$~yr~timescales. For example, Earth's pre-industrial partition of C is 60 parts in the ocean for one part in air, with an atmosphere-ocean exchange time of $\sim$10$^3$ yr. 

%Low values of $p_{\mathrm{CO2}}$ (including zero) permit global glaciation near the outer edge of the habitable zone, and very high values of $p_{\mathrm{CO2}}$ lead to uninhabitably high $T_{surf}$. 

\begin{figure}
%\epsscale{1.2}
\begin{centering}
\includegraphics[width=0.8\columnwidth,clip=true,trim={5mm 60mm 8mm 50mm}]{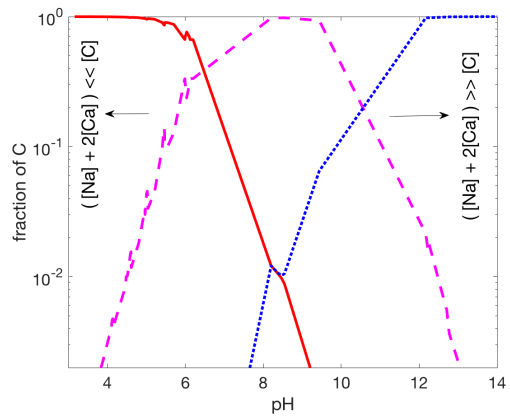}
\caption{Showing C distribution within the aqueous phase (Bjerrum plot). The solid red line corresponds to the fractional abundance of dissolved CO$_2$, the dashed magenta line corresponds to the fractional abundance of HCO$_3^-$, and the dotted blue line corresponds to the fractional abundance of CO$_3^{2-}$. Square brackets denote molalities; Ca and Na are examples of cations. Small wiggles in the curves are interpolation artifacts. $T$ = 0.1$^\circ$C.} 
\label{fig:bjerrum}
\end{centering}
\end{figure}

Therefore, $p_{\mathrm{CO2}}$ depends on C abundance, planet water mass fraction ($\equiv$ dilution), and ocean pH.  pH matters because of the strong effect of H$^+$ activity on carbonate equilibria (Figure~\ref{fig:bjerrum}). pH $\approx$ -log$_{10}$[H$^+$], where the square brackets here denote molarity. (Despite the similar notation, pH has nothing to do with the partial pressure of H$_2$.) $p_{\mathrm{CO2}}$ in the atmosphere is in equilibrium with dissolved CO$_2$ in the ocean:

\begin{equation}
\mathrm{\emph{p}_{CO2}} = \frac{\mathrm{[CO_{2(aq)}]}}{k_H}
\end{equation}

\noindent where the square brackets here denote concentration, and $k_H$ is a solubility constant. However, $p_{\mathrm{CO2}}$ can be much less than expected by dividing the total inorganic C (i.e.,~[C]) in the ocean by $k_H$. Specifically, if the pH~is~high enough to form HCO$_3^-$ or CO$_3^{2-}$ at the expense of CO$_{2(aq)}$ (Figure~\ref{fig:bjerrum}),

%\begin{equation}
\begin{equation}
\mathrm{(CO_2(aq)\!+\!H_2O\!\rightleftharpoons\!H_2CO_3)
\!\rightleftharpoons\!H^+\!+\!HCO_3^{2-}
\!\rightleftharpoons\!2H^+\!+\!CO_3^{2-}}
 \end{equation}
%\end{equation}

\noindent then $p_{\mathrm{CO2}}$ can be very low.

In this paper, we will refer to the sum of the electrically neutral species $\mathrm{(CO_2(aq)}$ and $\mathrm{H_2CO_3)}$ as CO$_{2\emph{T}}$; in Earth seawater, the concentration of $\mathrm{H_2CO_3)}$ is $\lesssim$0.3\% that of $\mathrm{CO_2(aq)}$.

When ocean pH~is~high (H$^+$ concentration is low), then by Le~Chatelier's principle carbonic acid gives up its H$^+$. Thus, at high pH, C is hosted in the ocean as CO$_3^{2-}$ and HCO$_3^-$, $\mathrm{CO_{2(aq)}}$ is low, and by Equation 1, $p_{\mathrm{CO2}}$ is low. In this case, the~fraction of total~C in the atmosphere is small (Figure~\ref{fig:bjerrum}). This describes the modern Earth \citealt{ZeebeWolfGladrow2001,RidgwellZeebe2005,Butler1982,Zeebe2012}). In other words, high pH effectively sequesters CO$_2$ from the atmosphere by driving the carbonate equilibria to the right. This raises the ratio [$\mathrm{CO_3^{2-}}$]/[$\mathrm{CO_{2(aq)}}$]  (Equation 2), and so decreases the $p_{\mathrm{CO2}}$ in equilibrium with the ocean (Equation 1). For a fixed total atmosphere+ocean C inventory, $p_{\mathrm{CO2}}$ must fall. Conversely, when ocean pH~is~low (H$^+$ concentration is high), then by Le~Chatelier's principle, C in the ocean exists mostly as CO$_{2\emph{T}}$, and the~fraction of total~C in the atmosphere is large.

What sets pH? Ocean pH rises when dissolved rock (e.g, Ca$^{2+}$, Na$^+$) is added to the ocean. That is because charge balance is maintained almost~exactly in habitable oceans. Adding a~mole of Na$^{2+}$ to an ocean relieves one~mole of H$^+$ from their duty of maintaining charge balance, so they revert to H$_2$O. Loss of H$^+$ raises pH, and sucks C out of the atmosphere (Equation 1, Figure~\ref{fig:bjerrum}). This effect of dissolved rock on $p_{\mathrm{CO2}}$ does not require carbonate minerals to form. However, C can exist mostly as solid CaCO$_3$ if [Ca] is high (or even as solid Na$_2$CO$_3$ minerals, if [Na] is extremely high). The relationship between pH and $p_{\mathrm{CO2}}$ and pH is an equilibrium, and so it equally true to think of $p_{\mathrm{CO2}}$ as driving pH. 

\begin{figure}
   \includegraphics[width=1.0\columnwidth,clip=true,trim={15mm 50mm 15mm 50mm}]{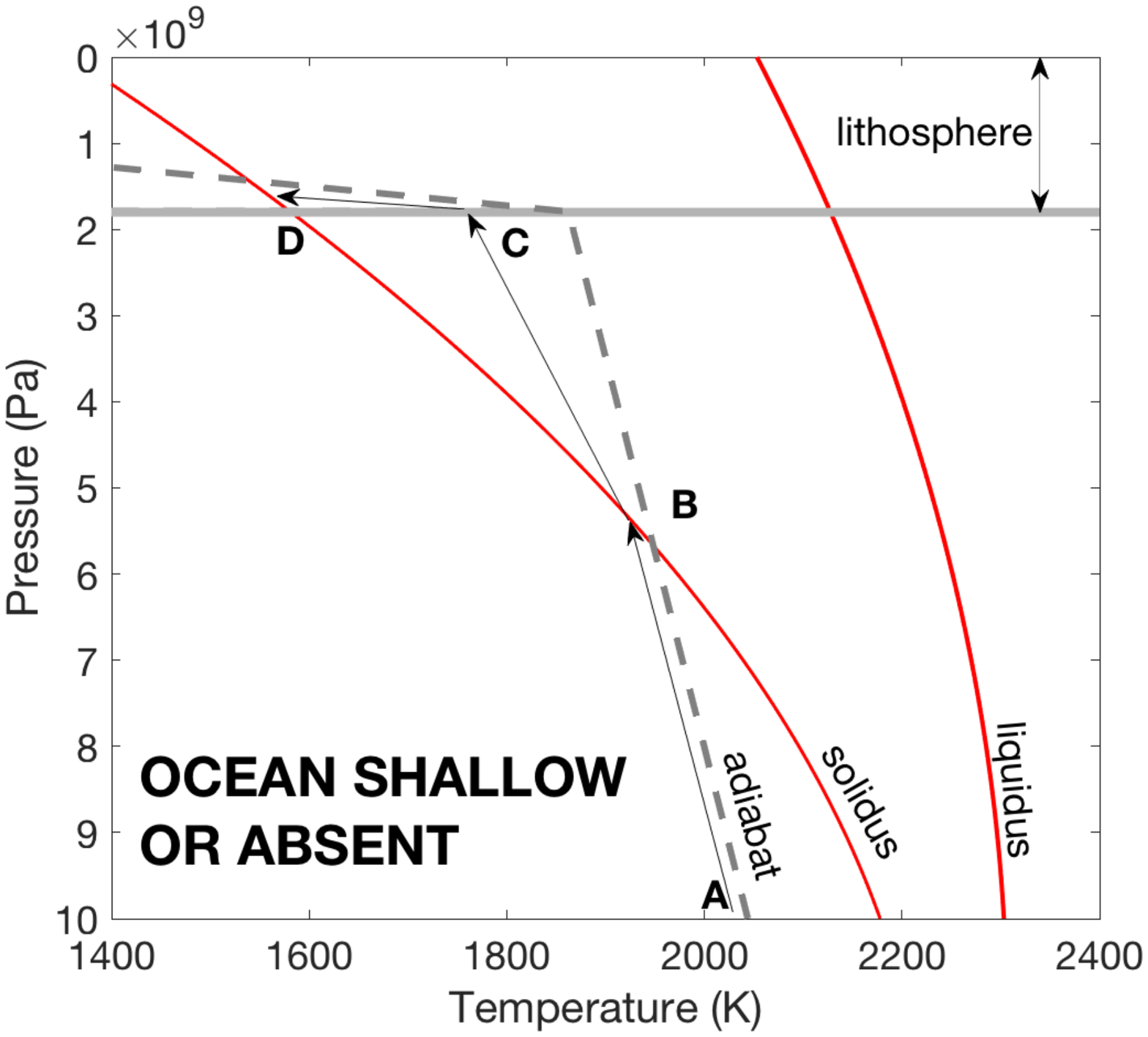} %
     \includegraphics[width=1.0\columnwidth,clip=true,trim={15mm 50mm 15mm 50mm}]{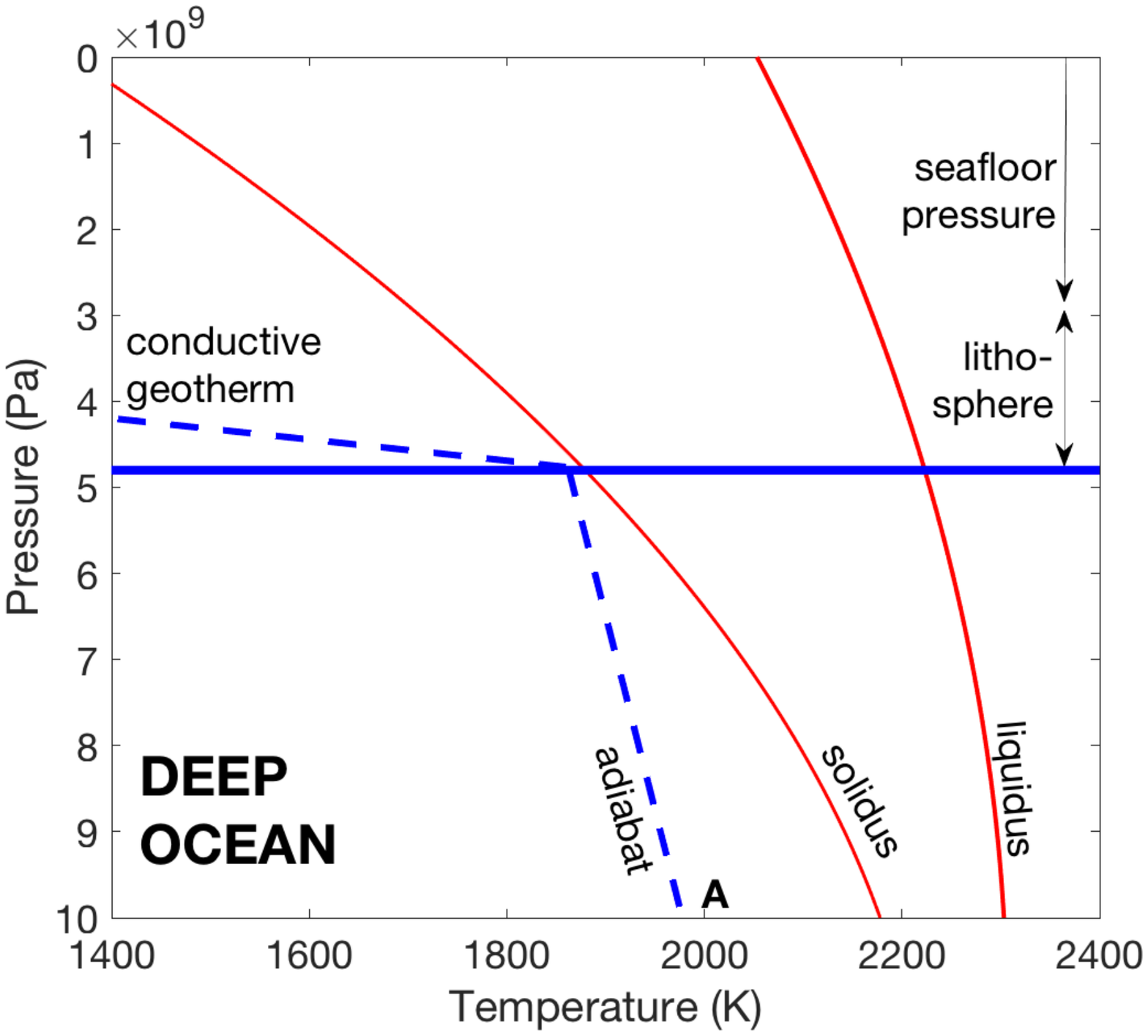} %\\
\caption{How the weight of water inhibits silicate volcanism on stagnant-lid waterworlds (and thus inhibits C exchange between the convecting mantle and the water ocean) \citep{Kite2009}. This figure is for stagnant-lid worlds, but similar arguments apply for plate tectonic worlds. In the \emph{upper panel}, the arrowed thin line A$\rightarrow$B$\rightarrow$C$\rightarrow$D tracks the partial (batch) melting of a parcel of initially solid mantle rock in the rising limb of a solid-state mantle-convection cell beneath a stagnant-lid lithosphere on a shallow-ocean world. From A$\rightarrow$B, the path parallels the solid-state adiabat (dashed gray line). At B, melting begins, and increases until (at C) the parcel reaches the base of the 60~km-thick stagnant~lid (thick gray horizontal line). Further ascent (if any) is slow, and conductive cooling shuts down further melting (at D). For a stagnant-lid planet with a 3~GPa ocean (\emph{bottom panel}), no melting occurs. The dashed blue line is an adiabat displaced downwards, such that the sub-lithospheric temperature is the same as in the no-ocean case. For details, see Appendix~A.
}
\label{fig:volcaniccutoff}
\end{figure}

Assuming constant atmosphere+ocean C content, when $T$ rises ocean pH will fall. This fall is due to changes in $k_H$, the analogous equilibrium constants in Equation 2, and increasing self-dissociation of water.

A general model for $p_{\mathrm{CO2}}$($t$) would be complicated \citep{Holland1984,HayesWaldbauer2006,KrissansenTotton2018}. Complexity occurs because $p_{\mathrm{CO2}}$ is coupled to pH, pH~is affected by rock-water reaction, and the controls on rock-water reactions are complex and evolve with time. Fortunately, for waterworlds, the problem is simpler. 

 \begin{figure*}
\epsscale{1.2}
\plotone{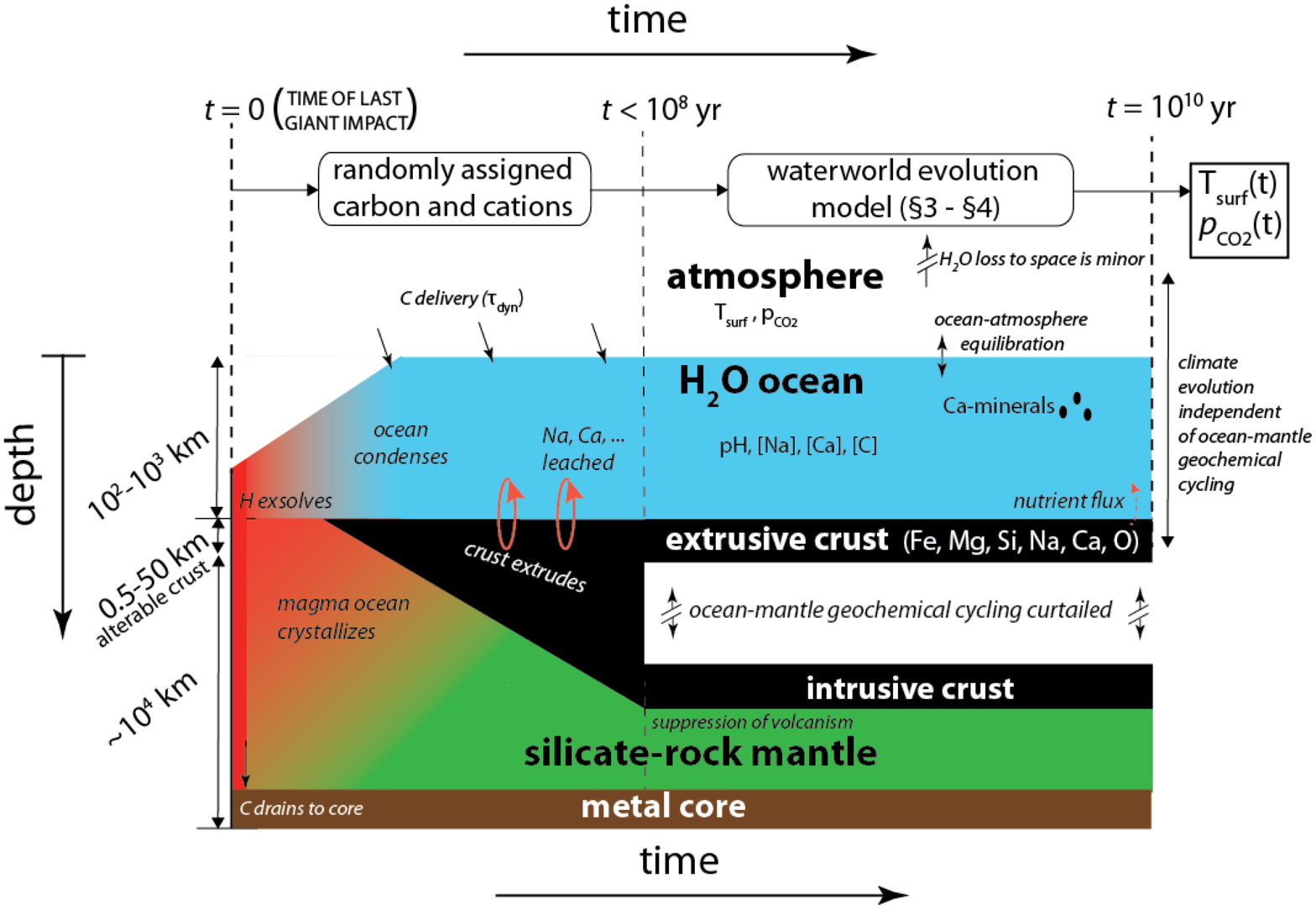}
\caption{The flow of our waterworld evolution model. Seawater composition is set in the first 10$^8$~yr (Section 3.1). Then, the atmosphere+ocean+(shallow crust) are isolated and the climate evolution is calculated (Section 3.2). Reservoirs (\textbf{bold}), processes (\emph{italics}), and variables (regular text) that are discussed in the text are shown. Cycle-independent planetary habitability is enabled by cessation of ocean-mantle geochemical cycling after $\sim$10$^8$~yr on waterworlds. See Section 6.4 for discussion of nutrient flux.}  %Compare Figure~5 in \citet{Zahnle2007} and Figure~9 in \citet{Abe2000}.} % $a, f_W and M_{pl}$ can be imposed, or obtained from the waterworld assembly model (Section 5).
\label{fig:codeflow}
\end{figure*}

\subsection{$p_{\mathrm{CO2}}$\rm{(}$t$) \emph{modeling is simplified by the waterworld approximation}}

The waterworld climate evolution problem is simplified by five factors. Each factor tends to reduce the number of fluxes that we have to model. The cumulative effect of these five factors is to uncouple climate evolution from ocean-mantle geochemical cycling (Figure~\ref{fig:codeflow}). The most important factor is that C exchange between the convecting mantle and the water ocean is limited on waterworlds -- Factor \#4. This is because seafloor pressure on  waterworlds curtails adiabatic decompression melting, which is the dominant mechanism of volcanism on rocky planets \citep{Kite2009} (Figure~\ref{fig:volcaniccutoff}) (Appendix~A). The five factors, which we collectively term the ``waterworld approximation,'' are as follows.

\begin{enumerate}

\item \textbf{H$_2$ is negligible.}
\vspace{-0.075in}

\item \textbf{Water loss to space is small.} %Following on from (3), $\partial  / \partial t \approx$ 0. 
\vspace{-0.075in}

\item \textbf{H$_2$O in the atmosphere+ocean greatly outweighs H$_2$O species in the silicate mantle.}
\vspace{-0.075in}

\item \textbf{C exchange between the deep mantle and the water ocean shuts down within \emph{O}(10$^8$) yr after the last giant impact.} 
\vspace{-0.075in}

\item \textbf{No land.}  

\end{enumerate}

\noindent We explain each factor below.

\renewcommand{\arraystretch}{0.99}% Tighter
\begin{deluxetable*}{lllrrcrrrrr}
\tabletypesize{\scriptsize}
\tablewidth{0pt}
\tablecaption{Parameters and variables.}
\tablehead{
\colhead{Parameter}              &
\colhead{\pbox{7.7cm}{Description}}           &
\colhead{Value}          & 
\colhead{\pbox{3.0cm}{Units}} }
\startdata
$a$ & semimajor axis & & AU\\
$a_{sl}$ & anchor for the evolving snowline & 1.6 & AU\\
$C$ & atmosphere+ocean C in CO$_2$-equivalent & & kg m$^{-2}$\\
$c_C$ & CO$_2$-equivalent C mass in atmosphere+ocean & & fraction of planet mass \\
$c_{p,r}$ & specific heat of rock & & J kg$^{-1}$ K$^{-1}$\\
$c_{p,f}$ & specific heat of aqueous fluid & 3800 & J kg$^{-1}$ K$^{-1}$\\
$\mathrm{[C],[Na],[Ca]}$ & aqueous concentrations (molalities) of C; Na; Ca\tablenotemark{*}& & mol (kg H$_2$O)$^{-1}$ \\
$f_W$ & \parbox[t]{8cm}{H mass fraction of the atmosphere+ocean+crust+ \\(silicate mantle), excluding any H in the core, \\ multiplied by 9 to get H$_2$O-equivalent mass fraction} & & fraction of planet mass \\
$f_{W,max}$ & maximum $f_W$ in planetary embryos & & fraction of embryo mass\\
I/E & intrusive:extrusive volume ratio (lavas vs. sills/dikes) &  & - \\
$k_{cond}$ & thermal conductivity of ocean & &  W m$^{-1}$ K$^{-1}$ \\
$L_*$ & stellar luminosity at planet & & W m$^{-2}$ \\
$M_{pl}$ & planet mass & & Earth masses \\
$P$ & pressure & & Pa, or bars  \\
$P_{lith}$ & pressure at the base of the lithosphere& &\\
$p_{\mathrm{CO2}}$ & partial pressure of atmospheric CO$_2$ & & bars  \\
%$Q_s, Q_r$ & energies & & J/kg \\
$R_{pl}$ & planet radius & & km \\
$T$, $T_{surf}$ & temperature; ocean-surface temperature & & K, or $^\circ$C &\\
$T_p$ & silicate-mantle potential temperature & & K\\
$t$ & time &  & Gyr \\
$v_\infty$ & velocity at infinity (pre-giant-impact) &  & m s$^{-1}$ \\
W/R & effective water/rock ratio of water/rock reactions  &  & - \\
$z_{cr}$ & crust thickness  & & \\
$\Delta Z$ & ocean depth & & km \\
$\Delta z_{bl}$ & thickness of conductively cooling layer(s) within ocean & & km \\
%$\mu$ & reduced mass & &  \\
$\tau_c$ & timescale for steam-atmosphere cooling & & Gyr \\
$\tau_x$ & timescale for crust formation & & Gyr \\
$\tau_{dyn}$ & timescale for C delivery & & Gyr\\
$\tau_{hab}$ & duration of habitable surface water &  & Gyr \\
$\tau_{unmix}$ & mixed-layer exsolution timescale & & Gyr \\
$\rho_{cr}$ & crust density & 3000 &  kg m$^{-3}$\\
$\rho_{w}$ & ocean density & 1300 & kg m$^{-3}$\\
%\hline \\
\tablenotetext{*}{We use mol/kg interchangeably with mol/(kg H$_2$O), which introduces small errors that are acceptable for our purposes.}
%\tablerefs{}
\end{deluxetable*}
\renewcommand{\arraystretch}{1.0}% Tighter

%\subsection{Justification of the waterworld approximation.}

 \vspace{0.1in}

\noindent 1. \textbf{H$_2$/He is negligible.} H$_2$ blankets of the appropriate thickness to prolong habitability within the HZ of \citet{Kopparapu2013} require fine-tuning to form. They also swiftly escape to space (e.g. \citealt{Wordsworth2012,OwenMohanty2016,Odert2018}; but see \citealt{RamirezKaltenegger2017}). While H$_2$-rich low-density planets likely exist in the HZ \citep{Rogers2015,vanEylen2018}, we expect that they will have uninhabitably hot rock-volatile interfaces (``surfaces'') \citep{Rogers2012} and we do not model them. We do not track H$_2$ outgassing by Fe~+~H$_2$O~$\rightarrow$~FeO~+~H$_2$. In the context of modern planet formation models, this reaction requires mixing of Fe with the volatile envelope during collisions between planets and planetary embryos. However, planetary embryos contain metal cores of $>$500~km diameter, which merge quickly (\citealt{DahlStevenson2010, Jacobson2017}, but see also \citealt{Genda2017}). We assume that the mantle is sufficiently oxidized that any volcanically outgassed C is in the form of CO$_2$, and that H$_2$ outgassing is minor. Therefore, we consider an \mbox{H$_2$-free}, CO$_2$+H$_2$O($\pm$N$_2$) atmosphere.
 
 \vspace{0.1in}
 
 \noindent 2. \textbf{Water loss to space is small,} relative to the initial water complement, for waterworlds orbiting FGK stars. This is explained in Section 2.1.
 
 \vspace{0.1in}

\noindent 3. \textbf{H$_2$O in the atmosphere+ocean outweighs ``H$_2$O'' in the silicate mantle.} Upon magma-ocean crystallization shortly after the last giant impact, the silicate-magma ocean exsolves H$_2$O \citep{ElkinsTanton2011}. The H$_2$O concentration that is retained in the now-frozen magma (= silicate rock) is limited by the upper mantle silicate minerals' H storage capacity.\footnote{Assuming that the solid mantle convects sufficiently vigorously that each parcel of mantle passes through the upper mantle at least once (which is reasonable; \citealt{Schubert2001}), H$_2$O contained in deeper hydrous phases will be rejected from the crystal during phase transition on ascent.} This storage capacity is small relative to the total water on waterworlds  (\citealt{Hirschmann2006,CowanAbbot2014}, but see \citealt{Marty2012} for a contrary view). As a result, the water in the ocean greatly exceeds the water in the silicate mantle, so any subsequent cycling (e.g. \citealt{SchaeferSasselov2015,KomacekAbbot2016,Korenaga2017}) will have little (fractional) effect on ocean depth. Hypothetical rock-hydration feedbacks (e.g. \citealt{KastingHolm1992}), if they are real, would also be limited by the small mantle storage capacity for H. Moreover, rock-hydration feedbacks do not reduce ocean depth on worlds that have oceans much deeper than on Earth \citep{Korenaga2017}. For seafloor pressure $>$~10~GPa, the H$_2$O storage capacity of mantle rock just beneath the stagnant~lid jumps to $\sim$0.5-1.0 wt\% \citep{Ohtani2005}. Even for such high seafloor pressures, however, the dominant H reservoir is still the ocean; this is because the H$_2$O storage capacity of average mantle rock is still $\ll$0.5~wt\%. Moreover, in this paper we only consider $<$~8~GPa seafloor pressures. To sum up, after the magma ocean has crystallized, ocean depth can be considered as a  constant.
 \vspace{0.1in}

\noindent 4. \textbf{C exchange between the convecting mantle and the water ocean shuts down within \emph{O}(10$^8$) yr after the last giant impact.}
To shut down C~cycling between a planet's mantle and ocean, it is necessary to shut down volcanic outgassing. For this to occur, it is sufficient to shut down volcanism. (Other modes of outgassing such as geysers are ultimately driven by volcanism). Volcanism is curtailed by seafloor pressure (Figure~\ref{fig:volcaniccutoff}) \citep{Kite2009}. The seafloor pressure threshold needed to curtail volcanism depends on how the mantle convects, and is smaller for stagnant-lid convection  ($\sim$1-3~GPa). We analyze this effect in Appendix~A. For simplicity, we impose an abrupt 1~GPa cutoff on volcanism. In reality some volcanism probably continues above this seafloor pressure, but (we assume) not enough to buffer atmosphere+ocean C. Because volcanism is intimately connected to plate~tectonics, it is plausible that with little or no volcanism there is little or no tectonic resurfacing \citep{Sleep2000}. Without tectonic resurfacing, the planet's mantle convects the same way as the mantles of Mars, Mercury, and the Moon:  \mbox{stagnant-lid}~worlds \citep{ORourkeKorenaga2012,Noack2017,Dorn2018}. In \mbox{stagnant-lid} mode, the mantle's convecting interior is cooled from above by conduction across a stable, very viscous boundary layer (the stagnant~lid) \citep{Stevenson2003}. We assume stagnant-lid volcanism in this paper, with little or no role for plate~tectonics. Plate tectonics is considered in Section 6.5. %\footnote{It is possible that volcanic outgassing is shut down for pressures lower than those needed to shut down volcanism. Pressures~$>$1~GPa~prevent exsolution of CO$_2$ from magma. Instead, CO$_2$ is trapped upon crystallisation in glass. Glass does not persist for Gyr and devitrifies, but devitrification might result in crystals with as much or more C than in the parent magma.}.  

After volcanic outgassing has shut down, C is not released from the mantle into the ocean; neither is C~subducted into the deep mantle, because no-volcanism stagnant-lid tectonics lacks subduction. Therefore, we treat the atmosphere+ocean+(shallow lithosphere) as a closed system with respect to C after $10^8$ yr (Figure \ref{fig:codeflow}).

Even without sustained volcanism, water and rock still react on waterworlds. Most importantly, volcanism will occur $\lesssim$10$^8$~yr after the last giant impact, for example as the dregs of the magma ocean extrude. This volcanism is key to setting ocean chemistry, and is discussed in Section 3. Even after $\sim$10$^8$ yr, slow alteration of porous seafloor rocks can continue, due to diffusion and/or changes in seafloor $T$ that shift mineral equilibria. This slow process, which we do not model, supplies chemical energy and nutrients (Section 6.4). Finally, carbonate precipitation and dissolution at (or just below) the seafloor may occur. 

%Molality of Na for complete leaching

 \vspace{0.1in}
 
\noindent 5. \textbf{No land.}  Mountain-peak height is set by tectonic stress in competition with gravity-driven crustal spreading \citep{Melosh2011}. The tallest peaks in the solar system are $\sim$20~km above reference level. Such peaks are drowned on waterworlds, so there is no land. With no~land, there is no continental weathering.

\vspace{0.1in}

\noindent Now, for a given planet mass ($M_{pl}$), we can write

\begin{equation}
p_{\mathrm{CO2}}\!=\!p_{\mathrm{CO2}}(f_{W,t=t_0}, c_{C,t=t_0},  {\mathrm{[}X\mathrm{]}_{t=t_0}}, L_*(t),t)
\end{equation}

\noindent where $f_W$ is planet water mass fraction, $L_*$ is stellar luminosity, $[X]_{t=0}$ is initial ocean chemistry (including mineral phases in equilibrium with the ocean, if any), and $t$ is time in Gyr. The start time $t_0$ = 0.1 Gyr; our simulations have 0.1 Gyr timesteps (Figure~\ref{fig:codeflow}). Continental weathering flux does not appear in Equation 3, because we assert ``no land.'' Ocean-interior exchange flux does not appear in Equation 3, because we assert ``no \mbox{post-0.1 Ga} C exchange between the deep mantle and the water ocean.'' (Figure~\ref{fig:codeflow}). The only time dependence in Equation 3 is warming as the star brightens. If we can solve Equation (3) to get $p_{\mathrm{CO2}}(t)$, we can find the corresponding surface temperature (from Figure 2) and thus find the duration of habitable surface water as the star brightens."

\section{Description of the Waterworld evolution model}

We set out to solve Equation 3. The overall flow of~our~code is shown in Figure~\ref{fig:codeflow}. Our code has two main stages:

\vspace{-0.05in}
\begin{itemize}
\item set initial seawater composition: we assign ``initial'' C abundance and cation abundance at random from cosmochemically/geophysically reasonable ranges ($t$~$<$0.1~Gyr) (Section 3.1);
\vspace{-0.03in}
\item compute $T_{surf}(t)$ and $p_{\mathrm{CO2}}(t)$ ($<$0.1~-~10 Gyr) (Section 3.2).
\end{itemize}
\vspace{-0.1in}

%Waters interacting with rock are isolated from the ocean during hydrothermal activity, and the resulting fluids are diluted into the global ocean. Crystals forming within the ocean are not allowed to settle out (there is no equivalent to Earth's biological pump). To set internal structure, we let 

%\textbf{Add veins and label lower section (with braces) as veins and/or pervasive alteration}} 

%
%We provide a capsule summary -- the minimum needed to understand our main results.
%Readers familiar with ocean chemistry may skip to Section 3.2.

\subsection{Step 1: Set initial seawater composition}

In this subsection, we justify (1)~treating the atmosphere+ocean CO$_2$ mass fraction ($c_C$) as a free parameter, varying between planets from 10$^{-5}$~to~10$^{-3}$~of planet~mass, such that $c_C$/$f_W$ varies from 10$^{-6}$ to 10$^{-1}$; and (2)~using salinities $O$(0.1-1)~mol/kg. We define $c_C$ to be the mass of C in the atmosphere+ocean after the magma ocean has crystallized but before any carbonate formation (excluding C in the metal core or silicate mantle), divided by the mass of the planet, and multiplied by 44/12 to get the CO$_2$-equivalent mass fraction. In this paper we do not consider compositions that are very C-rich -- comet-like. Comets made only a minor contribution to Earth's volatile budget \citep{Marty2016}, but exoplanets might have comet-like compositions. Comet-like compositions require a more sophisticated treatment including clathrates and other C-rich phases \citep{Bollengier2013,Levi2014,Marounina2017,Levi2017,RamirezLevi2018}. Busy readers may skip to Section 3.2.

Seawater composition on stagnant-lid waterworlds is set by (1) delivery of CO$_2$, and (2) supply of cations via water-rock reaction in the 10$^8$ yr after the last giant impact.\footnote{Volatiles initially in the steam atmosphere are less important. To see this, consider that immediately after the last giant impact, volatiles and magma are in equilibrium \citep{Zahnle1988,Abe2000,Zahnle2015}. When the surface cools below $\sim$1600K, the surface can crust over. Volatiles that are in the atmosphere at this point might be taken up by the H$_2$O$_{(l)}$ ocean after further cooling. The most abundant non-CO$_2$ non-H$_2$O volatile in a 100-bar, 1600~K steam-atmosphere is HCl according to the calculations of \citep{Lupu2014}. HCl volume mixing ratio $\approx$~10$^{-2}$, i.e., small. If this atmosphere condenses into a hypothetical 100~km-deep ocean (the shallowest we consider), we obtain [Cl]~$\approx$~10$^{-2}$~mol/kg, which is small. Na species are even less abundant in the steam atmosphere (e.g. \citep{Schaefer2012}), and Na is more abundant in the planet as a whole, and is readily leached into the ocean.} 

\vspace{0.05in}
\noindent(1) \emph{C delivery}. C dissolves into liquid~Fe-metal in the immediate aftermath of giant impacts, when liquid~silicate and liquid~\mbox{Fe-metal} equilibrate \citep{KuramotoMatsui1996,Hirschmann2012,Dasgupta2013}. \mbox{C is so iron-loving} tfhat, even if 2\% of a 1~$M_\earth$~planet is composed of C (equivalent to 7~wt\%~CO$_2$!), only $<$100~ppm escapes the core (Figure~2~in~\citealt{Dasgupta2013}). The lure of the Fe-metal core for C was so strong that most of the C that existed in the Earth prior to the Moon-forming impact is believed to have been trapped in the core during the magma ocean era associated with giant impacts. Therefore, the C in our bodies (and in Earth's ocean) is thought to represent C delivered after the last giant impact \citep{Bergin2015}. C's behavior under giant impact conditions contrasts with that of H \citep{KuramotoMatsui1996}. H does not partition as strongly into liquid Fe-metal under giant impact conditions as does C. Therefore, in contrast to C, H now in Earth's oceans could have been delivered to the Earth prior to the last giant impact. Indeed, the H in Earth's oceans is thought to have arrived during the main stage of planet assembly (\citealt{OBrien2014,Rubie2015,OBrien2018}; see \citealt{Albarede2009} for an opposing view). Giant impacts are stochastic, so the ratio of CO$_2$-equivalent C mass in the atmosphere+ocean to the planet water content ($c_C$/$f_W$) is expected to vary stochastically \citep{Hirschmann2016} (Section 6.5). Another source of variation in $c_C$/$f_W$ is the varying extent (which we do not track) of the reaction Fe~+~H$_2$O~$\rightarrow$~FeO~+~H$_2$. Given these uncertainties, we vary~$c_C$ from 10$^{-5}$ to 10$^{-3}$, such that $c_C$/$f_W$ varies from~10$^{-6}$~to~$\sim$10$^{-1}$. $c_C$~=~10$^{-5}$ corresponds to an Earth/Venus-like C abundance, and we find that $c_C$~=~10$^{-3}$~all~but~ensures a $p_{\mathrm{CO2}}$~$>$~40~bar atmosphere, which we deem uninhabitable.

% In the immediate aftermath of giant impacts, the iron-nickel core is a sump for C \citep{Dasgupta2013,Bergin2015}.  For Earth, $c_C$ likely was supplied by a late veneer. Because the details of late-veneer dynamics and distribution-with-semimajor-axis are poorly known even for the solar system, we model $c_C$~as independent of $f_{W,max}$ %%\footnote{Just to simplify calculations, and not because we think it is cosmochemically required, we deliberately avoid $c_C$/$f_W$ ratios high enough that a C-rich fluid phase will separate from H-rich fluid ($>$5\% C; \citealt{Bollengier2013}).}

\vspace{0.05in}
\noindent (2) \emph{Supply of cations via water-rock reaction:} Rock with the composition of Earth's oceanic crust (CaO $\approx$ MgO $\approx$ FeO $\approx$ 10 wt\%, \citealt{Gale2013}) can form \{Ca,Mg,Fe\}CO$_3$ minerals, and potentially trap up to 70~(bars CO$_2$)/(km crust). To what extent is this potential realized? The fluid composition resulting from water-rock reactions $<$10$^8$~yr~after the last giant impact can be estimated (a)~by analogy to similar worlds, (b)~with reaction-transport models, and (c)~by geophysical reasoning about the mass of rock available for water-rock reactions.

(2a) Previous work on extraterrestrial oceans -- such as Europa and Enceladus -- suggests salinities $O$\mbox{(0.1-1)~mol/kg}. For example, \citet{Glein2015} and \citet{HandChyba2007} estimate salinity using spacecraft data; and \citet{McKinnonZolensky2003}, \citet{Zolotov2007}, and \citet{ZolotovKargel2009}, estimate salinity using models. 

(2b) We use the aqueous geochemistry thermodynamics solver CHIM-XPT \citep{Reed1998} to explore reactions between water and rock (Appendix~D). Our rock input for these runs has the composition of Earth's oceanic crust \citep{Gale2013}. Rocks release cations into the ocean that -- even when not forming carbonates -- can control ocean chemistry. Rock-sourced cations (e.g. Na$^+$ and Ca$^{2+}$)  substitute in the ocean charge balance for H$^+$. H$^+$ removal raises pH, which in turn raises the fraction of dissolved inorganic carbon that is stored as HCO$_3$ and CO$_3^{2-}$ rather than as dissolved CO$_2$ (Figure~\ref{fig:bjerrum}). As dissolved CO$_2$ decreases, $p_{\mathrm{CO2}}$~decreases in proportion (by~Henry's~Law). Two cations, Na$^+$ and Ca$^{2+}$, are used in our waterworld evolution model. Other cations contribute little to the charge balance of outlet fluid in our \mbox{CHIM-XPT}~runs (Appendix~D), and so are ignored. CHIM-XPT runs (for~$T \leq$ 600K) indicate that Mg and Fe likely form silicates, not carbonates, and so Mg and Fe do not draw down CO$_2$. Non-participation of \{Mg,~Fe\} cuts C uptake by a factor~of~$\sim$3. CaCO$_3$~forms, and buffers [C] in outlet fluid to $\lesssim$0.1 mol/kg. Complete use of Ca to form CaCO$_3$, with only minor MgCO$_3$/FeCO$_3$, matches data for carbonated Archean seafloor basalts \citep{NakamuraKato2004,Shibuya2012,Shibuya2013}. 

(2c) Geophysical arguments about the mass of rock available to react with fluid may be divided into:  (i)~eruption history arguments, and (ii)~pervasiveness arguments. We assume that rock-water reactions occur at~$T_{seafloor}$~$<$~650~K (for reasons given in Appendix~C).

(i) \emph{Eruption history}: Alteration by water is much easier for extrusive rocks (lavas) than for intrusive rocks (sills and dikes). That is because intrusives have \mbox{$\gtrsim$10$^3$-fold lower} permeability, which denies entry to altering fluids \citep{Fisher1998,Alt1995}. Intrusive:extrusive~ratios (I/E) on Earth are scattered around an average of $\sim$5:1 (6.5:1 for oceanic crust) \citep{White2006,WhiteKlein2014}. Extrusives less than 0.5~km below the~seafloor -- corresponding to the highest permeabilities in oceanic crust -- are the site of most of the alteration within the oceanic crust on Earth. Altered crust on an exoplanet could be much thicker than 0.5~km if the extrusive pile were thicker. A thicker extrusive pile is likely for stagnant~lid mode. In this mode, each lava erupts at the surface, cools and is rapidly altered, and then is buried by later lavas. Extrusives whose alteration leaves them completely leached of an element will give a waterworld ocean with an aqueous molality (denoted by square brackets) of 

\begin{equation}
\frac{23}{1000} \mathrm{[Na]} =   \mathrm{Na'}  \frac{z_{cr} \rho_{cr}}{ z_{oc} \rho_{w}}  (\mathrm{I/E})^{-1}
\end{equation}

\noindent where $z_{oc}$ is ocean thickness, $z_{cr}$ is crust thickness, $\mathrm{Na'}$~=~0.028 is the Na mass concentration in the crust \citep{Gale2013}, $\rho_{cr}$~=~3~$\times$~10$^3$ kg~m$^{-3}$ is crust density, and $\rho_{w}$~$\sim$~1300~kg~m$^{-3}$ is ocean density (this is the density of water at 1.8~GPa and 373K; \citealt{AbramsonBrown2004}). For $z_{oc}$~=~100~km, $z_{cr}$~=~50~km, and I/E~=~5, this gives [Na]~=~0.3~mol/kg.

%Eruption history is also discussed in \citep{McKinnonZolensky2003}.

(ii) \emph{Pervasiveness}:  Earth's modern oceanic crust is altered mainly near veins and cracks \citep{Alt1995}. However, more pervasive alteration of seafloor extrusive rocks took place under C-rich, perhaps warmer climates $\sim$3.4 Ga on Earth \citep{NakamuraKato2004}. Moreover, Earth's ocean pH is thought to have been moderate for 4 Ga (pH~=~7.5$\pm$1.5; \citealt{HalevyBachan2017,KrissansenTotton2018}), and either higher pH or lower pH would be more corrosive. 
Finally, water is highly corrosive as a supercritical fluid, and seafloor $T$ and $P$ on waterworlds can reach supercritical conditions (Figure \ref{fig:H2Oadiabats}).
Therefore, a wide range of alteration -- including pervasive alteration -- is possible on other worlds \citep{Vance2007,Vance2016,KelemenHirth2012,Neveu2015}. 

\vspace{0.07in}

Taken together, (2a)--(2c) suggest that, for waterworlds,

\begin{itemize}
\item both Na and Ca, but not Fe and Mg, participate in CO$_2$ drawdown if rocks interact with water;
\item the reasonable range of rock-layer thicknesses for complete/pervasive interaction with water is 0.1~km (Earth-like) to 50~km. The biggest value corresponds to an unusually low I/E = 1:1, pervasive alteration of extrusives, and a crust thickness of 100~km, which is large relative to models and relative to solar system data \citep{TaylorMcLennan2009,ORourkeKorenaga2012,Plesa2014,Tosi2017}. The largest values might be thought of as including weathering of impactors and meteoritic dust. For all but the very largest impacts,  impact-\emph{target} materials \citep{SleepZahnle2001} will not contribute cations, because the ocean will shield the seafloor from impacts. The only ejecta will be water and impactor materials \citep{Nimmo2008, Marinova2008}. 
\end{itemize}
The corresponding cation abundances (for a 100~km-deep ocean, and neglecting Ca-mineral formation) are 0.003~--~1.4~mol/kg Na, and 0.008~--~4~mol/kg~Ca. Lumping Na and Ca as moles of equivalent charge (Eq), we obtain 0.02~--~9~Eq/kg. 

\subsection{Step 2: Compute $T_{surf}(t)$ and $p_{\mathrm{CO2}}(t)$}
 %Water in the air is not included in the ocean mass balance, because it is small.

$p_{\mathrm{CO2}}(t)$ is set by sea-surface equilibration between the well-mixed ocean and the atmosphere (Figure~\ref{fig:codeflow}). This equilibrium is rapid compared to the time steps in our model.

%This is because CO$_2$ solubility increases with pressure along an adiabat, and the ocean is mixed on a timescale $<<$1 Gyr. 
To find pH($T_{surf}$) and $p_{\mathrm{CO2}}(T_{surf})$, for each of a wide range of prescribed seawater cation content and dissolved inorganic C content, we use CHIM-XPT. We found (batch) equilibria in the system H$_2$O-HCO$_3^-$--Ca$^{2+}$-Na$^+$, including formation of calcite (CaCO$_3$) and portlandite (Ca(OH)$_2$). 

We consider the full range of geophysically plausible cation contents. In Section 4, we initially describe results for 3~cation~cases: (1)~[Na]~$\approx$~0.5 mol/kg, [Ca]~$\approx$~0 -- higher-pH, but lacks minerals; (2)~[Ca]~$\approx$~[Na]~=~0 -- a lower-pH case; (3)~[Ca]~=~0.25 mol/kg, [Na]~$\sim$~0 -- higher-pH. Many of the runs in case (3) form calcite (or, for pH $>$ 9, portlandite). We do not track the fate of Ca-mineral grains, which store most of the Ca in our higher-pH equilibrium model output. Depending on ocean depth and grain size, minerals might stay suspended in the water column, sink and redissolve, or pile up on the seafloor. Whatever the fate of individual grains, equilibrium with calcite buffers shallow-ocean (and thus atmosphere) CO$_2$ content for ocean chemistries that form calcite.

%[Update on how we calculate pCO2. Say that we do not consider the effect of salt on CO2 solubility, Perhaps use Lucile et al. J Chem Eng Data 2012 for NaOh solubility.] check that P does not affect (much) the~fraction of C in CO2aq.}.

%, where measurements are less unreliable according to \citet{DiamondAkinfiev2003}
To find CO$_2$ solubilities, we used tables of CO$_2$ solubility in pure H$_2$O, fit to experimental data. Specifically, we interpolated and extrapolated in Table 3 of \citet{Carroll1991} and Table 3 of \citet{DuanSun2003}, also using the constraint that CO$_2$~solubility~=~0 when  $p_{atm}$~=~$p_{\mathrm{H2O}}$. For $p_{\mathrm{H2O}}$, we used the formulation in Appendix~B of \citet{DuanSun2003}. CO$_2$ solubility is reduced by salts. For a 1~mol/kg solution at 10~bar and 303K, salting-out reduces solubility by 18\%  \citep{DuanSun2003}. However, we do not include the direct effect of Na$^+$ or Ca$^{2+}$ on CO$_2$ solubility in our model, reasoning that the pH effect is more important and that the range of C considered in our model is so large that the effect of the ions on CO$_2$ solubility is less important. Using the CO$_2$ solubilities and the carbonate system equilibria from the \mbox{CHIM-XPT} calculations, we compute $p_{\mathrm{CO2}}(T_{surf}$,~[Na],~[Ca],~[C]). Equipped with $p_{\mathrm{CO2}}(T_{surf},$~[Na],~[Ca],~[C]), we next calculate (for a given planet surface gravity, and fixing atmospheric molecular mass~=~44 g/mole) $p_{\mathrm{CO2}}(T_{surf}$,~[Na],~[Ca],~$c_C$). Here, $c_C$ is the atmosphere+ocean column C abundance and can (in principle) greatly exceed the ocean C-storage capability. To ameliorate interpolation artifacts, we smooth out each $p_{\mathrm{CO2}}(T_{surf})$ curve using the MATLAB \texttt{smooth} function with a 30 $^\circ$C full-width bandpass. (Appendix~E has more details about how ocean-atmosphere equilibration is calculated).

\begin{figure*}[ht]
\includegraphics[width=1.00\columnwidth,height=8.5cm,keepaspectratio,clip=true,trim={15mm 45mm 5mm 45mm}]{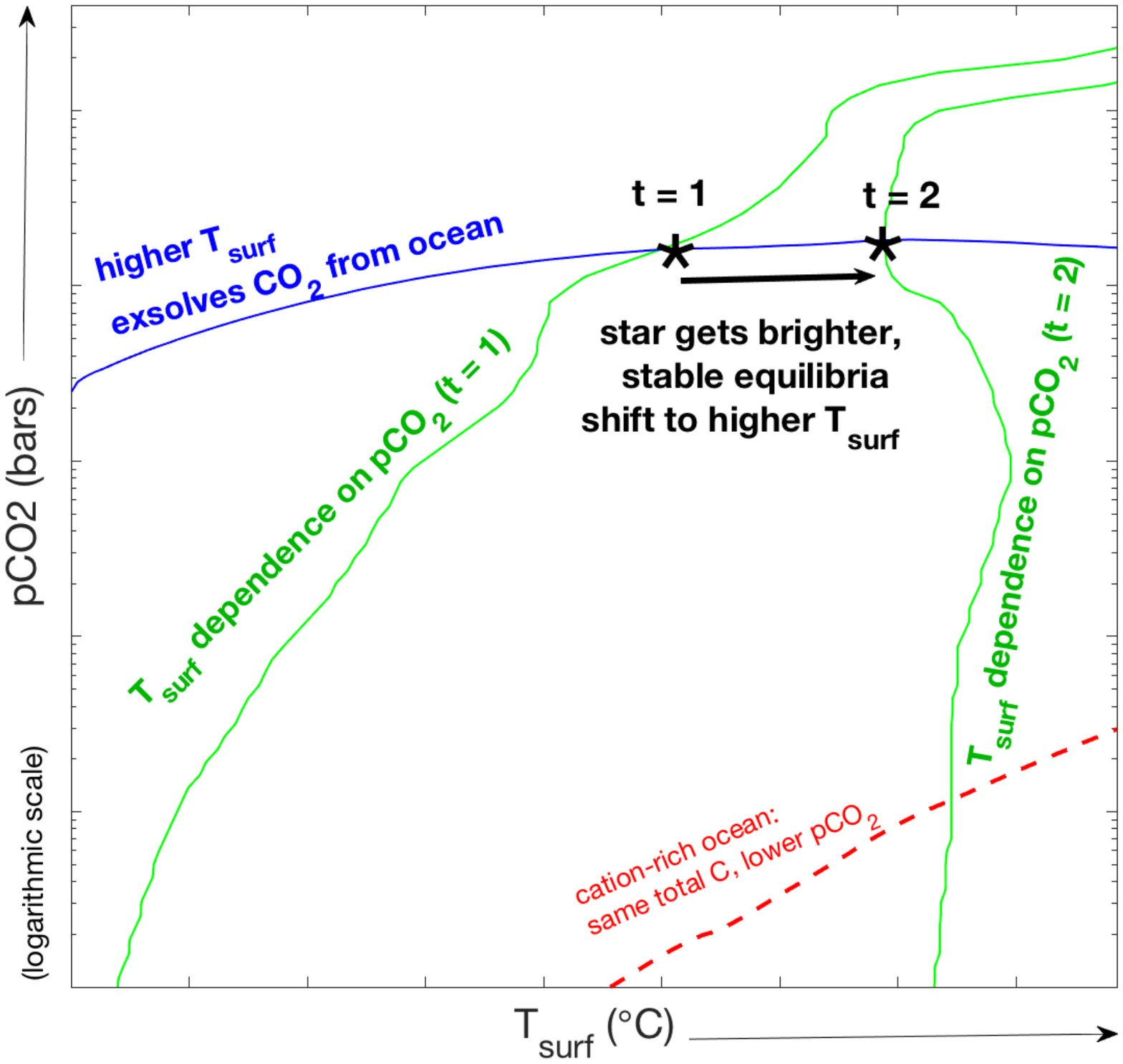} %was _AC
\includegraphics[width=1.00\columnwidth,height=8.5cm,keepaspectratio,clip=true,trim={0mm 35mm 5mm 30mm}]{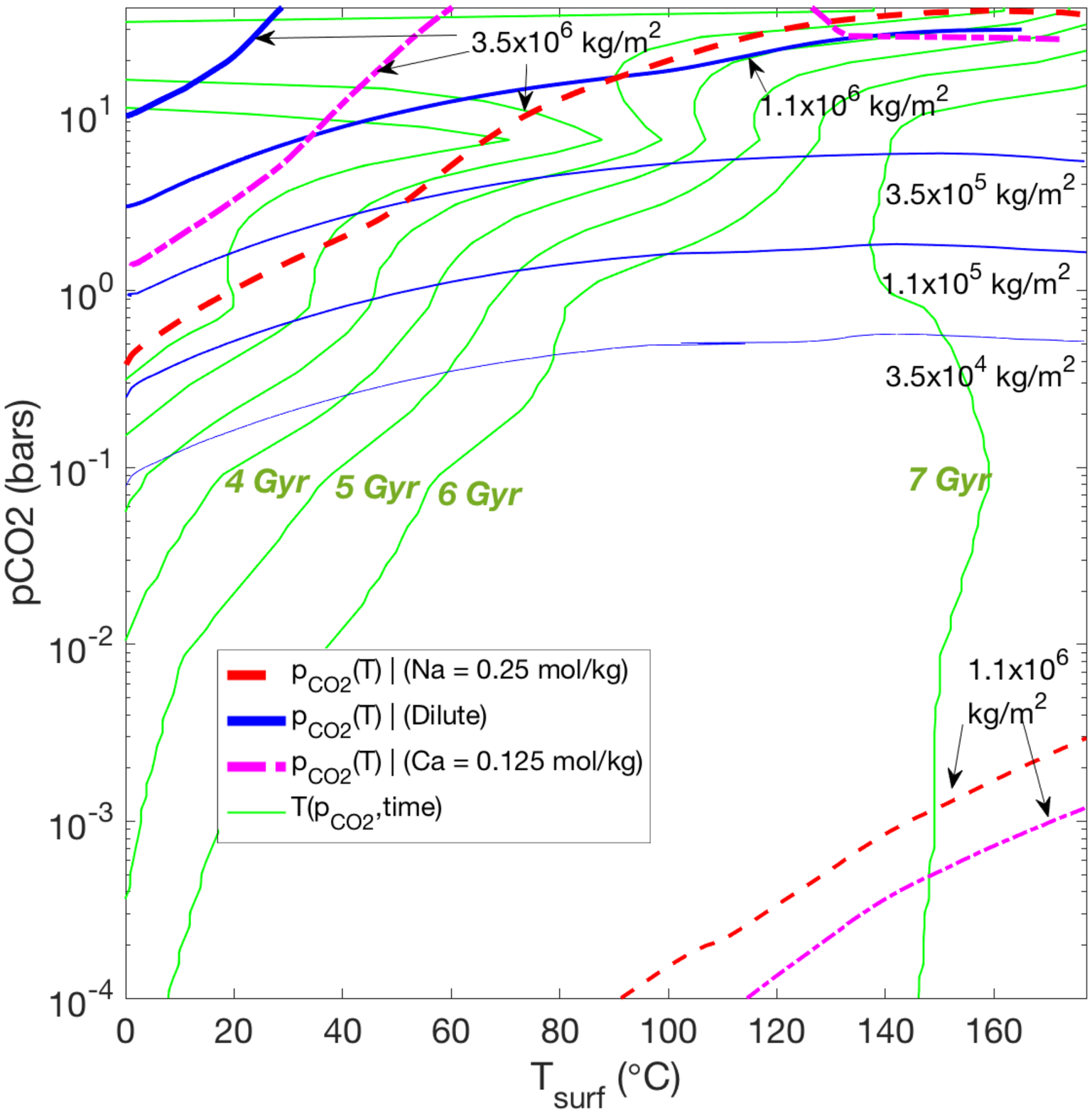} %was _AC
\caption{\emph{Left panel:} scheme for tracking equilibria between ocean chemistry and atmospheric $p_{\mathrm{CO2}}$. Equilibria shift as the insolation increases with host-star age on the main sequence ($t$). \emph{Right panel:} chart for finding equilibria between ocean chemistry and atmospheric $p_{\mathrm{CO2}}$ (planet water mass fraction~=~0.01, $a$~=~1.1~AU). Green lines correspond to $p_{\mathrm{CO2}}$($T_{surf}$) for fixed $L_*$, with $T_{surf}$ increasing as $L_*$ rises with star age on the main sequence (interpolated from \citealt{WordsworthPierrehumbert2013}; small wiggles are interpolation artifacts). The green lines are seperated at intervals of 1 Gyr in star age on the main sequence. The blue, red, \& magenta lines correspond to $p_{\mathrm{CO2}}$ as a function of~$T_{surf}$, for different assumptions about ocean cation content. The solid blue lines correspond to a cation-poor ocean, the dashed red lines correspond to a [Na]~=~0.25~mol/kg ocean, and the dash-dot magenta lines correspond to a [Ca]~=~0.125~mol/kg ocean. Line thickness increases with equal amounts of total atmosphere+ocean~C (including C in Ca-minerals, if any). Lines of the same thickness have the same total atmosphere+ocean C: \{3.5 $\times$10$^4$, 1.1 $\times$10$^5$, 3.5$\times$10$^5$, 1.1$\times$10$^6$, 3.5$\times$10$^6$\}~kg~m$^{-2}$ CO$_2$-equivalent~C are considered.}  
\label{fig:findequilibria}
\end{figure*}

Next, we interpolate the $T_{surf}(L_*, p_{\mathrm{CO2}})$ results of a 1D radiative-convective climate model \citep{WordsworthPierrehumbert2013}. Their study neglects clouds, assumes a fixed relative humidity of 1.0, and adjusts the surface albedo to a value (0.23) that reproduces present-day Earth temperatures with present-day CO$_2$ levels. (Comparison to the results of models that include clouds, and omit the albedo contribution of bright continents, indicates that this assumption will not affect the qualitative trends presented in the current paper; see Section 6.5.) We extrapolate the log-linear fit to $p_{\mathrm{CO2}}$~=~3~$\times$~$10^{-6}$~bar. $T_{surf}$ always increases with $L_*$. $T_{surf}$ usually increases with $p_{\mathrm{CO2}}$ as well, except near the outer edge of the HZ ($p_{\mathrm{CO2}}$ $>$ 8 bars in Figure~\ref{fig:findequilibria}). Near the outer edge of the habitable zone, adding CO$_2$ leads to cooling because the Rayleigh scattering of incoming shortwave radiation from extra CO$_2$ exceeds the greenhouse warming from extra CO$_2$.
Where \citet{WordsworthPierrehumbert2013} find multiple stable equilibria in their 1D atmospheric radiative-convective model, we use the warmer of their two solutions, on the assumption that impacts intermittently allow the atmosphere+(wave-mixed ocean) (which have relatively low thermal inertia) to jump onto the warm branch. Once the atmosphere+(wave-mixed ocean) have reached the warm branch, the deep ocean will gradually adjust to the new, higher temperature.
 To find ocean-atmosphere equilibria, we find \{$T_{surf}$, $p_{\mathrm{CO2}}$\} combinations that satisfy $L_*(t)$, [Na], and [Ca] (Figure~\ref{fig:findequilibria}), as explained in Section 4.1. Although the ocean $T$ varies with depth within the ocean (Figure~\ref{fig:H2Oadiabats}), the CO$_2$ solubility increases with depth, and so the ocean surface temperature $T_{surf}$  is the temperature that matters for the purpose of finding ocean-atmosphere equilibria. (Appendix~E gives more details, and shows the results of a sensitivity test using the 1D radiative-convective climate model of \citealt{Ramirez2014}).

At each timestep, we check for ice. Ice at the ocean surface (ice I) will have steady-state thickness $<$20~km (due to geothermal heat). We assume initially ice-covered surfaces have an albedo that declines due to dark exogenic contaminants (meteoritic dust) on a timescale $\ll$~10~Gyr, back to albedo~=~0.23. This value is used for convenience, as it is the surface+clouds albedo used by \citep{WordsworthPierrehumbert2013}. This lowering of initially high albedo is reasonable because, in our model, planets develop surface ice not at all, or very early. Early-developed ice cover will sweep up dark planet-formation debris \citep{Lohne2008}. The low albedo of debris is one of the reasons that all icy worlds in the solar system have dark surfaces except terrains which have been tectonically resurfaced and/or created after the stage of high impact flux in the early solar system. Lower albedo allows stellar luminosity increase to cause ice-covered surfaces to melt (Section 6.5). Melting with these assumptions always leads to a habitable state (this need not be true if CO$_2$ outgassing is allowed; \citealt{Yang2017,Turbet2017}). If ice cover develops later in planetary history, then initially high albedos can stay high, but this is very uncommon in our model. In addition to checking for surface ice, we also evaluate whether high-pressure ice phases (ice~VI, ice~VII) are stable at any depth between the ocean surface and the seafloor along a pure-liquid-H$_2$O adiabat (Appendix~B, Figure~\ref{fig:H2Oadiabats}). We do not track the climate or pH in detail when HP-ice is present (see \citealt{LeviSasselov2018}), and in particular we do not redistribute CO$_2$ between clathrates, the ocean, and the atmosphere. This is because we conservatively assume that if \mbox{HP-ice} is present, the world is not habitable.

\section{Results of Waterworld evolution model}

All of the results presented here are for a \mbox{Sun-like} (G-type) star with (log luminosity)-versus-time fit to a standard solar model \citep{Bahcall2001}. For this model, $L_*$~increases by $\sim$8\% Gyr$^{-1}$. We find that:

\vspace{-0.05in}
\begin{itemize}[leftmargin=0.5em]
\item waterworld climate is usually stable when $T$ is perturbed (Section 4.1);
\vspace{-0.07in}
\item habitable surface water can persist for many Gyr (Section 4.2);
\vspace{-0.07in}
\item using geologically and cosmochemically plausible priors in a parameter sweep, we estimate that $\sim$\nicefrac{1}{4}~of~waterworlds have habitable surface water for $>$1 Gyr (Section 4.3).
\end{itemize}
\vspace{-0.05in}

\subsection{How individual waterworlds evolve}

\vspace{0.07in}
\noindent \emph{Charting planet evolution.} With the waterworld approximation, the cations available to the ocean are constant with time after 0.1 Gyr. Thereafter,  $T_{surf}$ and the total amount of C in the atmosphere+ocean (including C in Ca-minerals, if any) are the sole controls on $p_{\mathrm{CO2}}$. $p_{\mathrm{CO2}}$~usually~increases with $T_{surf}$, due to the low solubility of CO$_2$ in warm water, and related $T$-dependent shifts in the carbonate system equilibria (Figure~\ref{fig:findequilibria}). $p_{\mathrm{CO2}} = p_{\mathrm{CO2}}(T_{surf})$ is a single-valued function for any cation and total-C concentration (Figure~\ref{fig:findequilibria}). $T_{surf} = T_{surf}(p_{\mathrm{CO2}})$ greenhouse curves (green lines) can be computed using a climate model given semimajor axis $a$ and $L_*(t)$. To find the $p_{\mathrm{CO2}}$ and $T_{surf}$ for a given time, we locate the green line corresponding to that star age, then find its lowest-$T_{surf}$ stable intersection with the appropriate $p_{\mathrm{CO2}}$($T_{surf}$) line (Figure~\ref{fig:findequilibria}).

As $L_*$ increases with time \citep{Bahcall2001}, $T_{surf}$~=~$T_{surf}$($p_{\mathrm{CO2}}$,~$L_*$) increases for a given $p_{\mathrm{CO2}}$. This is shown by the rightward drift of the green lines (greenhouse curves) in Figure~\ref{fig:findequilibria}. The green lines are closely spaced at $p_{\mathrm{CO2}}$ $\sim$0.2--20~bars, corresponding to a small $T_{surf}$ rise for a given increase in $L_*$ (Figure~\ref{fig:co2matters}). This prolongs the duration of habitable surface water (longer $\tau_{hab}$). By contrast the green lines are widely spaced at low $p_{\mathrm{CO2}}$, so a small increase in $L_*$ corresponds to a rapid $T_{surf}$ rise (Figure~\ref{fig:co2matters}). The $T_{surf}$ increase is usually more than would occur without ocean chemistry \citep{Kitzmann2015}, because the greenhouse gas CO$_2$ is usually exsolved with increasing $T_{surf}$. For a given total atmosphere+ocean C inventory (e.g., 1.1~$\times$~10$^6$ kg/m$^2$ CO$_2$-equivalent), the $p_{\mathrm{CO2}}$ is much higher (for a given $T_{surf}$) for the cation-poor case relative to the cation-rich cases. This is due to the higher pH of the cation-rich cases (Figure \ref{fig:bjerrum}). However, $p_{\mathrm{CO2}}$ increases more steeply with $T_{surf}$ for the cation-rich cases. %This can be understood as follows. In the dilute ocean, the C in the ocean is only in the form of CO$_{2(aq}}$. As $T$ rises, $K_H$ shifts to favor $CO2(g)$ at the expense of $CO2(aq)$, so $p_{\mathrm{CO2}}$ increases. For the cation-rich cases, the same is true. But in addition, HCO3- and CO32- equilibrium constants change as T rises to favor $CO2(aq)$ at the expense of HCO3- and CO32-. So, for the cation rich cases, there is more CO2aq and there with increasing $T$, the CO$_{2(aq}}$

\begin{figure}
\includegraphics[width=1.02\columnwidth,clip=true,trim={13mm 60mm 13mm 50mm}]{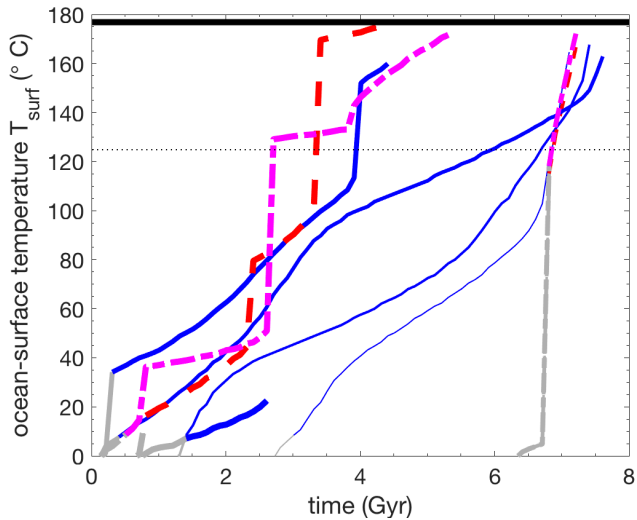} %was _AD
\caption{How ocean-surface temperature ($T_{surf}$) evolves with time for $a$~=~1.1~AU, planet water mass fraction~=~0.01, $M_{pl}$~=~1 $M$$_\earth$. Dashed red lines correspond to oceans with 0.5~mol/kg~Na, dash-dotted magenta lines correspond to oceans with 0.25~mol/kg~Ca, and solid blue lines correspond to cation-poor~oceans. Grayed-out parts of the lines correspond to parts of the time evolution for which high-pressure ice occurs at depth within the ocean. For these grayed-out parts, $T_{surf}$ will be an underestimate to the extent that \mbox{HP-ice} excludes CO$_2$ from the matrix. The thickness of the line corresponds to~the CO$_2$-equivalent atmosphere+ocean C content ($c_C$), with thicker lines marking higher~$c_C$: \{3.5 $\times$10$^4$, 1.1 $\times$10$^5$, 3.5$\times$10$^5$, 1.1$\times$10$^6$, 3.5$\times$10$^6$\}~kg~m$^{-2}$ CO$_2$-equivalent~C  are considered. The dotted line marks 395K (highest temperature at which life has been observed to proliferate). The black bar marks 450K (end of habitability). The lines end if $p_{\mathrm{CO2}}$~$>$~40~bars.} 
\label{fig:tsurfevo}
\end{figure}

\begin{figure}
%\epsscale{1.2}
\includegraphics[width=1.02\columnwidth,,clip=true,trim={13mm 60mm 13mm 50mm}]{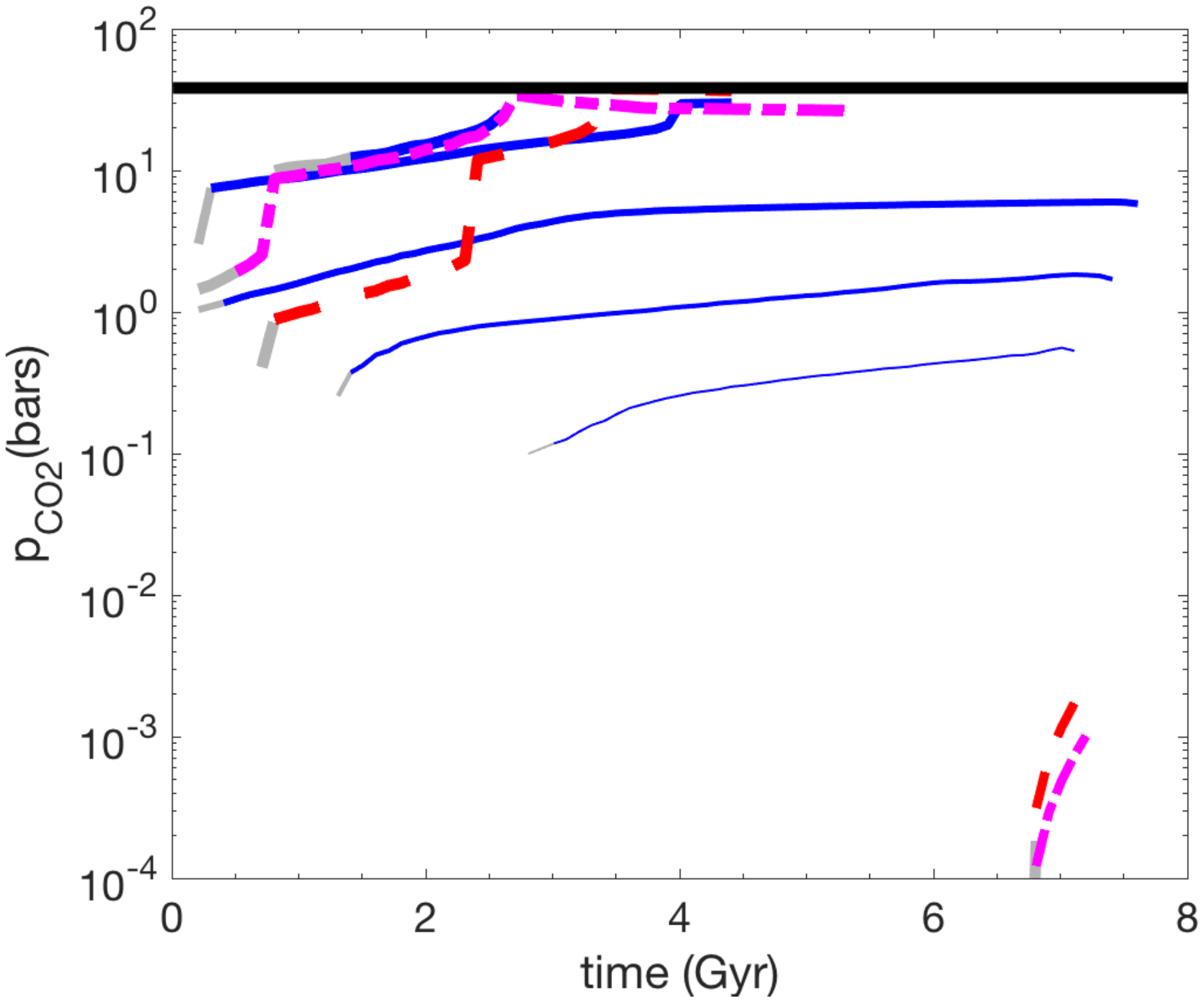} %was _AE
\caption{How $p_{\mathrm{CO2}}$ evolves with time for $a$~=~1.1~AU, planet water mass fraction~=~0.01, $M_{pl}$~=~1 $M$$_\earth$. Dashed red lines correspond to oceans with 0.5~mol/kg~Na, dash-dotted magenta lines correspond to oceans with 0.25~mol/kg~Ca, and solid blue lines correspond to cation-poor oceans. Grayed-out parts of the line correspond to parts of the time evolution for which \mbox{HP-ice} occurs within the ocean. For these grayed-out parts, $T_{surf}$ will be an underestimate to the extent that \mbox{HP-ice} excludes CO$_2$ from the matrix. The thickness of the lines correspond to the CO$_2$-equivalent atmosphere+ocean C content ($c_C$), with thicker lines marking higher~$c_C$: \{3.5 $\times$10$^4$, 1.1 $\times$10$^5$, 3.5$\times$10$^5$, 1.1$\times$10$^6$, 3.5$\times$10$^6$\}~kg~m$^{-2}$ CO$_2$-equivalent~C  are shown for the cation-poor-ocean case. For the 0.25~mol/kg~Ca and 0.5~mol/kg~Na cases, $c_C$ = 1.1$\times$10$^6$~kg~m$^{-2}$ CO$_2$-equivalent~C  plots on the left half of the figure and the smaller values of $c_C$ plot on top of one another as the near-vertical line around 7 Ga (only one representative low-$c_C$ track for each case is shown). The black bar marks 40~bars CO$_2$ (which we deem sufficient for the end of habitability). The lines end if $T$~$>$~450K.} 
\label{fig:pco2evo}
\end{figure}

\begin{figure}
%\epsscale{1.2}
\includegraphics[width=0.99\columnwidth,clip=true,trim={13mm 60mm 13mm 50mm}]{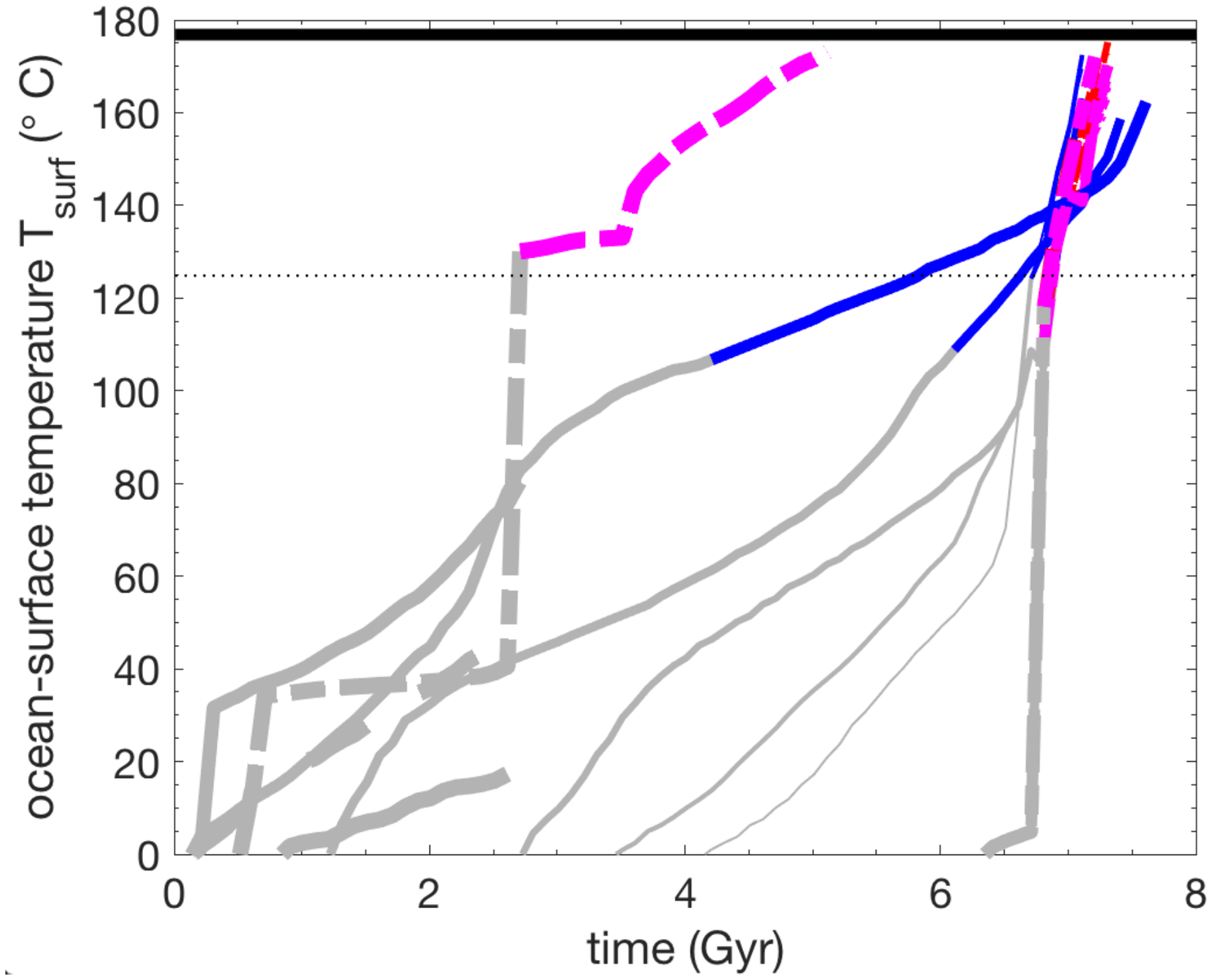} %was _EA
\caption{As Figure~\ref{fig:tsurfevo}, but for planet~water~mass~fraction~($f_W$)~=~0.1. Evolution of ocean-surface temperature $T_{surf}$, for semimajor~axis~=~1.1~AU, $M_{pl}$ = 1 $M_\earth$. Dashed red lines correspond to oceans with 0.5~mol/kg~Na, dash-dotted magenta lines correspond to oceans with 0.25~mol/kg~Ca, and solid blue lines correspond to cation-poor oceans. Note greater persistence  of high-pressure (HP) ice (gray shrouding) at depth within the oceans. The thickness of the line corresponds to CO$_2$-equivalent atmosphere+ocean C content ($c_C$), with thicker lines marking higher~$c_C$.  Tracks are generally rather similar to those for~$f_W$~=~0.01, except that the same climate track requires $\sim$ 10$\times$ more~$c_C$ due to dilution. The dotted line marks 395K (highest temperature at which life has been observed to proliferate). The black bar marks 450K (end of habitability). The lines end if $p_{\mathrm{CO2}}$~$>$~40~bars.} 
\label{fig:tsurfevo01}
\end{figure}

For most waterworld-evolution tracks, each time step is independent of the previous ones. Therefore, the colored portions of the waterworld evolution tracks (ice-free oceans) in Figures~\ref{fig:tsurfevo}-\ref{fig:tsurfevo01} should be valid whether or not HP-ice is present earlier in that world's history (gray portions of the lines). %There are cases where history-dependence does occur, which we now discuss.

\vspace{0.07in}
\noindent \emph{Stability and instability.} To see that $T_{surf}$~--~$p_{\mathrm{CO2}}$ equilibria are usually stable, consider the point around 50$^\circ$C, 1~bar. Suppose we add some energy to the system, perturbing $T_{surf}$. This drives CO$_2$ from the ocean to the atmosphere -- moving along the blue line -- which will cause more warming, leading to a positive feedback on the initial perturbation. However, at fixed $L_*$, the value of $p_{\mathrm{CO2}}$ that is in equilibrium with the increased surface temperature is even higher than the increased CO$_2$. Therefore, the climate is stable. 

For $P$~$>$~1~bar, exsolving CO$_2$ can be a negative (stabilizing) feedback on climate evolution. For example, consider the point at 40$^\circ$C, 10~bars in Figure~\ref{fig:findequilibria}. An increase in $L_*$ corresponding to 1 Gyr of stellar evolution at constant $p_{\mathrm{CO2}}$ would cause a $T_{surf}$ increase of $\sim$40K. However, the climate is constrained to evolve along a curve of constant seawater cation abundance. Therefore, CO$_2$ is exsolved from the ocean. Cooling from Rayleigh scattering per unit of added CO$_{2\mathrm{(g)}}$ exceeds greenhouse warming per unit of added CO$_{2\mathrm{(g)}}$, so adding CO$_{2(\mathrm{g})}$ cools the planet \citep{Ramirez2014}. For the purple dashed line, the $T_{surf}$ increase is only $<$~10~K; a negative feedback. This negative feedback is not a geochemical cycle.

CO$_2$'s solubility in water (for fixed  $p_{\mathrm{CO2}}$) increases for $T_{surf}$ $>$ (110-160)$^\circ$C. This feedback is mildly stabilizing, but we neglect $\partial p_{\mathrm{CO2,GH}} / \partial T$ $<$ 0 in the following discussion as these cases are rarely important.

%are uncommon in Figure~\ref{fig:findequilibria} \citep{Kite2011}. 
For a \{$p_{\mathrm{CO2}}$, $T_{surf}$\} point to undergo runaway warming in response to small $T_{surf}$ increases, two conditions must be satisfied. (1) $\partial p_{\mathrm{CO2,OC}} / \partial T$~$>$~$\partial p_{\mathrm{CO2,GH}} / \partial T$, where the subscript $_{\mathrm{OC}}$ refers to ocean chemistry curves, and the subscript $_{\mathrm{GH}}$ refers to greenhouse curves. Otherwise, the positive feedback has finite gain, and there is no runaway. (2) $\partial p_{\mathrm{CO2,GH}} / \partial T$ $>$ 0. If (2) is not satisfied, then there is a stabilizing, negative feedback. %Therefore, counterintuitively, instabilities are relatively likely in the vicinity of areas where $T_{surf}$ increases \emph{weakly} with increasing$p_{\mathrm{CO2}}$ -- because these are the areas where 

Exsolution-driven climate instabilities (runaway feedbacks) are possible in the model. For example, consider the red thick-dashed line corresponding to Na~=~0.25~mol/kg, $c_C$~=~3.5~$\times$~10$^6$~kg~m$^{-2}$ in Figures~\ref{fig:tsurfevo} and \ref{fig:pco2evo}. At 2.0~Gyr and~1.1 AU, two stable states are possible for these inputs: \{37~$^\circ$C, $\sim$2~bars\} and \{75~$^\circ$C, $\sim$10~bars\} (Figure~\ref{fig:findequilibria}). As $L_*$~increases, both stable states get warmer. Then, at 2.3 Gyr, the GH curve no longer intersects the thick red dashed line for $T_{surf}$~$<$~60~$^\circ$C. The cool branch has disappeared; as a result, the climate suddenly warms. (\emph{Emergence} of cold solutions as solar luminosity increases is much less common in our model.) Because of interpolation artifacts and our very approximate treatment of CO$_2$ solubility, one should be careful about reading out quantitative gradients for specific data points from the plots. The rate of change associated with these instabilities is unlikely to risk whole-ocean microbial extinction (in our opinion), provided that both the initial climate and final climate are habitable. The model-predicted instabilities are discussed further in Appendix~F.

\vspace{0.07in}
\noindent \emph{Duration of habitable surface water ($\tau_{hab}$).} In some cases, habitability can persist for many Gyr (Figures \ref{fig:tsurfevo} and \ref{fig:pco2evo}). As $L_*$~rises, either $T_{surf}$ eventually exceeds 450K, or CO$_2$ exceeds $\sim$40~bars, either of which we deem sufficient to extinguish surface habitability (for a discussion, see Section 6.4). Different cation cases with the same CO$_2$-equivalent atmosphere+ocean C content ($c_C$) have different $p_{\mathrm{CO2}}$($t$) trajectories  (different colors of the same thickness in Figure~\ref{fig:pco2evo}). This is because, for the same total amount of C, the fraction~of~C stored in the~ocean differs between cation cases. However, the trajectories for different cation cases bunch together at high $c_C$ (thick lines of all~colors in Figure~\ref{fig:pco2evo}). That is because, for high~total~atmosphere+ocean C inventories, the ocean is saturated and most of the~C is in the~atmosphere. When (C~in~atmosphere)~$\gg$~(C~in~ocean), C$_{\mathrm{atm}}$~$\approx$~C$_{\mathrm{total}}$, independent of how much C is in the ocean. The colors also bunch together for very low~$c_C$. For such worlds, the $p_{\mathrm{CO2}}$ is too~low to affect climate until $T_{surf}$~$>$~100~$^\circ$C. As a result, such worlds have H$_2$O as the~only greenhouse~gas. Then, very~low~$c_C$~worlds deglaciate at the same time (for a given planet orbital semimajor~axis), trek~quickly across the narrow ``neck'' at the bottom of the diagram on Figure~\ref{fig:co2matters}, and then suffer the runaway greenhouse.

In Figure~\ref{fig:tsurfevo}, cases with high CO$_2$-equivalent atmosphere+ocean C content have longer durations with habitable surface water ($\tau_{hab}$). The longest-lived worlds maintain $p_{\mathrm{CO2}}$ in the 0.2-20 bar range for Gyr (Figure \ref{fig:pco2evo}).

Worlds with planet~water~mass~fraction~=~0.1 (Figure~\ref{fig:tsurfevo01}) show greater persistence of \mbox{HP-ice}. This is because $f_W$~=~0.1 seafloor pressure is 12~GPa (for $M_{pl}$~=~1~$M_\earth$), and these high pressures favor \mbox{HP-ice} (Figure \ref{fig:H2Oadiabats}). For a given CO$_2$-equivalent atmosphere+ocean C content, the climate is colder, because the bigger ocean dilutes the CO$_2$. For high-$f_w$ worlds, because the ocean is the dominant reservoir of C, small fractional changes in the storage of C in the ocean can lead to large fractional changes in $p_{\mathrm{CO2}}$. 

% Because $\partial p_{\mathrm{CO2,OC}} / \partial T$ is larger, the condition $\partial p_{\mathrm{CO2,OC}} / \partial T$~$>$~$\partial p_{\mathrm{CO2,GH}} / \partial T$ occurs for a wider range of \{$p_{\mathrm{CO2}}$,$T_{surf}$\} parameter space and thus exsolution-driven climate instabilities can have larger amplitude.

 \begin{figure*}
 \begin{centering}
%\epsscale{1.2}
\includegraphics[width=0.68\columnwidth,clip=true,trim={23mm 70mm 23mm 65mm}]{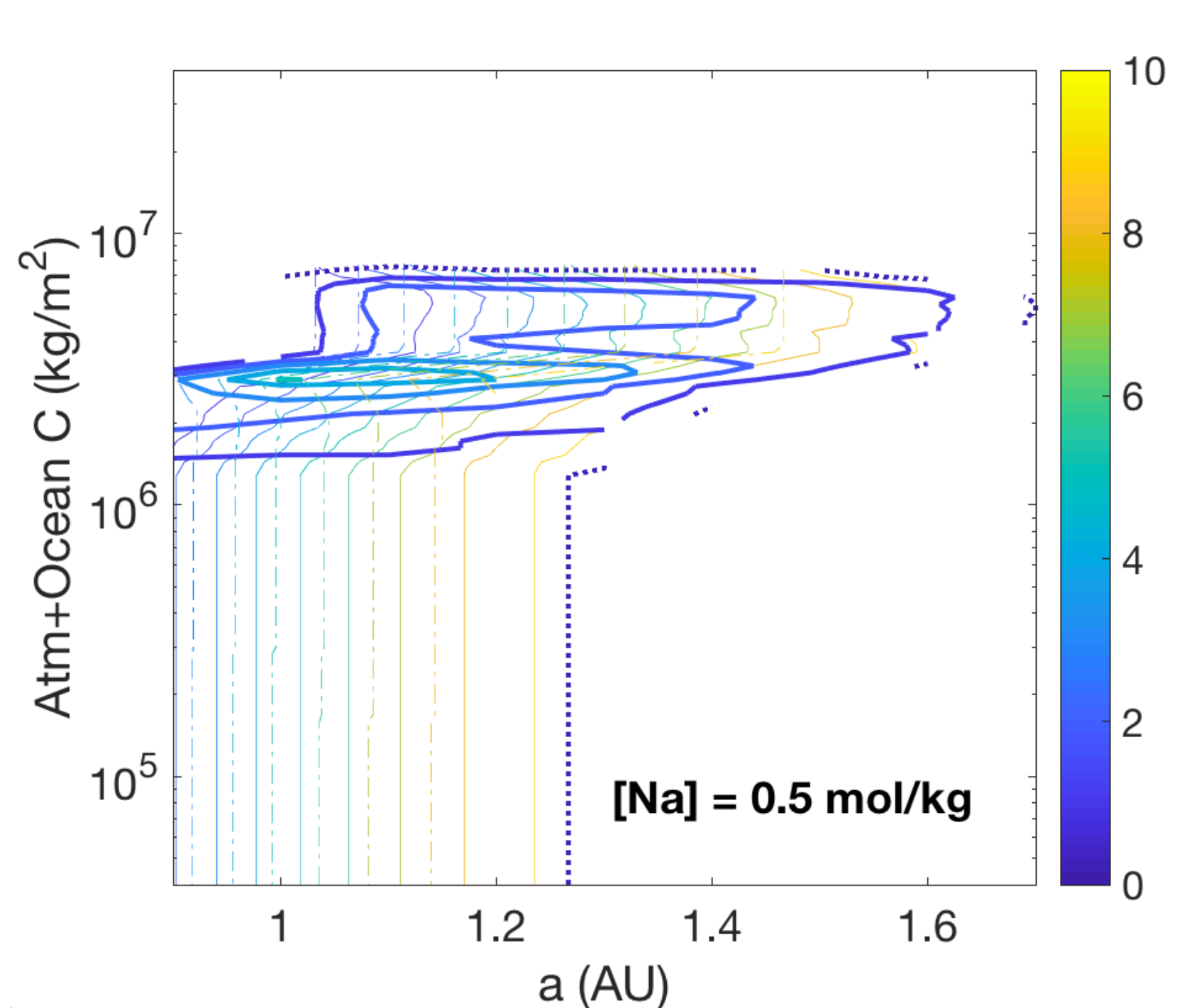}
\includegraphics[width=0.68\columnwidth,clip=true,trim={23mm 70mm 23mm 65mm}]{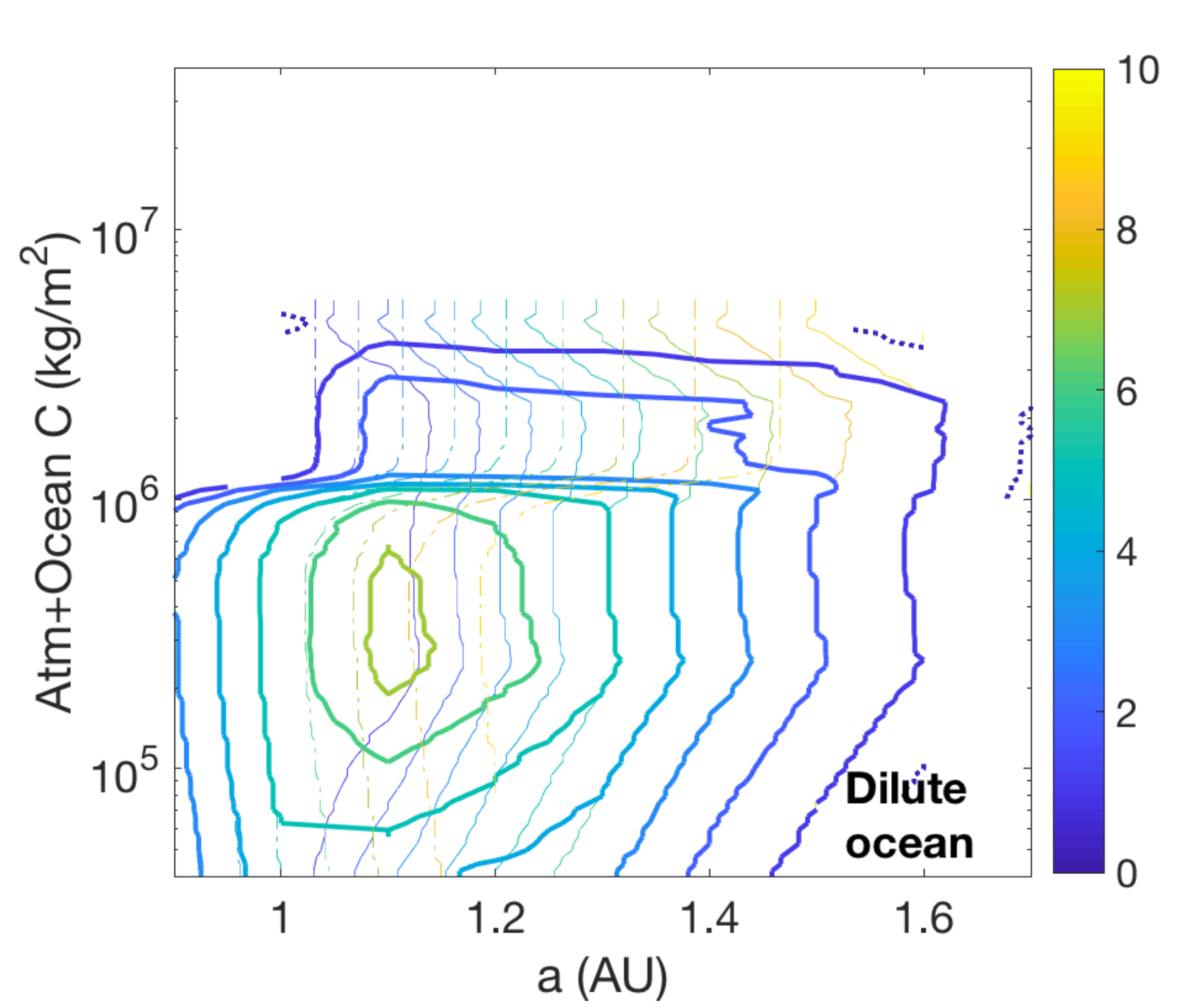}
\includegraphics[width=0.68\columnwidth,clip=true,trim={23mm 70mm 23mm 65mm}]{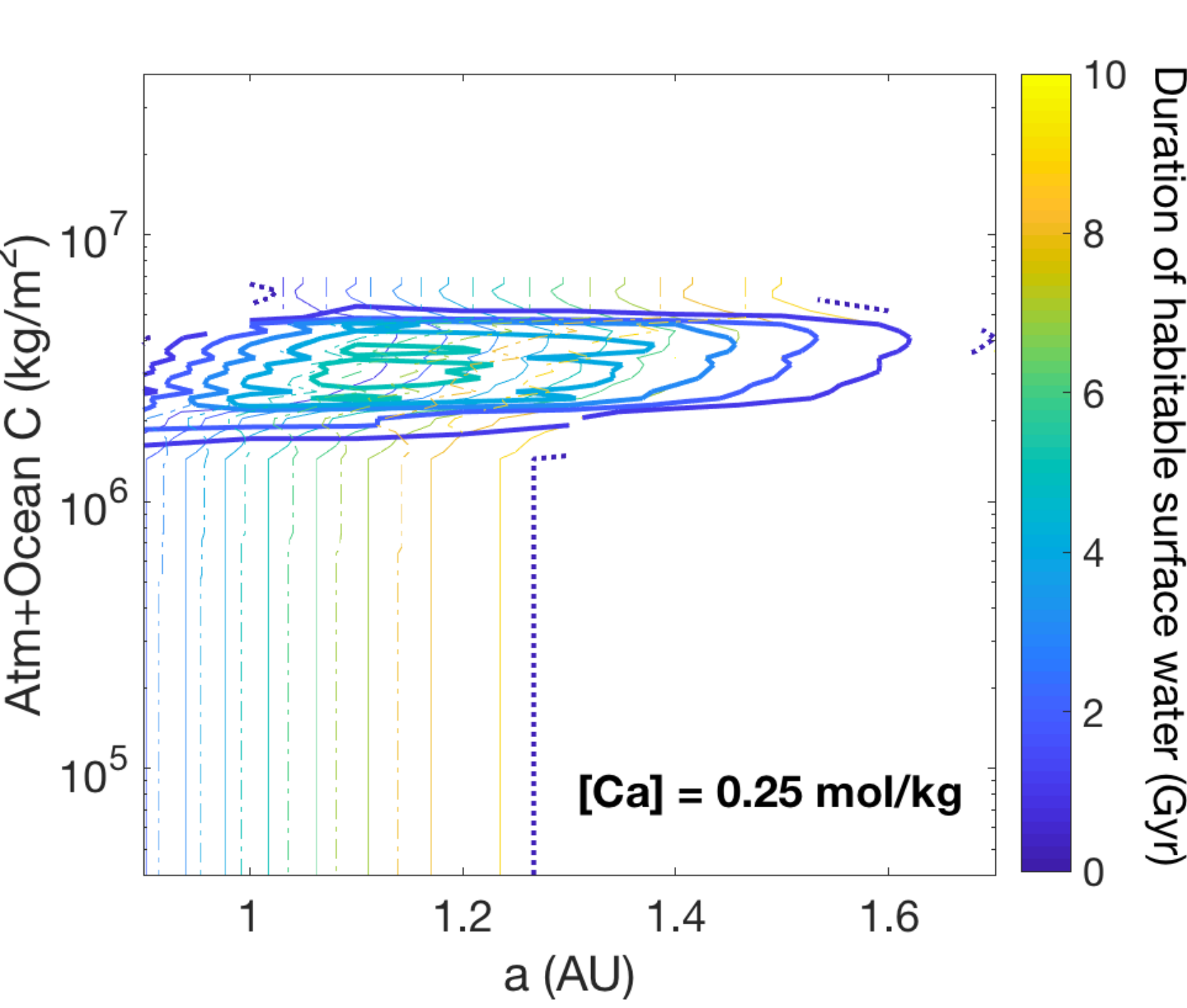}
\caption{Waterworld $\tau_{hab}$ (duration with habitable surface water) diagrams for planet water mass fraction ($f_w$)~=~0.01, $M_{pl} = 1 M_\earth$. Column-masses of C are in~CO$_2$~equivalent. Thick contours correspond to the duration with habitable surface water, in Gyr. They are spaced at intervals of 1 Gyr, except for the outer dotted contour, which corresponds to a duration of 0.2 Gyr. Thin solid contours correspond to the time of onset of habitable surface water (i.e., time after which ocean has neither \mbox{HP-ice} nor surface ice), spaced at intervals of 1 Gyr. Worlds to the right of the leftmost thin solid contour in each panel are ``revival worlds'' (worlds that start with surface ice, and subsequently deglaciate). Thin dashed contours mark the time when the planet is no longer habitable (i.e., either $p_{\mathrm{CO2}}$~$>$~40~bars, or $T_{surf}$~$>$~450K), spaced at intervals of 1 Gyr. The small wiggles in the contours, and the indentations at $\sim$3$\times$10$^7$ kg~m$^{-2}$, are interpolation artifacts.} 
\label{fig:lifetimes001}
\end{centering}
\end{figure*}
%\textbf{Add horizontal lines showing the C consumption potential of the crust as a function of I/E}. 

 \begin{figure*}
%\epsscale{1.2}
 \begin{centering}
\includegraphics[width=0.68\columnwidth,clip=true,trim={23mm 70mm 23mm 65mm}]{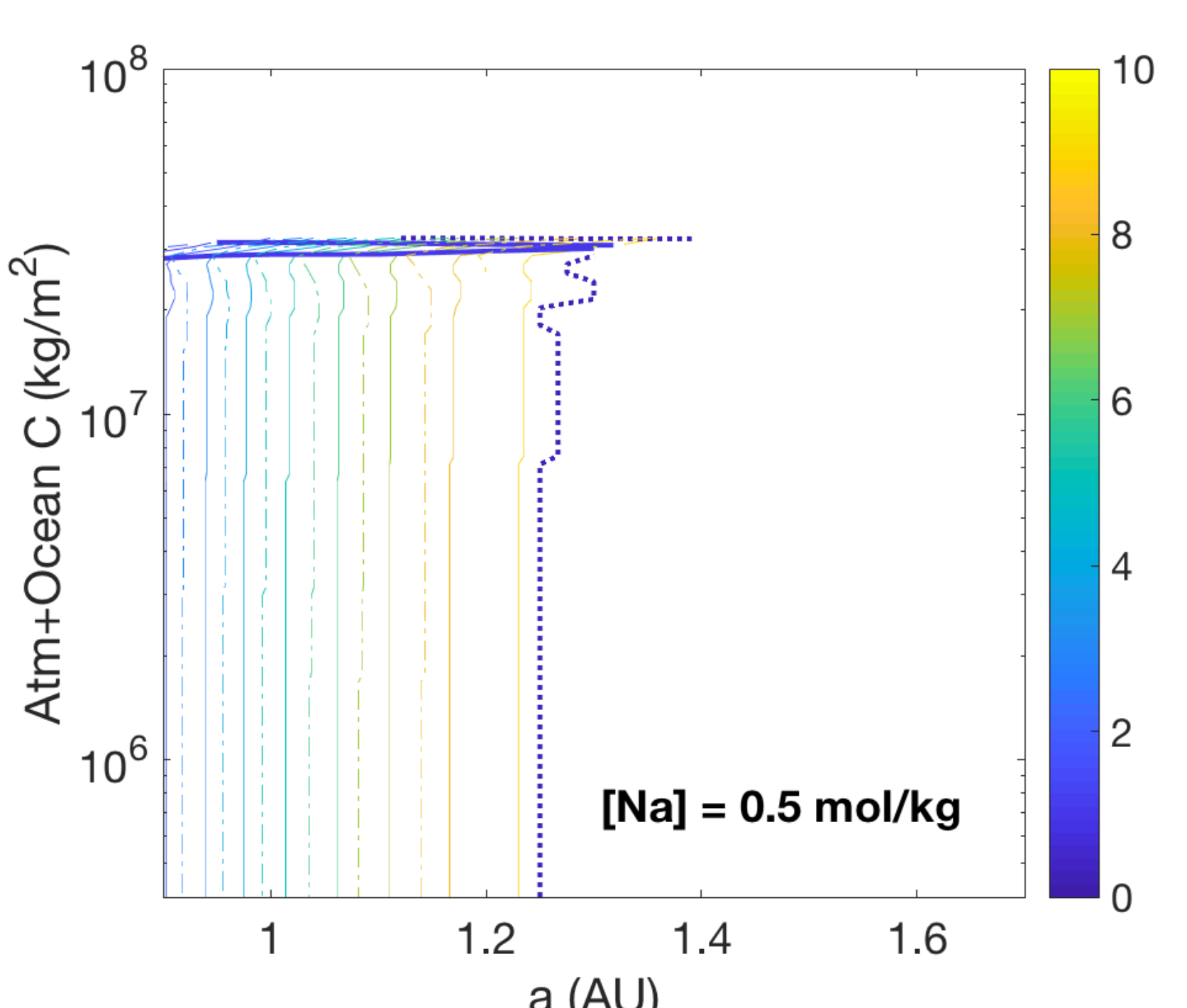}
\includegraphics[width=0.68\columnwidth,clip=true,trim={23mm 70mm 23mm 65mm}]{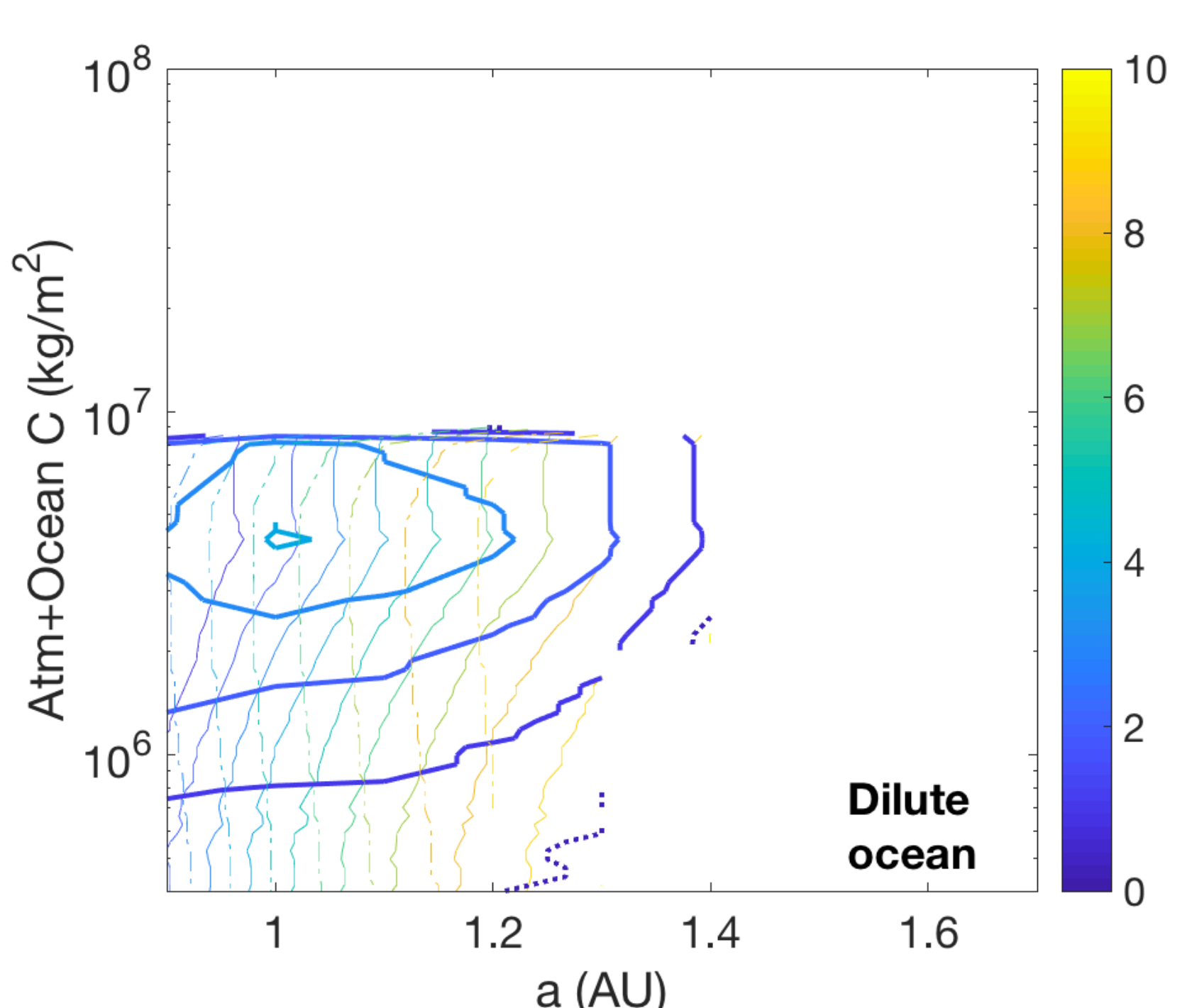}
\includegraphics[width=0.68\columnwidth,clip=true,trim={23mm 70mm 23mm 65mm}]{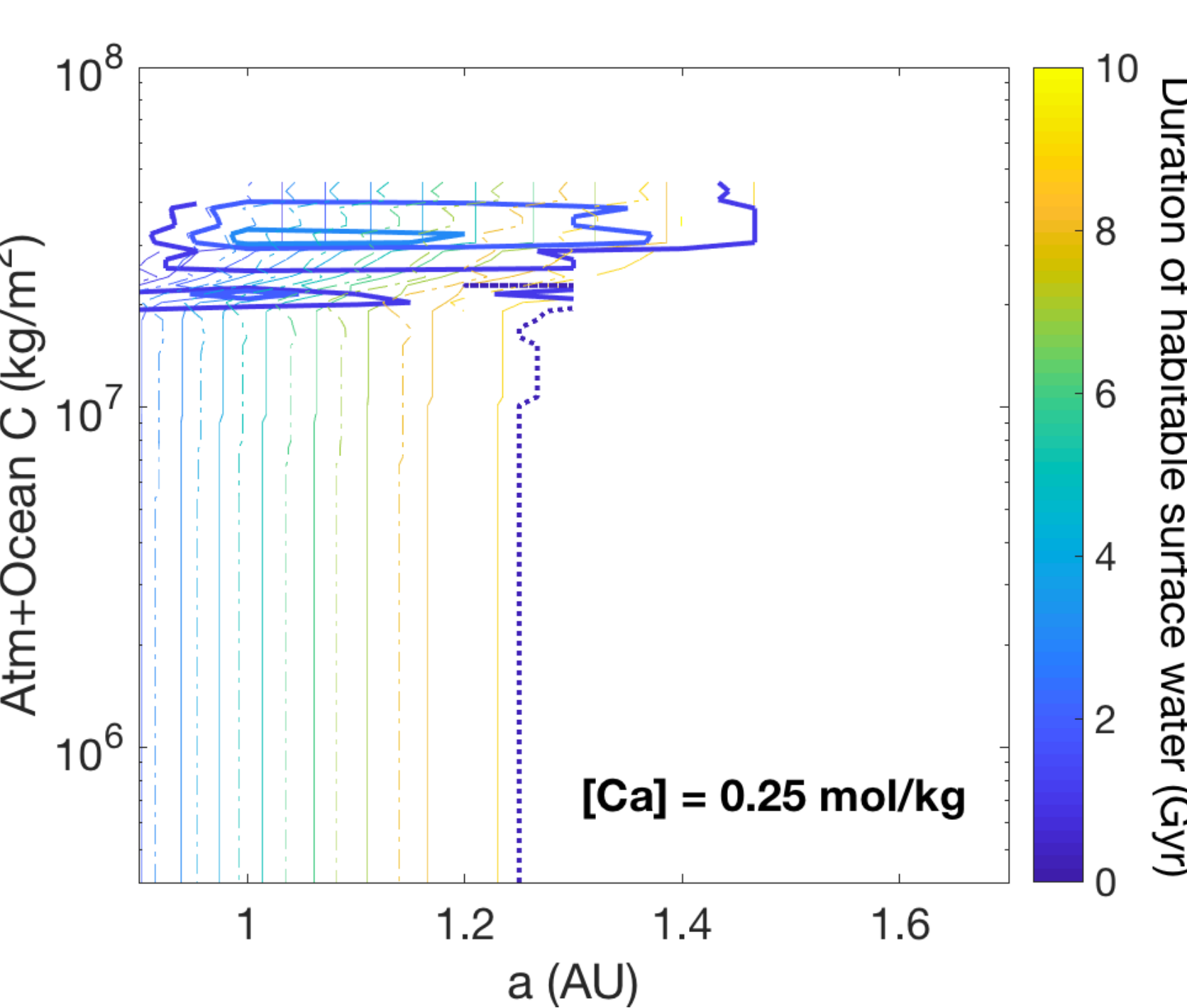}
\caption{As Figure~\ref{fig:lifetimes001}, but for planet water mass fraction~=~0.1.} 
\label{fig:lifetimes01}
 \end{centering}
\end{figure*}

The initial pH of the habitable surface water ranges from~3~to~$>$11. In our model, as atmosphere $p_{\mathrm{CO2}}$ rises during the interval of habitable surface water, ocean pH falls \citep{Zeebe2012}, typically by $<$~1~pH unit. For \{$p_{\mathrm{CO2}}$,~$T_{surf}$\} values that allow habitable surface water, pH is higher for the `0.25 mol/kg~Ca'~case and highest of all for the `0.5~mol/kg~Na'~case. 

In summary, ocean chemistry is key to waterworld climate. This is because of the following effects. (1) Ocean pH sets the ocean-atmosphere partitioning of C. This in turn sets the sensitivity of $p_{\mathrm{CO2}}$ to increases in total~C~abundance (the Revelle factor). The Revelle factor corresponds to the differences in spacing between the y-axis positions of lines of the same color (cation abundance) but different thickness ($c_C$) in Figures \ref{fig:findequilibria}-\ref{fig:tsurfevo01}. (2) Carbonate-system equilibria are $T$-dependent. This causes the upwards slope of the colored lines in Figure \ref{fig:findequilibria}-\ref{fig:tsurfevo01}.

 \begin{figure}
%\epsscale{1.2}
\includegraphics[width=0.99\columnwidth]{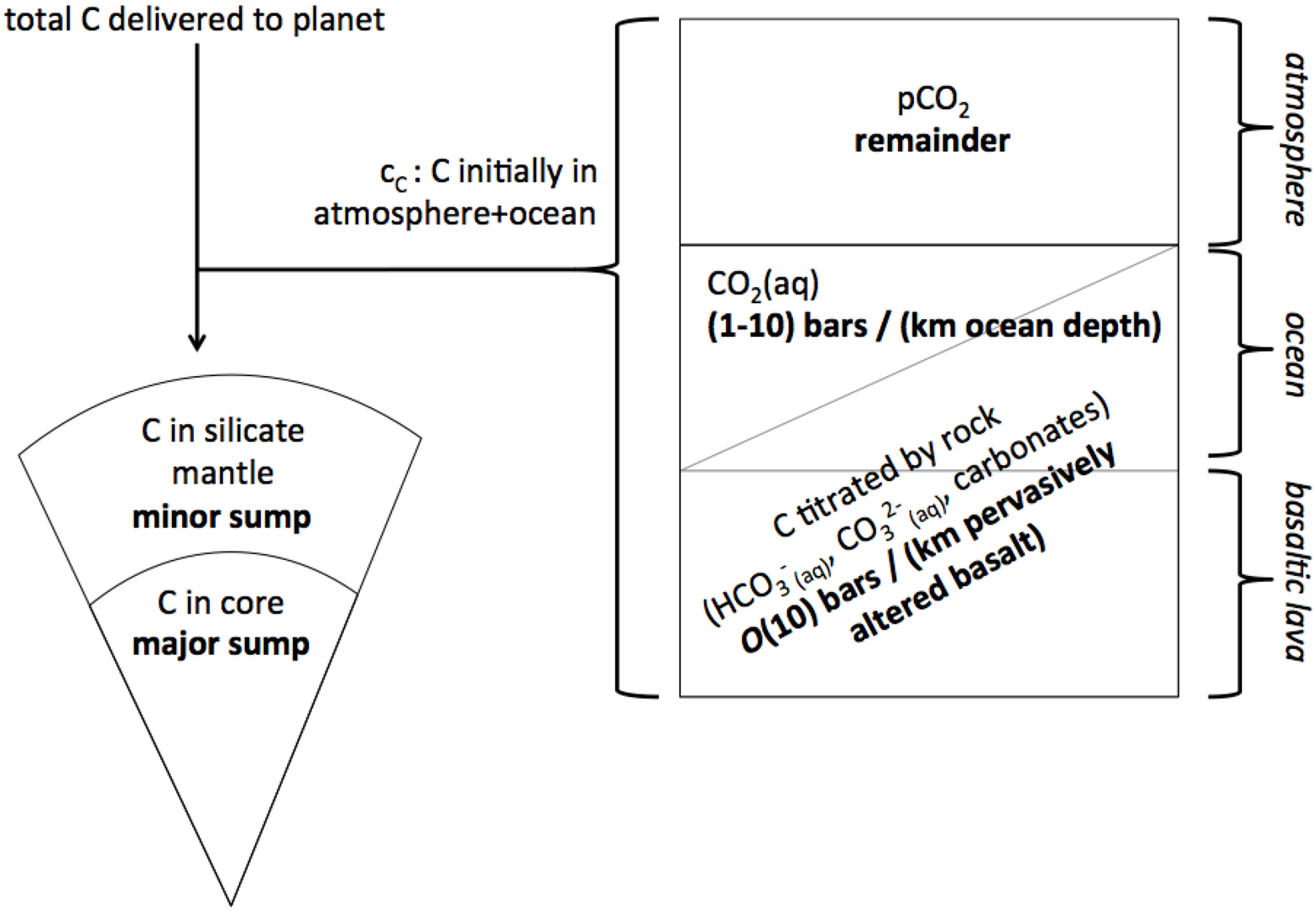}
\caption{A simplistic ``bucket'' model of $p_{\mathrm{CO2}}$. } 
\label{fig:bucket}
\end{figure}

\subsection{Controls on the duration of \\ habitable surface water ($\tau_{hab}$)}
Figure~\ref{fig:lifetimes001} shows, for $f_W$~=~0.01, $M_{pl}~=~1\,M_\earth$ waterworlds, the onset time, shutoff time, and duration of habitable surface water. In each case, nonzero onset times (thin solid lines in Figure~\ref{fig:lifetimes001}) correspond to the meltback time for an early-established ice lid on the ocean. Shutoff times (thin~dash-dot~lines in Figure~\ref{fig:lifetimes001}) correspond to either the runaway greenhouse or to excessive ($>$40 bar) $p_{\mathrm{CO2}}$. With increasing CO$_2$-equivalent atmosphere+ocean C content, for all panels in Figure~\ref{fig:lifetimes001}, $\tau_{hab}$ (thick solid lines in Figure~\ref{fig:lifetimes001}) rises and then falls over an interval of $\sim$~10$^6$~kg~m$^{-2}$, i.e. \emph{O}(10$^2$)~bars of \mbox{CO$_2$-equivalent}. This exceeds the $O$(10) bar $p_{\mathrm{CO2}}$ range for optimum habitability in Figure~\ref{fig:co2matters}. That is because around the habitability optimum, as C is added to the system, most of the extra C is partitioned into the ocean and does not contribute to $p_{\mathrm{CO2}}$ \citep{Archer2009}. 

Too much C makes the planet uninhabitably hot. Figure~\ref{fig:bucket}, which is a simplistic cartoon of C partitioning, shows why. Starting with a small CO$_2$-equivalent atmosphere+ocean C content, $p_{\mathrm{CO2}}$ is initially low. Adding C will eventually overwhelm rock-sourced cations. The ocean then becomes acidic. Once the ocean is acidic, extra C spills into the atmosphere and $p_{\mathrm{CO2}}$ becomes high. Supposing the ocean storage capacity for C to be fixed at $\sim$0.5 mol/kg~C (in reality capacity will rise with $p_{\mathrm{CO2}}$; Equation 1), and for pH low enough that all aqueous C is stored as CO$_{2\emph{T}}$, then ocean CO$_2$-equivalent content is 20~g/kg -- for a 100~km~deep~ocean, 3~$\times$~10$^6$~kg~m$^{-2}$. If extra C goes into the atmosphere, then an additional 4~$\times$~10$^5$~kg~m$^{-2}$ leads to an uninhabitably hot surface. Although this ``bucket model'' (over-)simplifies carbonate system chemistry, it explains the $\sim$10$^6$ kg~m$^{-2}$ upper limit for habitability for the Na-free, Ca-free case (middle panel in Figure~\ref{fig:lifetimes001}). The C threshold for planet sterilization is raised if cations are available (left panel and right panel of Figure~\ref{fig:lifetimes001}), because they neutralize the carbonic acid. For example, adding 0.5~mol/kg~of~positive~charges (0.25~mol/kg of Ca$^{2+}$, or 0.5~mol/kg of Na$^+$) means that an extra 0.5 mol/kg~of~C must be added to get nontrivial CO$_2$ in the atmosphere. For a 100~km deep ocean, this is 200~bars~=~2~$\times$~10$^6$~kg~m$^{-2}$. This quantity corresponds to the vertical offset between the habitability optimum in the middle panel, versus the habitability optimum in the two side panels, of Figure~\ref{fig:lifetimes001}. This quantity of 2~$\times$~10$^6$~kg~m$^{-2}$ also corresponds to the vertical offset between the upper limit for habitability in the middle panel, versus the upper limit for habitability in the two side panels, of Figure~\ref{fig:lifetimes001}. Neutralization by cations also explains why $\tau_{hab}$ is brief (and uniform) for low C content when cation content is high. For these worlds, $p_{\mathrm{CO2}}$ is so low (because pH~is~so~high) that these planets effectively have H$_2$O~+~N$_2$ atmospheres. Examples of evolutionary tracks for such worlds include the nearly vertical tracks to the right of Figure \ref{fig:tsurfevo}. Pure-H$_2$O atmospheres lead to planets having habitable surface water for $<$ 1 Gyr if they orbit G-stars, according to the 1D model of \citet{Goldblatt2015}. 

$\tau_{hab}$ is longest for semimajor axes $a$ $\sim$1.1~AU. For $a$~$<$1.1~AU habitable surface water is available from $t$~=~0 at the inner edge of the HZ, but the runaway greenhouse comes soon. For $a$~$\gg$1.1~AU, the surface is frozen until near the end of the model run (and the end of the main~sequence, which we decree to occur at 10 Gyr for the G-type host~star, arrives swiftly). For $a \sim$ 1.5~AU, only a narrow range of high CO$_2$ permits habitability, corresponding to the wing-shape extending to the top right on all panels. CO$_2$ condensation \citep{Turbet2017} might clip this wing of high-semimajor-axis habitability. Only around 1.1~AU does habitable surface water both begin early and end late. The 1.1 AU maximum in $\tau_{hab}$ corresponds to the ``bull's-eyes'' in Figure~\ref{fig:lifetimes001}.

On a 10$\times$-deeper ocean ($f_W$~=~0.1; Figure~\ref{fig:lifetimes01}), the habitability optimum shifts to 10$\times$ larger values of C. This shift is because more C is needed to overwhelm the ocean sinks (Figure~\ref{fig:bucket}). The range of C that gives long durations of surface liquid water appears narrower on a log scale. That is because the range from ``minor~$p_{\mathrm{CO2}}$ to 40~bars $p_{\mathrm{CO2}}$'' is the same ($\sim$4 $\times$ 10$^5$ kg~m$^{-2}$ C for fixed cations) in absolute terms, so smaller in fractional terms. The upper limit for habitability has also moved higher, but only by a factor of two. This is because [C]-rich tracks that reach 40~bars $p_{\mathrm{CO2}}$' at $T$~$<$~100~$^\circ$C always have HP ice for $f_W$~=~0.1, and so do not count as habitable (thick stubby gray lines that terminate within the bottom left of Figure \ref{fig:tsurfevo01}).  The strong dependence of the lifetime of habitability on semimajor~axis for $f_W$~=~0.01 is more subdued for $f_W$~=~0.1. This is because almost all $f_W$~=~0.1 worlds start with \mbox{HP-ice} due to the deeper ocean (Figure~\ref{fig:H2Oadiabats}), and such worlds are not counted as habitable in Figure~\ref{fig:lifetimes01} until temperature has risen sufficiently to melt all the high-pressure ice. %This is shown in Figure~\ref{fig:lifetimes01}. %Seafloor pressures are higher, and thus \mbox{HP-ice} is harder to avoid; thus, the $Tfso far w_{surf}$ range for habitability is narrower for higher~$f_W$, \mbox{$\sim$100$^\circ$C-180$^\circ$C}, and thus durations with habitable surface water are shorter. Dilution moves the optimum $c_C$ for habitability to higher column abundances of CO$_2$.

 \begin{figure}
%\epsscale{1.2}
\begin{centering}
\includegraphics[width=0.90\columnwidth,clip=true,trim={3mm 25mm 3mm 15mm}]{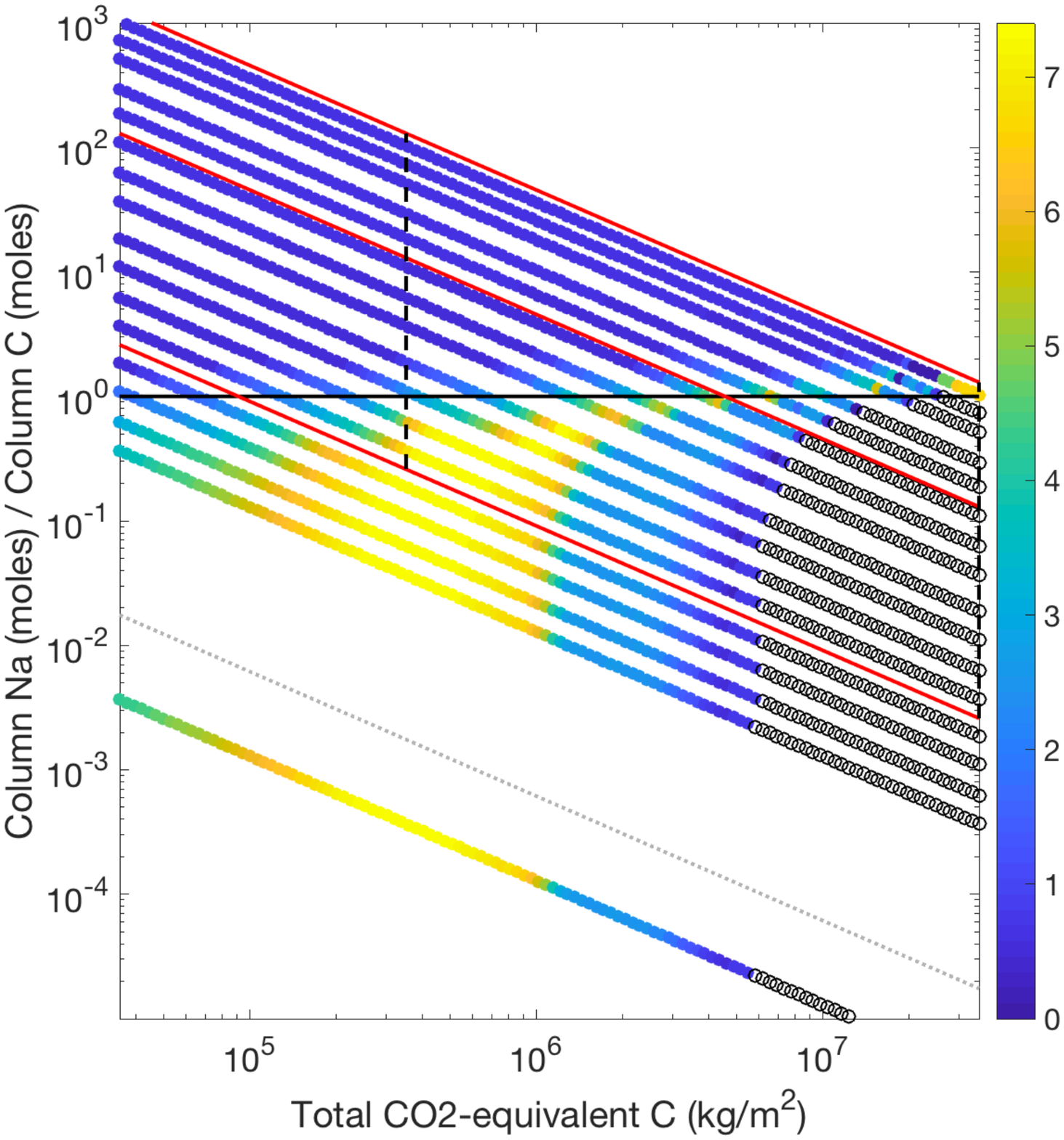}
\caption{To show controls on the duration of habitable surface water ($\tau_{hab}$) on waterworlds. Color scale corresponds to $\tau_{hab}$ (Gyr) for $a$~=~1.1~AU, $f_W$~=~0.01, $M$~=~1 $M_\earth$. The black circles are never-habitable worlds. Red lines show the cation content (mEq/kg)  for complete leaching of both Na and Ca -- with no secondary-mineral formation -- from a basalt column 50~km~thick (top line), 5~km~thick (middle line), or 0.1~km~thick (bottom line).  The top and bottom red lines are the bounds for cation content used in the Figure \ref{fig:mcuninformed} parameter sweep. The vertical dashed lines at 3.5~$\times$~10$^5$~kg/m$^2$ and 3.5~$\times$~10$^7$~kg/m$^2$ are the bounds for C abundance used in the Figure \ref{fig:mcuninformed} parameter sweep. The dotted grey line is the Na from condensation of a 100-bar steam atmosphere, assuming a volume mixing ratio in that steam atmosphere of 0.002 NaCl$_{\mathrm{(g)}}$ \citep{Lupu2014}.} 
\label{fig:chvsccat}
\end{centering}
\end{figure}

\begin{figure}
\epsscale{1.2}
\begin{centering}
\includegraphics[width=0.8\columnwidth,clip=true,trim={12mm 80mm 10mm 80mm}]{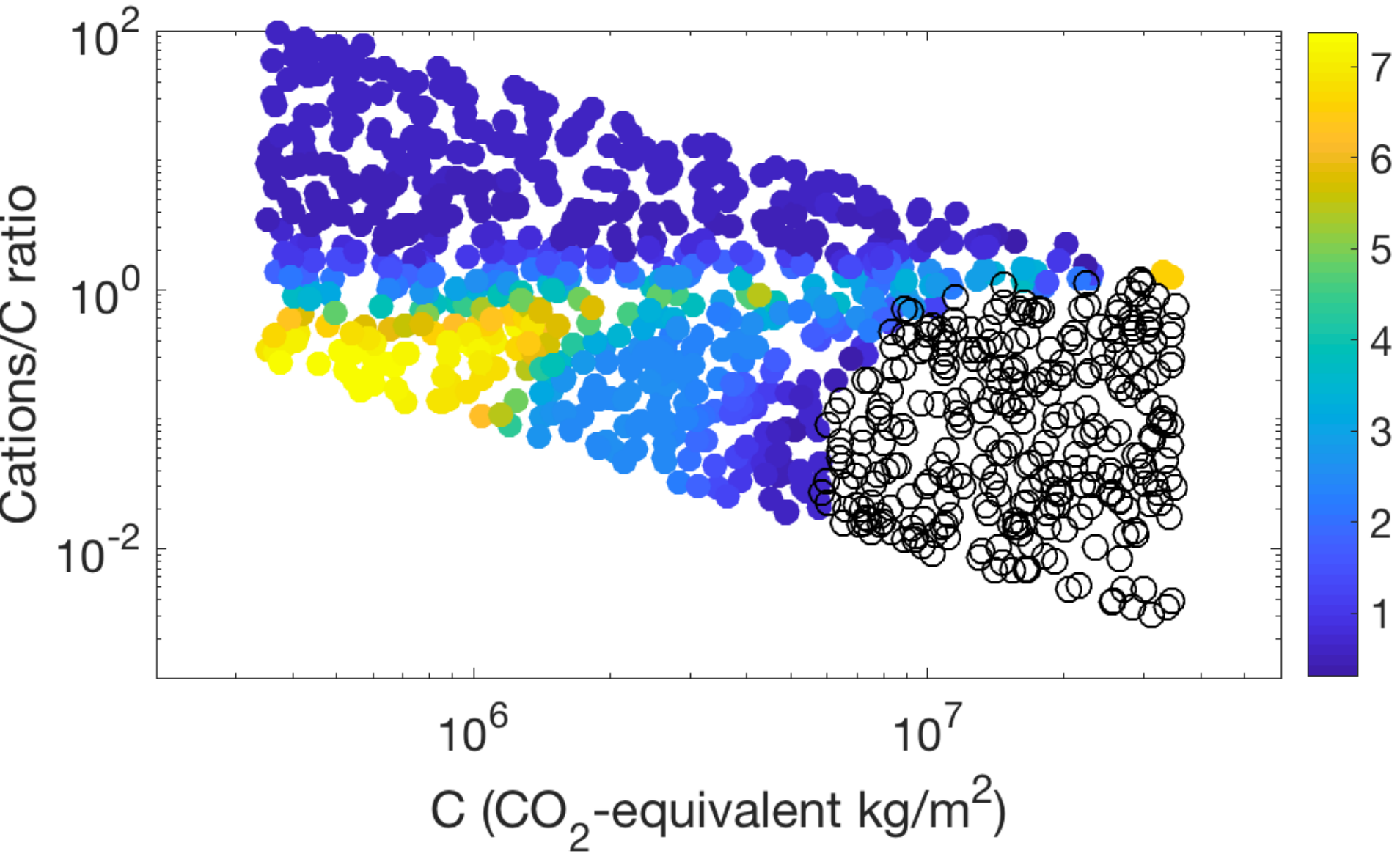}
\includegraphics[width=0.85\columnwidth,clip=true,trim={8mm 80mm 10mm 80mm}]{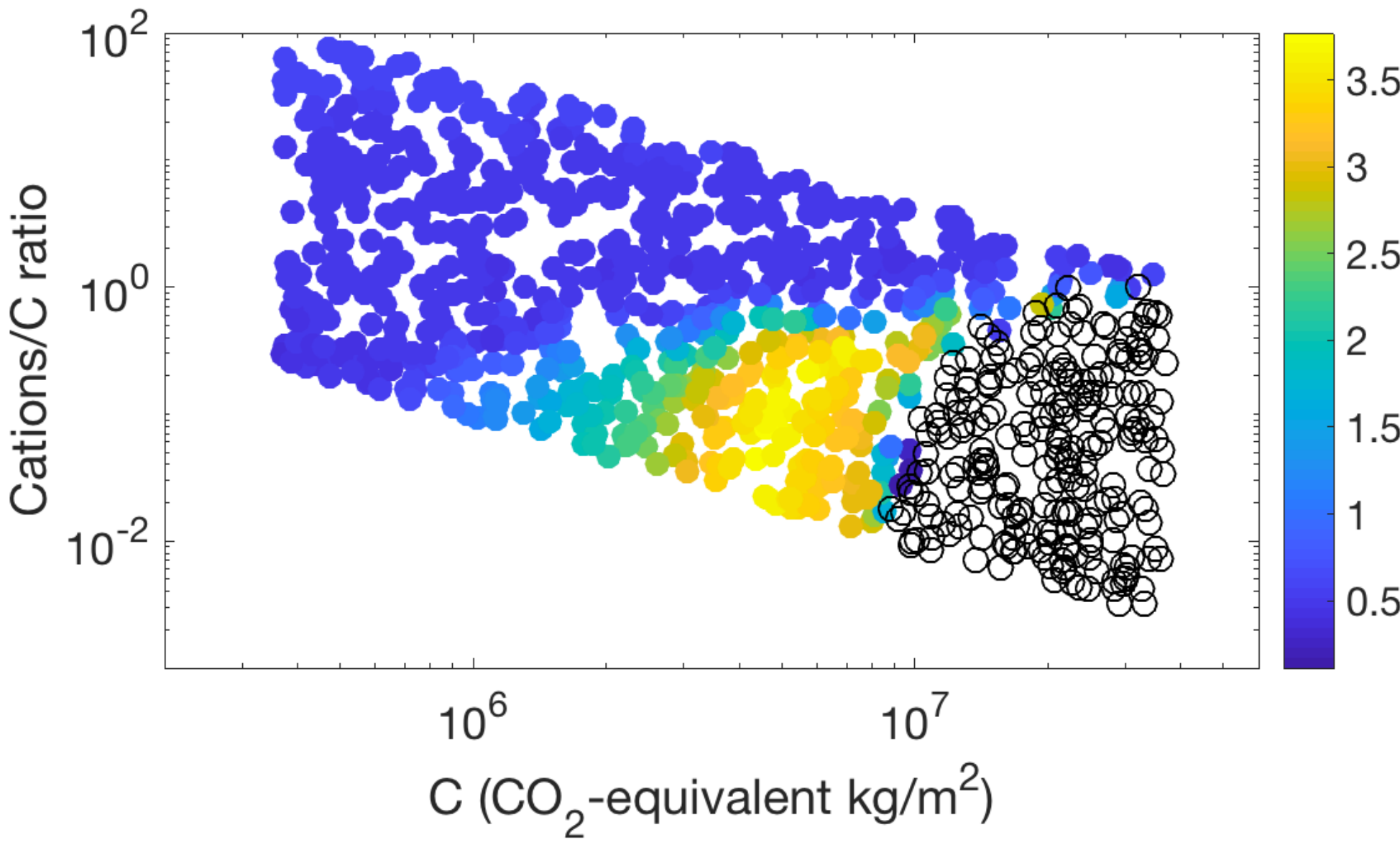}
\includegraphics[width=1.03\columnwidth,clip=true,trim={20mm 80mm 10mm 80mm}]{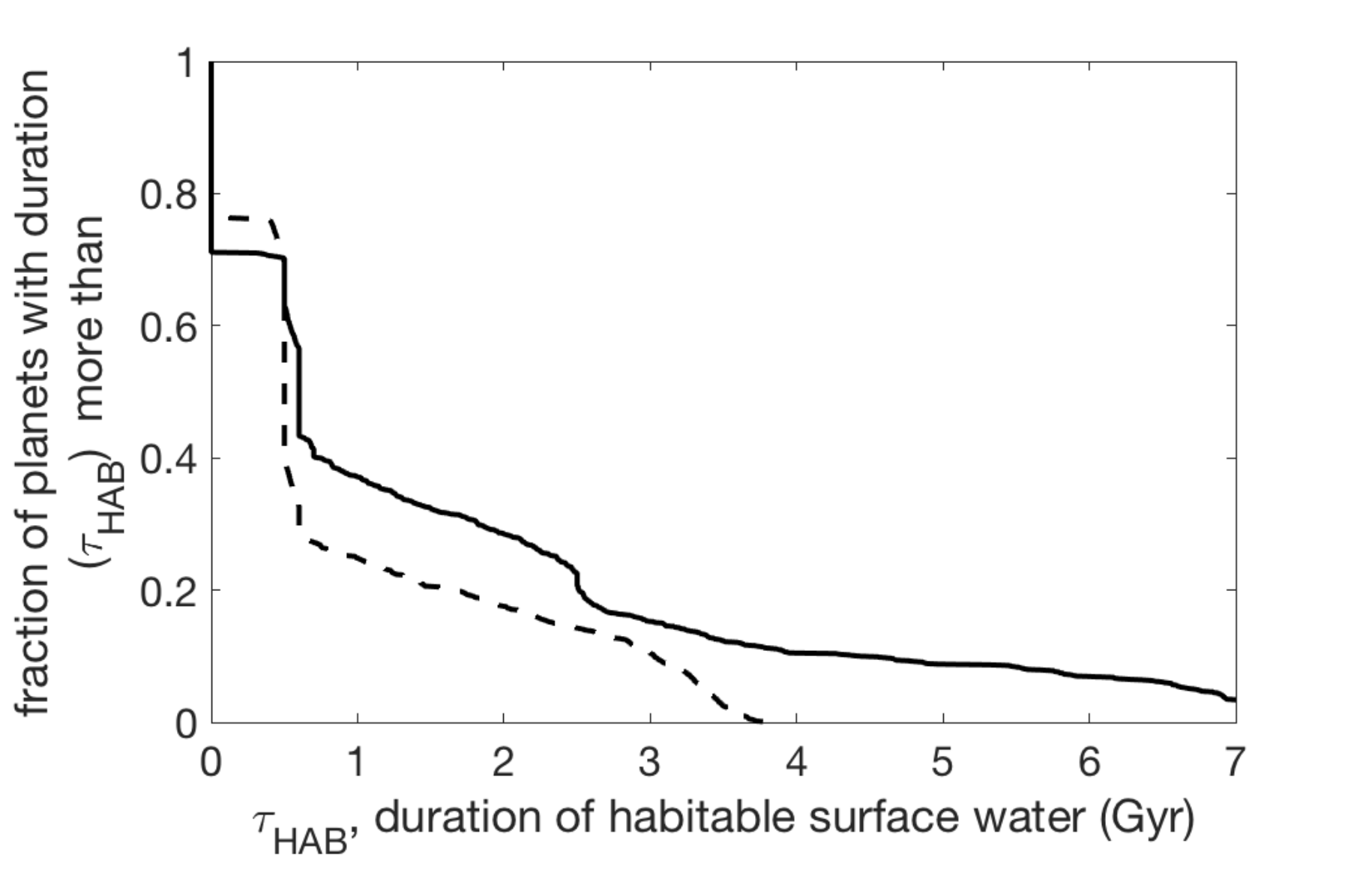}
\caption{Duration of habitable surface water on waterworlds: results of a parameter sweep. \emph{Upper~panel:} the color scale corresponds to the duration of habitable surface water, $\tau_{hab}$~(Gyr). Limits of the parameter sweep are explained in text. Open black circles are never--habitable worlds. For $M$~=~1~$M_\earth$, semimajor axis $a$~=~1.1~AU, planet water mass fraction ($f_W$)~=~0.01. \emph{Middle~panel:} the same, but for $f_W$~=~0.1. Note the change in color scale.  \emph{Lower~panel:} output sorted by the duration of habitable surface water. Solid line, $f_W$~=~0.01; dash-dot line, $f_W$~=~0.1.}
\label{fig:mcuninformed}
\end{centering}
\end{figure}

\subsection{Summary and results of parameter sweep}

%
% \begin{figure*}
%%\epsscale{1.2}
%\includegraphics[width=0.99\columnwidth]{neutral_planetary_habitability_figure_C_H.pdf}
%\includegraphics[width=0.99\columnwidth]{neutral_planetary_habitability_figure_C_Cat.pdf}
%\caption{Diagrams of c, h and cations against one another (2 diagrams) } 
%\label{fig:cvshvscat}
%\end{figure*}

So far we have considered three specific cation-abundance cases. Now we consider a much wider range of cation abundances.

Figure~\ref{fig:chvsccat} shows how the duration of habitable surface water ($\tau_{hab}$) depends on CO$_2$-equivalent atmosphere+ocean C content ($c_C$) and cation abundance (i.e. the extent of rock-water reaction) for semimajor~axis~=~1.1~AU, planet water mass fraction~=~0.01.  For high $c_C$ (black~circles in Figure~\ref{fig:chvsccat}), C swamps all available oceanic and crustal sinks, and so C piles up in the atmosphere (Figure~\ref{fig:bucket}). This leads to sterilizing surface temperatures. When cations released from rock dominate ocean chemistry, pH~is~high, and almost all C is sequestered as CO$_3^{2-}$ ions, HCO$_3^-$ ions, or CaCO$_3$ \citep{KempeDegens1985,Blattler2017}. Therefore little C remains for the atmosphere, and so $p_{\mathrm{CO2}}$ is minimal and has little role in setting planet climate. The planet then has habitable surface water for only $<$1 Gyr (dark~blue~disks at the top of Figure~\ref{fig:chvsccat}). For cation-poor oceans (lower part of plot), $\tau_{hab}$ is longest for the~[C] which gives $p_{\mathrm{CO2}}$ $\sim$ 1 bar at habitable temperatures. This is as expected, because the range of $L_*$ that permits habitable surface water is widest for  $p_{\mathrm{CO2}}$ $\sim$ 1 bar (Figure~\ref{fig:co2matters}). Durations dwindle for lower~[C] and vanish entirely for higher~[C] (excessive $p_{\mathrm{CO2}}$ and/or $T_{surf}$). When cations in the ocean balance~[C] (black horizontal line), the maximum in habitable-surface-water duration occurs at larger C. We suppose that the altered rock layer has the composition of mid-ocean ridge basalt, that it is completely leached of the charge-equivalent of its Na and Ca content, and that the leached thickness is log-uniformly distributed from 0.1~km~to~50~km (Section 3.2). For the purposes of Figures~\ref{fig:chvsccat}-\ref{fig:mcuninformed} we assume that carbonates don't form; if they did form, then a factor~of~$\sim$2 more C would be needed to neutralize the cations. Equilibrium with~minerals (e.g.~CaCO$_3$, NaAlSi$_2$O$_6$) might buffer ocean cation content to $<$0.5 mol/kg, which roughly corresponds to the middle red line in Figure \ref{fig:chvsccat}. Most points above this line have short $\tau_{hab}$. Therefore, our decision to allow cation contents $>$~0.5~mol/kg is geochemically less realistic, but conservative for the purpose of estimating $\tau_{hab}$ on waterworlds.

To explore habitable-surface-water duration, we do a parameter sweep. We consider a hypothetical ensemble of $M_{pl}$~=~$M_\earth$, planet water mass fraction~=~0.01, semimajor axis~=~1.1~AU~worlds (Figure~\ref{fig:mcuninformed}).  Suppose that log($c_C$) is uniformly distributed (independent of the leached thickness) \mbox{from~-5~to~-3} \mbox{(0.3\%~-~30\%~of~water~content}; \mbox{C/H$>1$} is cosmochemically implausible) (Section 3.2). This prior likely overweights small values of $c_C$, because for small values of non-core planet C mass fraction ($c_C$), C~can~be stored mainly in the silicate mantle \citep{Hirschmann2016}. For these cosmochemically and geophysically reasonable priors, the duration of habitable surface water is shown in Figure~\ref{fig:mcuninformed}. The~lower~panel shows that C~has a negative or neutral effect on $\tau_{hab}$ for slightly more than half of the parameter combinations tested. About a~quarter of~worlds never have habitable surface water, due to excessive initial $p_{\mathrm{CO2}}$. Another quarter have $p_{\mathrm{CO2}}$ so low, due to efficient rock leaching, high-pH oceans, and storage of C as CO$_3^{2-}$ ion or HCO$_3^-$ ion, that CO$_2$ has minor climate effect. As a result, the (short) duration of habitable surface water is about the same as for an N$_2$+H$_2$O atmosphere, and $<$~1~Gyr. The remaining half~of~worlds have longer $\tau_{hab}$ due to CO$_2$. In $\sim$9\%~of~cases, the waterworlds stay habitable for longer than the age of the Earth. 

Figure~\ref{fig:mcuninformed} also shows the results for planet water mass fraction~=~0.1. Durations of habitable surface water are shorter, because \mbox{HP-ice} can only be avoided for a narrow (steamy) range of temperatures. Reducing the maximum temperature for life from 450 K to 400 K would reduce $\tau_{hab}$ to $<$~1~Gyr on $f_W$~=~0.1 worlds. For deeper oceans, fewer worlds are sterilized by too-high CO$_2$ values: for these worlds, the solution to pollution is dilution.

Results depend on the mean value of the ratio of CO$_2$-equivalent atmosphere+ocean C content to planet water content (i.e., $c_C/f_W$). If $c_C$ often exceeds 10$^{-3}$, then a greater fraction of waterworlds will have thick CO$_2$-rich atmospheres and be too hot for life.

Mean waterworld lifetime is $\sim$2 Gyr (area under the curves in the lowest panel of Figure~\ref{fig:mcuninformed}). This is not~very~ much less than the maximum planet lifetime for $a$~=~1.1~AU, $M_{pl}$~=~1~$M_\earth$, which is 7~Gyr. After 7~Gyr, the runaway greenhouse occurs even on planets where geochemical cycles aid habitability  \citep{CaldeiraKasting1992,WolfToon2015}. Therefore, cycle-independent planetary habitability is not much less effective at sustaining habitable surface water relative to a hypothetical geochemical cycle that maintains $p_{\mathrm{CO2}}$ at the best value for habitability.

The effect of C on habitability in our model is positive (Figure~\ref{fig:mcuninformed}, lowest panel). Our parameter sweep suggests that the median $\tau_{hab}$ is comparable in the \mbox{random-allocation-of-C case} relative to a C-starved case. Maximum $\tau_{hab}$ is greatly increased \citep{Carter1983}. This gain outweighs the losses due to planets that form with too much C for habitable surface water. Furthermore, life requires C, so very low [C] in the ocean could hinder life's origin and persistence \citep{KahRiding2007}.

\section{Planet assembly model}

We carry out \emph{N}-body simulations (Section 5.1) in order to get a physically well-motivated ensemble of \{$a$,~$M_{pl}$,~$f_W$\}-tuples as input for the waterworld evolution code.  We choose initial and boundary conditions that cause planets to migrate, as this is the easiest way to form HZ waterworlds.  In Section 5.2, we will apply a volatile tracking code to the mass and orbit histories output by the \emph{N}-body simulations.  Embryos are assigned an initial volatile mass percentage that ramps from zero well inside the snowline to $f_{W,max}$ well beyond the snowline. The embryos evolve via migration and mutual gravitational scattering, and develop into planets, some of which sit within the HZ. 
Pebbles \citep{Sato2016,Levison2015} and planetesimals are not tracked. 
The volatile content of the forming planets is tracked as the embryos merge. Readers interested only in geoscience may skip this section.

\subsection{\emph{N}-body model ensemble}

We perform \emph{N}-body simulations including simple gravitational interactions with a smooth, 1D gas disk with a surface density proportional to $r^{-3/2}$, but not including the back-reaction of disk perturbations on the planets.  
This results in a simple strong-migration scenario, similar to that in many other studies (e.g.~\citealt{Cossou2014,Ogihara2015,Sun2017,Izidoro2017,RoncodeElia2018,Raymond2018}).
For the purposes of this study, the \emph{N}-body simulations serve to provide illustrative migration and collisional histories.  Therefore, details of the \emph{N}-body simulations are not essential; some readers may choose to skip to Section 5.2.

We employ the REBOUND \emph{N}-body code \citep{ReinLiu2012} and its IAS15 integrator \citep{ReinSpiegel2015}, with additional user-defined forces to represent gravitational tidal drag from the disk \citep{ZhouLin2007}.  
Each simulation is initialized with dozens of planetary embryos near or beyond the H$_2$O snowline, orbiting a solar-mass star. 
No gas giants exist at the start of our simulation, consistent with the wetter-than-solar-system cases we model \citep{BatyginLaughlin2015,Morbidelli2016}. 
The following parameters are varied across simulations: the total mass in embryos (from~10~to~40~$M_{\earth}$); the outer edge of the disk (from~3~to~10~AU); and a scale factor for the mass of the gas disk relative to a fiducial minimum mass solar nebula (from ~2-to-16 times more massive than the minimum mass solar nebula). 

For each set of model parameters, we generate multiple sets of initial conditions, resulting in a total~of~$>$145~\mbox{\emph{N}-body}~runs. 
Embryo initial masses are $\sim$1~Mars~mass. 
The initial semi-major axes are drawn from a power-law distribution with a power-law index of -1.5.     
Orbits are initialized with modest eccentricities and low inclinations (relative to the plane of the gas disk). 
The initial orbital eccentricities and inclinations are drawn from Rayleigh distributions with a scale parameter of 0.1 for eccentricity, and 0.1~radians for inclination.  
In practice, the eccentricities and inclinations within a system rapidly damp to a level where damping is balanced by \emph{N}-body excitation from neighbors.  This~choice of initial conditions helps to avoid a very brief phase of rapid collisions before the system is able to relax to a physically plausible set of inclinations and eccentricities.  
The three remaining angles (argument of pericenter, longitude of ascending node, mean anomaly) were drawn from uniform distributions (0 $\leftrightarrow$ $2\pi$).  

While the gas disk is present, orbital migration is parameterized following Eqns.\ 19 and 20 from \citet{ZhouLin2007}. The gravitational accelerations due to tides from the gas disk are proportional to: (a)~the difference in the velocity of the planet (or embryo) from the circular velocity at the planet's current position (i.e., proxy for the gas velocity), (b)~the planet-star mass ratio, (c)~a 1/$a^2$ term that accounts for a $a^{-3/2}$ power-law dependence of the gas-disk surface density, and (d)~a time-dependent factor that exhibits first slow dispersal, and then rapid dispersal due to photoevaporation.  This prescription results in orbital migration, eccentricity damping, and inclination damping that have self-consistent timescales \citep{ZhouLin2007}.  
For each of our simulations, the gas-disk mass initially undergoes exponential decay with a characteristic timescale of 2 Myr.  
Starting at 3 Myr, the gas-disk decay accelerates due to photoevaporation, so the exponential decay rate is decreased to 50,000 years.  
After 4 Myr, the gas-disk mass has essentially dispersed, so the gas-disk mass is set to zero and the simulation becomes a pure \emph{N}-body integration.  

Following gas disk dispersal, planets continue to scatter and collide with one another. We consider planets as having collided when they come within a radius corresponding to a density of 0.125~$\times$~that~of~Mars. Collisions are assumed to be perfect mergers in the \emph{N}-body code. 
%Thus, we do not include the feedback of evolving volatile content (Section 4) on the masses of the planets nor on the probability of collision. This is acceptable because collisions are not very erosive in our simulations, so the masses are not much different from the masses with no erosion.

Of our simulations, 29\% of the runs resulted in one or more planets in the HZ, based on the zero-age main sequence luminosity of the Sun ($\sim$0.7$\times$~present~solar~luminosity).  We set aside \emph{N}-body simulations where no planets survived exterior to the HZ, as the results of those simulations may be impacted by edge effects (i.e., if embryos formed beyond the outer end of our initial set of embryos). In some cases, multiple planets in the HZ at the end of the simulation have semimajor axes differing by less than $<$0.05~AU from one another.  In these cases, we merge such planets, as they are expected to collide if we were able to extend each of the \emph{N}-body simulations. We do not include the loss of water due to these (final) giant impacts. After applying these filters, we are left with \mbox{$\sim$30~HZ~planets}.

 \begin{figure}
%\epsscale{1.2}
\includegraphics[width=0.99\columnwidth,clip=true,trim={7mm 60mm 7mm 60mm}]{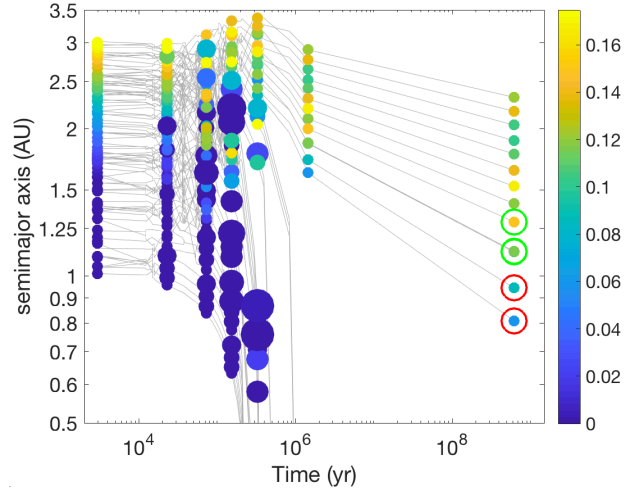}
\caption{Example output from our planet assembly model, showing stages in planet
growth. The color scale shows H$_2$O fraction ($f_W$). 
Red rings mark planets that initially have habitable surface water.
Green rings mark HZ planets with initially frozen surfaces. Disk size corresponds to planet
mass. Planets migrate inwards for the first few Myr of the simulation subject to parameterized,
imposed migration torques.} 
\label{fig:accretion}
\end{figure}

\subsection{Volatile-tracking model}

Embryo H$_2$O content is set as follows. We represent the cumulative effects of the evolving H$_2$O-ice snowline on the water mass fraction of planetary embryos $f_{W,o}$ \citep{Hartmann2017,Piso2015} by 

\begin{equation}
\begin{split}
f_{W,o} &= f_{W,max} \left(a - a_{sl} / a_{sl} \right) , a_{sl} < a < 2 a_{sl} \\
f_{W,o} &= f_{W,max}, a > 2 a_{sl} \\
f_{W,o} &= 0, a < a_{sl}   
\end{split}
\end{equation}

%\footnote{Although diffusion plus drift can lead to $f_W >$ 50\% \citep{Ciesla2015}, we find $f_{W,max}$ 20\% is sufficient for waterworlds, and so there is no need to consider $f_W >$ 50\% scenarios.}.
\noindent where $a_{sl}$~=~1.6~AU is an anchor for the evolving snowline \citep{Mulders2015}. For each \emph{N}-body simulation, we consider $f_{W,max}$~=~\{50\%,~20\%,~5\%,~1\%\}. %
The highest value, 50\%, corresponds to theoretical expectations from condensation of solar-composition gas \citep{Ciesla2015}. Intermediate values match density data for Ceres \citep{Thomas2005} and inferences based on Earth's oxidation state \citep{Rubie2015,Monteux2018}. 5\%~corresponds to meteorite (carbonaceous chondrite) data. Why are the carbonaceous chondrites relatively dry? One explanation involves heating of planetesimals by short-lived radionuclides (SLRs; $^{26}$Al~and~$^{60}$Fe), leading to melting of water ice. Melt-water then oxidizes the rocks, and H$_2$ returns to the nebula (e.g., \citealt{Young2001,Rosenberg2001,Monteux2018}). Because most protoplanetary disks probably had fewer SLRs than our own (\citealt{Gaidos2009,Gounelle2015,Lichtenberg2016,Dwardakas2017}; see also \citealt{Jura2013}), this process would operate differently (or not at all) in other planetary systems. Another explanation for the dryness of carbonaceous chondrites relies on Jupiter's early formation \citep{Morbidelli2016}; gas giants of Jupiter's size and semimajor axis are not typical. The~range of~$f_{W,max}$ we consider corresponds to processes occurring during growth from gas+dust to embryos. We neglect H$_2$O loss by XUV-stripping at the pre-embryo stage \citep{Odert2018}.

%Seperately, we assign to each final planet an initial atmosphere+ocean C inventory (following Section 3.1). 

Growing planets shed water during giant impacts. For volatile-rich planets, at the embryo stage and larger, the most plausible mechanism for losing many wt\% of water is giant impacts. The shift to giant-impact dominance can be seen by comparing Figures 14-15 of \citealt{Schlichting2015}, and extrapolating to the smallest volatile~layer considered in this~paper (10$^8$~kg~m$^{-2}$). Such deep volatile layers greatly inhibit volatile loss by $r$$<$50~km projectiles, and blunt the escape-to-space efficiency of larger projectiles. (For volatile-poor planets, by contrast, small impacts are extremely erosive; \citealt{Schlichting2015}.) Combined with the expectation that most of the impacting mass is in the largest projectiles, erosion by giant~impact is the main water loss process for waterworlds.

H$_2$O is attrited by individually-tracked giant impacts  \citep{Marcus2010,Stewart2014,GendaAbe2005}, using the relative-velocity vectors from the \emph{N}-body code.
% \begin{equation}
%\begin{split}
%\mu~=~M_1 M_2 / (M_1 + M_2) \\
%Q_r~=~0.5\mu V_i^2 / (M_1 + M_2) \\
%Q_s~=~Q_r (1 + M_p/M_t) (1 - b) \\
%\end{split}
%\end{equation}
The dependence on relative velocity of fractional ocean loss is taken from \citet{Stewart2014}. Because the parameterization of \citet{Stewart2014} is for $f_W$ $\sim$ 10$^{-4}$, and ocean loss efficiency decreases with ocean thickness \citep{InamdarSchlichting2016}, this is conservative in terms of forming waterworlds. More recent calculations find lower ocean-loss values \citep{Burger2018}. When embryos collide with other embryos, or collide with planets \citep{Golabek2018,Maindl2017}, Fe-metal cores are assumed to merge efficiently. We do not track protection of H$_2$O from giant impacts by dissolution within deep magma oceans formed during earlier giant impacts \citep{ChachanStevenson2018}. This is also conservative in terms of forming waterworlds.

 \begin{figure}
\epsscale{1.2}
\includegraphics[width=1.02\columnwidth,clip=true,trim={25mm 0mm 30mm 0mm}]{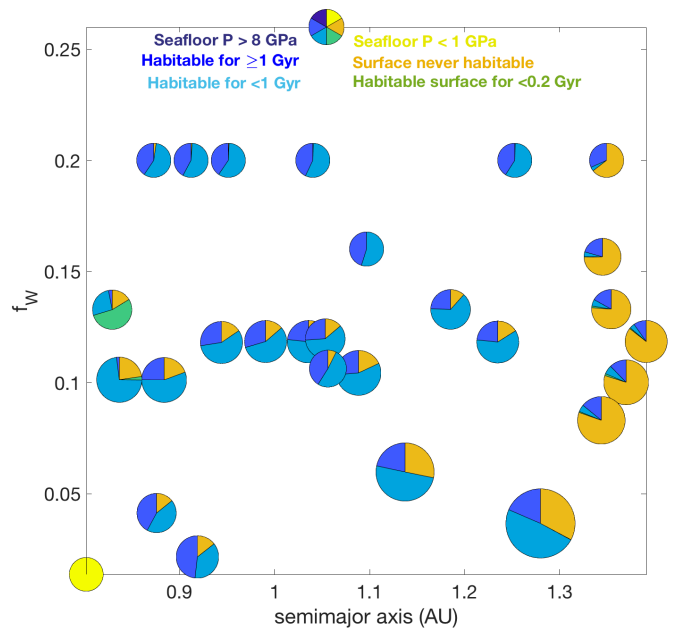}
\caption{Fate of the HZ planets generated by an ensemble of \emph{N}-body simulations. Model output for maximum embryo water content~($f_{W,max}$)~=~20\%. Disk size increases with $M_{pl}$, which ranges from 0.1 to 1.5 $M_\earth$. For each planet, 1000 random draws are made from a cosmochemically reasonable range of initial atmosphere+ocean C content ($c_C$) and (independently) from a geophysically reasonable range of cation abundance (Section 3.2), and the resulting durations of habitable surface water are computed using the methods of Section 4. The disk colors summarize the outcomes. The yellow planet at $a$~=~0.85~AU has seafloor pressures too low to suppress C exchange between the convecting mantle and the water ocean. Dark blue colors correspond to $\tau_{hab}$~$\ge$~1~Gyr. Zero worlds have seafloor~$P$~$>$~8~GPa, so no worlds are colored purple. See Section 5.3 for a discussion.} 
\label{fig:piecharts}
\end{figure}

\subsection{Results}

Planet assembly is relatively gentle in our model. Most planet-forming collisions occur while disk gas maintains low planet eccentricities and inclinations. As a result, $\left| v_\infty \right|$ is small, and because giant-impact loss is sensitive to $\left| v_\infty \right|$, these early collisions have individually-minor effects on planet mass. When $f_W$~$\ge$~0.15, an individual giant impact does not remove a large percentage of the initial planet volatile endowment (Figure~ \ref{fig:accretion}). %Later collisions are more erosive, but our habitable-zone planets typically suffer zero or one such collision.

Planet water mass fraction is anti-correlated with $M_{pl}$, due to the larger number of devolatilizing giant impacts needed to assemble larger planets. For a given $M_{pl}$, worlds assembled via collisions between equal-mass impactors have higher~$f_W$ than worlds that are built up piecemeal (one embryo added a time). This is because piecemeal assembly requires a greater number of giant impacts \citep{InamdarSchlichting2016}. There is no obvious trend of $f_W$~with~$a$ (nor $M_{pl}$~with~$a$) within the HZ. 

The final planet water mass fractions are sensitive to $f_{W,max}$, as expected \citep{Mulders2015,Ciesla2015}. For a given $f_{W,max}$, seafloor pressures are bunched because both the weight per kg of water, and the number of water-removing giant impacts, increase with planet mass. None of the $f_{W,max}$~=~0.05 or $f_{W,max}$~=~0.01~planets have seafloor pressures $>$1~GPa. By contrast, 97\% of the $f_{W,max}$~=~0.2~worlds have 1-8~GPa seafloor pressure. Finally, of the $f_{W,max}$~=~0.5~worlds, 7\% have seafloor pressure $<$1~GPa, 3\% have seafloor pressure $\sim$ 6~GPa, and 90\% have seafloor pressure $>$~8~GPa. These relatively low seafloor pressures are the consequence of low surface gravity: $M_{pl}$ averages 0.3 $M_\earth$ for this ensemble. To calculate the gravitational acceleration experienced by the ocean, we follow \citet{Valencia2007} and let $R_{seafloor}/R_{\Earth} = (M_{pl} / M_{\Earth}) \,^{0.27}$. If more massive planets had emerged in our simulations, then we would expect seafloor pressures to be higher for a given $f_{W,max}$.

We injected the \{$a$,~$M_{pl}$,~$f_W$\} tuples from the $f_{W,max}$~=~0.2 planet assembly run into the waterworld evolution code, varying the CO$_2$-equivalent atmosphere+ocean C content and varying cation-leaching using the same procedure and the same limits as in Section 4. The results (Figure~\ref{fig:piecharts}) show that habitable surface water durations~$>$~1~Gyr occur for $\sim$20\% of the simulated parameter combinations. $\tau_{hab}$~$>$~4.5~Gyr occurs for $\sim$5\% of the simulated parameter combinations. Only around a third of trials never have habitable surface water. Never-habitable outcomes can correspond to always-icy surfaces (especially near the outer edge of the HZ), or to $p_{\mathrm{CO2}}$ $\gtrsim$ 40~bars. Never-habitable outcomes become less likely as planet water mass fraction increases, mainly because of dilution. These results are illustrative only because mean (and median) $M_{pl}$ for this ensemble is 0.3 $M_\earth$. The lower gravity of 0.3 $M_\earth$ worlds relative to our 1 $M_\earth$ reference will move the inner edge of the HZ further out than is assumed in Figure~\ref{fig:piecharts} \citep{Kopparapu2014}.

\section{Discussion}

\subsection{All small-radius HZ planets are potentially waterworlds}
Waterworlds cannot be distinguished from bare rocks by measuring radius and density. Even if radius and density measurements were perfect, a small increase in core mass fraction could offset a 100~km increase in ocean depth \citep{RogersSeager2010}. Indeed, scatter in core mass fraction is expected for planets assembled by giant impacts \citep{Marcus2010b}. The best-determined planet radius is that of Kepler-93b; 1.481$\pm$0.019~$R_\earth$ \citep{Ballard2014}, an error of 120~km in radius. Yet a~120~km~deep~ocean is enough to be in the cycle-independent planetary regime discussed in this paper, and a~0~km~deep~ocean is a bare rock. Small uncertainties in star radius, or in silicate-mantle composition (e.g.~\citealt{Dorn2015,StamenkovicSeager2016}) can also affect retrievals.  Therefore, all small-radius HZ planets are potentially waterworlds, and this will remain the case for many years \citep{Dorn2017,Simpson2017,Unterborn2018}.

\subsection{Climate-stabilizing feedbacks?}
The main climate feedback in our code is the decrease in CO$_2$ solubility with increasing temperature, which is usually destabilizing (Section 4.1). We now speculate on possible feedbacks, not included in our code, that might stabilize climate.

\begin{enumerate}[leftmargin=0.9em]

\item When carbonates dissolve, CO$_2$ is consumed (sic; \citealt{Zeebe2012}). This is because (for circumneutral pH) one~mole of Ca$^{2+}$ charge-balances two~moles of HCO$_3^-$, whereas CaCO$_3$ pairs one~mole of Ca with only one~mole of C (Figure~\ref{fig:bjerrum}). Carbonates dissolve with rising $T$ \citep{DolejsManning2010}, so the corresponding drawdown of CO$_2$ is a climate-stabilizing feedback. This feedback is curtailed on Earth by the build-up of an lag of insoluble sediment, for example continent-derived silt and clay. Insoluble sediments would be in short supply on waterworlds, and so the carbonate-dissolution feedback could be stronger on waterworlds. 
%Solubility data for aragonite at $\gtrsim$GPa~pressure could test this hypothesis. 

\item Pervasive aqueous alteration lowers the density of the upper crust. Low density of altered layers may cause rising magma to spread out beneath the altered layer, failing to reach the surface as an extrusive lava, and instead crystallizing in the subsurface as an intrusion. Intrusions have low permeability and so are resistant to alteration (in contrast to extrusive lavas that have high permeability). This negative feedback on aqueous alteration might limit cation supply.

\item When high-pressure ice forms, it excludes C from the crystal structure. This raises the C concentration in the ocean, and thus $p_{\mathrm{CO2}}$. This negative feedback on cooling will have a strength that depends on the extent to which C~is~taken~up by clathrates \citep{Levi2017}. Moreover, according to Levi et al., the solubility of CO$_2$ in water in equilibrium with CO$_2$ clathrate decreases with decreasing temperature. This solubility behavior is an additional negative feedback.

%Salts precipitating from the ocean as it cools after the magma ocean stage (including carbonates) may isolate the rocks from the ocean, due to the low permeability of salts \citep{Kargel2000,McKinnonZolensky2003}. This will moderate cooling by inhibiting the leaching that draws down CO$_2$.

\item High C concentration lowers pH. This favors rock leaching, which is a negative feedback on $p_{\mathrm{CO2}}$ \citep{SchoonenSmirnov2016}. But because water/rock~ratio is effectively a free parameter over many orders of magnitude, the effect of water/rock mass~ratio is more important in our model (Figure \ref{fig:sfb}). Moreover, high-pH NaOH solutions corrode silicate glass, a positive feedback.

\item It has been suggested that carbonate in seafloor basalt is an important C sink on Earth, and has a temperature-dependent formation rate, and so might contribute to Earth's climate stability \citep{Coogan2016}. 

\end{enumerate}
%Cracks are sealed more easily at high temperature. By~limiting leaching, this is a positive feedback on a warming climate.

%We do not track the accumulation of O$_2$ in the atmosphere-ocean \citep{Tian2015}, although O$_2$ will compete with CO$_2$ for crustal cations and thus reduce carbonate formation. This is likely a positive feedback on $T$ rise.

\noindent These feedbacks might extend $\tau_{hab}$ beyond the already-long durations calculated in Section 4. 

\subsection{Cycle-dependent versus cycle-independent planetary habitability}

The number of planet-years of habitable surface water enabled by cycle-independent planetary habitability on waterworlds ($T_{nocycle,ww}$) can be estimated as follows:

\begin{equation}
T_{nocycle,ww} = \sum\limits_{i=1}^{\eta_\earth N_* } \tau_{hab,i}(f_{W,i}, c_{C,i}, M_{pl,i}, ...)  
\end{equation}

\noindent where $N_*$ is the number of Sun-like stars in the galaxy, and $\eta_\earth$ ($\approx$~0.1) is the number of HZ small-radius exoplanets per Sun-like star. Given a $\tau_{hab}$ look-up table (e.g.~Figure~\ref{fig:mcuninformed}), Equation 6 can be evaluated (e.g.~Figure~\ref{fig:piecharts}) given estimates (from plate formation models constrained by data) of the distribution of $f_W$, $c_C$, $M_{pl}$, and so on. How does $T_{nocycle,ww}$ compare to the number of planet-years of habitable surface water maintained by geochemical cycles ($T_{cycle}$)? We have

\begin{equation}
\frac{T_{cycle}}{N_* \eta_\earth  } \sim   \,x_{IC} \, x_{evo}\,  \,5 \,\mathrm{Gyr} 
\end{equation}

\noindent Here,  $x_{IC}$ is the~fraction of small-radius HZ exoplanets with suitable initial conditions for habitability maintained by geochemical cycles, $x_{evo}$ accounts for planets where geochemical cycles initially maintain habitability but subsequently fail to do so, and 5~Gyr is a typical maximum duration of habitable surface water (before the runaway greenhouse) for a planet initialized at random log($a$) within the HZ of a Sun-like star (Figure~\ref{fig:co2matters}).

Despite the familiarity of habitability maintained by geochemical cycles, it is possible that $T_{cycle}$~$\ll$~$T_{nocycle,ww}$. This is because $x_{IC}$ and $x_{evo}$ are both very uncertain. This uncertainty is because predicting geochemical cycles involving habitable environments using basic models is difficult. This difficulty can be traced to the low temperatures of habitable environments. These low temperatures bring kinetic factors -- such as grain size, permeability, catalysts, and rock exposure mechanisms -- to the fore \mbox{\citep{WhiteBrantley1995}}. It is also difficult to forensically reconstruct the processes that buffer Earth's $p_{\mathrm{CO2}}$ \citep{BroeckerSanyal1998,Beerling2008}. Negative feedbacks involving geochemical cycles probably buffer the post-0.4~Ga atmospheric partial pressure of CO$_2$~($p_{\mathrm{CO2}}$) in Earth's atmosphere (e.g., \citealt{ZeebeCaldeira2008,Kump2000,Stolper2016}). The effectiveness and underpinnings of these feedbacks on Earth are poorly understood \citep{EdmondHuh2003,MaherChamberlin2014,Galy2015,KrissansenTottonCatling2017}. Therefore, it is hard to say how often these feedbacks break down or are overwhelmed due to the absence of land, supply limitation / sluggish tectonics, inhibition of carbonate formation due to low pH or to SO$_2$, contingencies of tectonic or biological evolution, and so on \citep{Abbot2012,Kopp2005,EdwardsEhlmann2015,Flament2016,BullockMoore2007,Foley2015,FoleySmye2018,HalevySchrag2009,Genda2016, GalbraithEggleston2017}.
Without a mechanistic understanding of planet-scale climate-stabilizing feedbacks on Earth, we cannot say whether these feedbacks are common among planets, or very uncommon \citep{Lacki2016}.

%
%\begin{deluxetable*}{lrrrrrrrrrcrl}
%\tabletypesize{\scriptsize}
%\setlength{\tabcolsep}{0.02in} 
%%\rotate
%\tablecaption{Predictions for selected small-radius habitable zone planets.}
%\tablewidth{0pt}
%\tablehead{
%\colhead{Planet} 
%& \colhead{$p$ (d)} 
%& \colhead{\pbox{1cm}{R \\ (R$_\earth$)}} 
%&  \colhead{\pbox{2.2cm}{M \\ (M$_\earth$)}} 
%& \colhead{\pbox{2.2cm}{Optimal CO$_2$}} 
%& \colhead{Prediction2} 
%& \colhead{ \pbox{1.2cm}{Prediction3}} 
%& \colhead{\pbox{1.2cm}{Prediction4}}} 
%\startdata
%\\
%\multicolumn{7}{l}{\emph{Orbiting G stars (evolutionary predictions included):}} \\
%Kepler-452b  \\
%$\tau$ Cet e  \\
%Kepler-1638b \tablenotemark{a}  \\
%Kepler-1606b \\
%Kepler-1090b \\
%Kepler-22b \\
%\\
%\multicolumn{7}{l}{\emph{Orbiting K/M stars (initial conditions only):}} \\
%Proxima Cen b\\
%TRAPPIST-1 e	\\
%GJ 667 C c	\\
%Kepler-442 b	\\
%GJ 667 C f	\\
%Kepler-1229 b	\\
%TRAPPIST-1 f	\\
%Ross 128b \\
%LHS 1140 b	\\
%Kepler-62 f	\\
%Kepler-186 f	 \\
%GJ 667 C e \\
%TRAPPIST-1 g \\
%% Text for table notes should follow after the \enddata but before
%% the \end{deluxetable}. Make sure there is at least one \tablenotemark
%% in the table for each \tablenotetext.
%\tablecomments{The applicability of our model to any particular planet depends on several assumptions that are not obviously correct and which should be tested in future work.}
%%\tablenotetext{a}{Test footnote.}
%\label{table:specific}
%\end{deluxetable*}
%
%

\subsection{Limits to life}

Is 450K a reasonable upper $T$ limit for habitability? A temperature of 450K exceeds the highest temperature for which life has been observed to proliferate (395K, \citealt{Takai2008}). However, no fundamental barriers have been identified to life at $>$400K \citep{Committee2007}. Theory tentatively suggests 423-453K as the upper $T$ limit for life in general \citep{Bains2015}. If we lower the limit for life to 400K, then our model worlds with planet water mass fraction ($f_W$)~=~0.1 will almost~always have \mbox{HP-ice} at times when their surface temperature is habitable (Figure \ref{fig:tsurfevo01}). It is conceivable that Earth's $T_{surf}$ was $>$350K at $>$3 Ga based on isotopic data (e.g.~\citealt{Tartese2017}, but see also \citealt{Blake2010}) and on the $T$ tolerance of resurrected ancestral proteins  \citep{Akanuma2013,Garcia2017}.  %Because culturing hyperthermophilic chemoautotrophic archaea that are adapted to pressures $>$10 MPa is quite difficult, the absence of definite observations of life at $>$400K is not strong evidence that such life does not exist on Earth

Is 40 bars a reasonable upper $p_{\mathrm{CO2}}$ limit for habitability? We use this upper limit mainly because, above this value, fine-tuning of $L_*$ is increasingly required to avoid uninhabitably high $T_{surf}$ (Figures~\ref{fig:co2matters}~and~ \ref{fig:ramirezlozenge}). Therefore, 40 bars is a conservatively low upper limit. Above 70 bars and 300K,  CO$_2$ is a supercritical fluid. Supercritical CO$_2$ is an excellent organic solvent and bactericidal agent. However, within the ocean, life might persist. We consider near-sea-surface habitability in this paper, but note that life proliferates at 1~GPa  \citep{Sharma2002}.

Extreme pH values destabilize simple biological molecules, but life on Earth solves these problems \citep{Committee2007}. The pH requirements for the origin of life might be more restrictive.

Life as we know it requires nutrients other than C, e.g.~P. Yet it is unclear to what extent life requires \emph{resupply} of such nutrients \citep{Moore2017}. To the contrary, given an initial nutrient budget and a sustained source of photosynthetic energy, a biosphere might be sustained by heterotrophy, i.e. biomass recycling.  Biomass recycling is easier for waterworlds that lack land and lack ocean-mantle geochemical cycling, because biomass cannot be buried by siliciclastic sediment, nor subducted by tectonics. If biomass recycling is incomplete \citep{RothmanForney2007}, then to stave off productivity decline will require nutrient resupply. Nutrient resupply will be slower (relative to Earth) on waterworlds that lack ocean-mantle geochemical cycling, because land is absent (so sub-aerial weathering cannot occur), and volcanism is absent or minor. Moreover, the larger water volume will dilute nutrients. In our model, nutrients are supplied by water-rock reactions at the seafloor early in planet history, with limited nutrient resupply thereafter (by slow diffusive seafloor leaching, or minor ongoing volcanism). Resupply might be shut down entirely by high-pressure (HP) ice. However, if \mbox{HP-ice} both convects and is melted at its base (as on Ganymede?), then nutrient supply to the ocean may continue (e.g., \citealt{Kalousova2018,KalousovaSotin2018}). Alternatively, nutrients can be supplied to habitable surface water by bolides and interplanetary dust, just as Earth's seafloor is nourished by occasional whale-falls. On a planet with habitable surface water, chemosynthesis (which requires an ongoing supply of rock) is greatly exceeded as a potential source of energy by photosynthesis. If photosynthesis occurs, then (given the likelihood of nutrient recycling), it has not clear how potentially observable parameters would scale with geological nutrient flux. Therefore, it is hard to make a testable prediction about the nutrient budgets of water-rich exoplanets that have habitable surface water.

Low-temperature hydrothermal vents, which are implied by our models, are a frequently proposed site for the origin of life (e.g. \citealt{Russell2010}). %However, very high pH   \citep{Committee2007}.%Origin-of-life scenarios are poorly connected to observables. 

\subsection{Uncertainties}

\noindent \emph{Astrophysics.} It would be difficult to gather enough water to form a waterworld if habitable-zone planets form without a major contribution of material from beyond the nebula snowline. 

If Fe-metal and atmophiles do not mix during giant impacts, then it is difficult to trap C~in~the~core \citep{Rubie2015}. This would lead to an atmosphere+ocean that inherits the C/H anticipated for nebular materials, which is more \mbox{C-rich} than considered here. The reaction Fe$^0$ + H$_2$O $\rightarrow$ Fe$^{2+}$O + H$_2$, with subsequent escape-to-space of H, could also raise C/H.  %Treating $c_C$ as independent of $f_W$ is a simplification; when C is large, the metal-core sump spills over \citep{Dasgupta2013}, and so a positive correlation between H and C is expected for large C. 

%However, exploring a broad $c_C$-$f_{W,max}$ range is the right response to our gross uncertainty \citep{Hirschmann2016,Bergin2015} about the controls of $c_C$.% in real habitable planets. This could change with improved observational constraints on $c_C$-$f_W$ correlations

Our results are sensitive to the assumption that initially-ice-coated worlds gather enough meteoritic debris during the late stages of planet accretion to lower albedo. If we are wrong and albedos of initially-ice-coated worlds all stay so high that surface ice does not melt, then the only habitable waterworlds will be those that are unfrozen for zero-age main sequence stellar luminosity (i.e., $a$~$\lesssim$~1.1~AU). For $a$~$\lesssim$~1.1~AU, the runaway greenhouse occurs relatively early in stellar main sequence evolution, so if surface ice does not melt, then most waterworlds with habitable surface water will orbit stars younger than the Sun. Because many HZ planets have $a$~$>$~1.1~AU, then if surface ice does not melt that would lower the number of waterworlds with habitable surface water by a factor~of~$>$2. 

We assume that our worlds lack H$_2$, so it is reasonably self-consistent to assume waterworlds have CO$_2$ (and not~CH$_4$) atmospheres. If some H$_2$ remains, then it could fuel early hydrodynamic escape of H$_2$O, and potentially aid Fischer-Tropsch type synthesis of organics \citep{Gaidos2005}. We assume a fixed p$_{\mathrm{N2}}$, $\sim$0.8 bar, but variation in the abundance of N~volatiles may be important, especially if planets draw material from $a$~$\gtrsim$~10~AU \citep{Marounina2018,Atreya2010}. N$_2$ affects $T_{surf}$ by pressure broadening, Rayleigh scattering, e.t.c. \citep{Goldblatt2013}, and NH$_3$ lowers the freezing-point. On Earth, the air:ocean partition of N$_2$ is $\sim$10$^3$:1 \citep{Ward2012}; a~very~massive~ocean becomes a major N$_2$ reservoir.

We neglect tidal heating \citep{Henning2009}. There is no evidence that tidal heating exceeds radiogenic heating for any planet in the HZ of an FGK star older than 100 Myr.

\vspace{0.07in}
\noindent \emph{Geophysics.} In order to make our calculations simpler, we understate the pressure needed to completely suppress C exchange between the convecting mantle and the water ocean. This pressure could be closer to 3~GPa than 1~GPa \citep{Noack2016}. %Even on waterworlds, volcanism may continue due to mantle plumes. 
%Feedbacks between $T_{surf}$ and volcanism are minor \citep{Kite2009}.

%Tidal heating from the orbital expansion of a $M$ = 0.01$M_\earth$ moon may delay magma ocean crystallisation for $O$(100 Myr) \citep{Zahnle2015} and thus protect the water  from XUV stripping .

%The theory of mantle convection by solid-state creep is well-developed. Asthenosphere rheology uncertainties \citep{KorenagaKarato2008,Karato2013} are small for our purposes. Our~range of cations/C in Figure~\ref{fig:mcuninformed} brackets the uncertainties. %Since Ca$^{2+}$ in seawater is twice as effective at removing CO$_2$ from the atmosphere as is Ca$^{2+}$ in carbonate, the~key is how much rock interacts with water -- not whether or not carbonate forms on the seafloor. 

We assume stagnant-lid tectonics. We suspect stagnant~lid is the default mode of planetary tectonics, in part because stagnant-lid worlds are common in the solar system whereas plate tectonics is unique, and in part because of the difficulty of explaining plate~tectonics \citep{Korenaga2013}. Also, plate~tectonics might require volcanism \citep{Sleep2000}, and volcanism is curtailed on waterworlds \citep{Kite2009}. What if plate~tectonics nevertheless operates on waterworlds \citep{NoackBreuer2014}? In that case, relative to stagnant~lid mode the lithosphere might store less C. That is because plate~tectonics involves a thin carbonated layer that is continually subducted and decarbonated  \citep{KelemenManning2015}, whereas stagnant-lid mode slowly builds-up a potentially thick layer of thermally stable carbonated rocks \citep{Pollack1987,FoleySmye2018}. For a planet with plate tectonic resurfacing (or heat-pipe tectonics), the ocean may interact with a large (time-integrated) mass of rock \citep{SleepZahnle2001,Valencia2018}. Increased rock supply makes it more likely that ocean chemistry would reach equilibrium with minerals \citep{Sillen1967}. If plate tectonics operated without volcanism, then the seafloor would be composed of mantle rocks. These rocks, when aqueously alterated, produce copious H$_2$ \citep{Klein2013}. Copious H$_2$ production would also occur if the crust extruded at the end of magma ocean crystallization was very olivine-rich  \citep{Sleep2004}. The alteration of mantle rocks to form serpentine would lead to a Ca$^{2+}$-rich ocean, with high pH \citep{FrostBeard2007}. In addition, H$_2$ production would affect atmospheric composition and climate (e.g. \citealt{WordsworthPierrehumbert2013b}).

 %Waterworld evolution in plate~tectonics mode is potentially more complicated than in stagnant~lid mode. %Given our stagnant~lid assumption, we self-consistently assume that the areal exposure of mantle rock at the seafloor is small. Areally extensive outcrops indicate rifting and plate~tectonics. Mantle rock exposure, due to low permeability, would decrease cation availability for leaching. %In plate~tectonics mode, the  

\vspace{0.07in}

\noindent \emph{Geochemistry and climate.} We use results from the \mbox{1D} climate model of \citet{WordsworthPierrehumbert2013} in this study (e.g., Figure~\ref{fig:co2matters}), which uses HITRAN H$_2$O absorption coefficients and a fixed relative humidity of 1.0. Subsequent work by  \citep{Ramirez2014} uses improved (HITEMP) H$_2$O absorption coefficients and an adjustable relative humidity. The results of a sensitivity test using the \citet{Ramirez2014} work are shown in Appendix~E. Recent 3D GCM results validate the trends shown in Figure~\ref{fig:co2matters} (e.g. \citealt{Leconte2013,WolfToon2015,Popp2016,Wolfetal2017,Kopparapu2017}). However, the amplitudes, location, and even existence of the instabilities discussed in Section 4 may depend on the specifics of the climate model, including how relative humidity and cloud cover respond to changing $T_{surf}$ (e.g. \citealt{Yang2016}). %For example, planetary albedo declines with increasing temperature in \citet{WordsworthPierrehumbert2013} because pH2O increases and so the atmospheric absorption of H2O in the visible increasingly offsets the cooling effect of CO2-induced Rayleigh scattering at high $L_*$. This might not be the case in a model that included clouds \citep{Popp2016}. 
%%We can say little about the effect of planet mass because the gravity correction on the greenhouse effect is important (and not included in this study) \citep{Pierrehumbert2010}. %On planets bigger than Earth, the high-permeability fractured zone in the uppermost crust is smaller \citep{Sleep2000b}.

%The initial partitioning of C during and immediately after the magma ocean phase is uncertain, especially the extent to which Fe metal, silicates, and H$_2$O-rich layers equilibrate \citep{Rubie2015}. Thi 

Our treatment of both carbonate formation and CO$_2$~solubility in cation-rich fluids in Section 4 is crude and this could shift the optimum CO$_2$-equivalent atmosphere+ocean C content for habitability by a factor~of~$\sim$2. %\textbf{Maybe just treat Ca as a monovalent cation}. %On Earth, carbonate formation is effective at drawing down CO$_2$ because of ongoing \{Mg, Ca\} supply from volcano-tectonic recycling. For 100 bars of calcite,  In our model oceans, Ca is supplied once by seafloor weathering

%Whether the H$_2$O is exsolved depends on whether magma ocean crystallisation begins in the middle or at the bottom. We effectively assume that crystallisation begins at the bottom. A speculative, but reasonable, scenario is that magma ocean crystallisation begins at the middle. If magma ocean crystallisation begins at the middle, then water dissolved in melt below will be trapped and hosted in dense hydrous magnesium silicates (DHMS). These hydrous crystals would be less dense than overlying water-poor upper mantle, so following complete crystallisation a solid-state mantle overturn would be expected that would lead to decompression, melting and thus dewatering of the (now-near-surface) DHMS phases. Because the DHMS-capacitor scenario protects the water for longer than in the bottom-up crystallisation scenario, our neglect of the DHMS-capacitor possibility is conservative in that it provides a lower bound on the importance of plutonic shielding.

%Empirically, Earth's upper ocean crust contains , or 0.2 wt\% crust average. The flux of subducted oceanic crust is about X km3/yr. Therefore, the CO$_2$ in Earth's atmosphere is subducted in basalt-hosted carbonate every X Myr. Thus 

We neglect Cl. We do so because Cl is geochemically much less abundant than Na and C \citep{Sleep2001,GrotzingerKasting1993,SharpDraper2013,Clay2017}. Because we neglect Cl, our model understates the acidity of low-C-content worlds. Therefore, the vertical stripe of ``optimal $c_C$'' may be drawn at too high a value of CO$_2$-equivalent atmosphere+ocean C content. The H$_2$O/Cl ratio depends on the mechanism for water loss, if any. Production of H$_2$ on planetesimals by oxidation of Fe$^0$ and Fe$^{2+}$, followed by escape-to-space of H$_2$, gets rid of water but not Cl. Mechanical ejection of water to space during impacts gets rid of both. 

 %Offsetting this uncertainty, carbonaceous chondrites represent an aqueously-processed subset of the rocks that formed the Earth. So the condensations of C, H2O, and other soluble components should be taken with grain of salt. Salt is soluble so the chlorine concentration is just a crude guide. 
%For these worlds, our model duration of habitable surface water is an underestimate. 
%Although other atmophiles in the post-giant-impact steam atmosphere \citep{Lupu2014,Fegley2016} would also condense with the ocean, none would contribute significantly to the charge balance for the most-habitable oceans we consider.

Although basalt with an elemental composition similar to that of Earth's oceanic crust -- as used in Appendix~D -- is ubiquitous in the solar system \citep{TaylorMcLennan2009}, the silicate Earth is depleted in Na relative to meteorites that are believed to represent planetary building blocks (the chondrites) \citep{PalmeONeill2014}. The chondrites are themselves depleted in Na relative to equilibrium condensation calculations. A waterworld formed from low-$T$ condensates, or formed from chondrites but without Na depletion, would have Na:Al~$\sim$1. Peralkaline magma would be common, stabilizing Na silicates. Na-silicates dissolve readily, releasing aqueous Na. A Na-rich ocean would have high pH and low $p_{\mathrm{CO2}}$. With Na-carbonates exposed at the surface, the dwarf planet Ceres may be a local example of a Na-rich ocean world \citep{Carrozzo2018}. 

Large-amplitude stochastic differences in planet composition can result from a collision (versus a near miss) with an individual planetary embryo.

Our modeled water-rock reactions, which use freshwater as the input fluid, raise pH due to release of (for example) Na$^+$ and Ca$^{2+}$ (Appendix~D) \citep{Macleod1994}. Na is released from seafloor rocks \citep{StaudigelHart1983} in our model, consistent with experiments \citep{GysiStefansson2012} and Na-leaching of preserved $~$3.4~Ga seafloor crust \citep{NakamuraKato2004}. Moreover, on waterworld seafloors, jadeite is the stable Na-hosting phase, and (at least for $T$~$>$~400C) the jadeite-albite transition is a maximum in [Na] \citep{Galvez2016}. Na release into waterworld oceans is consistent with Na uptake at Earth hydrothermal vents \citep{Rosenbauer1988,Seyfried1987}, because Earth seawater is Na-rich (0.5~mol/kg), which favors Na uptake in hydrothermal systems. In summary, waterworld ocean Na concentrations $O$(0.1-1) mol/kg are geologically reasonable.

%Hydrothermal reaction $T$ is also uncertain. 

%A lower bound on ocean cation abundance the is the K concentration in altered rocks, because K is effectively incompatible in alteration minerals \citep{Seyfried1987}.

%How does this map onto the future exoplanet habitability-testing procedure of Bean et al. 2017?

%High surface temperatures probably better than entomb your only nutrient and energy source (via \mbox{HP-ice}: Figure~\ref{fig:H2Oadiabats}). 

\subsection{Opportunities}

Our analysis suggests the following science opportunities for waterworld research.

\vspace{-0.08in}

\begin{itemize}[leftmargin=0.75em]

\item More detailed modeling of water-rock reaction throughout waterworld evolution, using newly available constraints (e.g., \citealt{Pan2013}).
\vspace{-0.06in}
\item New measurements (via numerical and/or lab experiments) of solubilities, etc., for high-pressure, low-temperature conditions relevant to waterworlds. % but not sampled on the Earth. \mbox{Solubility} data above 0.5~GPa are particularly valuable . %Experimental data is limited for aragonite, ice VII and ikaite, in part because Earth seafloor environments have pressures too low to stabilize these phases, but all are of interest for waterworlds. 
\vspace{-0.06in}
\item Map out rocky-planet atmospheres and atmosphere-ocean equilibria in volatile-abundance space, taking into improving constraints on the galactic range of H/Cl, H/N, and H/C \citep{Bergin2015}. Our results are sensitive to the mean value of the ratio ($c_C/f_W$) of~CO$_2$-equivalent atmosphere+ocean C content to planet water mass fraction.
\vspace{-0.06in}

\item Seek to confirm or reject hints of life proliferating at $T$~$>$~400K on Earth \citep{Schrenk2003}. If life at $T$~$>$~400K is confirmed, this would underline the habitability of worlds with planet water mass fraction ($f_W$)~=~0.1, regardless of whether or not \mbox{HP-ice} is sufficient to prevent habitability (Figure \ref{fig:tsurfevo01}).
\vspace{-0.06in}

\item Investigate the long-term stability of rock-water mixtures on waterworlds. During giant impacts, rock and water are fully miscible. On planets with $f_W$~$\ll$~0.01, the mixture cools quickly, and the rock settles out, because rock is insoluble in cool water. Rock-water-mixture lifetime cannot exceed the conductive cooling timescale,

\vspace{-0.1in}
\begin{equation}
\tau_{unmix} \lesssim \frac{\rho c_{p,f} \Delta Z \Delta z_{bl} }{(2.32)^2 k_{cond} } \sim1 \,\mathrm{Gyr} \left( \frac{ \Delta Z } { 200 \mathrm{km} } \right)^2
\label{eqn:tauslurry}
\end{equation}

\noindent where $\rho$ is characteristic fluid density ($\sim$1100~kg~m$^{-3}$), $c_{p,f}$ is fluid heat capacity ($\sim$4~$\times$~$10^3$~J~kg$^{-1}$~K$^{-1}$), $\Delta Z$ is ocean depth, $\Delta z_{bl}$ is the thickness of the conductive boundary layer, $k_{cond}$ is characteristic fluid thermal conductivity (1~W~m$^{-1}$~K$^{-1}$), and we have made the upper-limit assumption  $\Delta Z~=~\Delta z_{bl}$ \citep{TurcotteSchubert2011}. On worlds with large planet water mass fraction, a parcel of the dense, hot, rock+water layer is stable to rapid convective swapping with the overlying cool rock-poor aqueous fluid. Convective inhibition slows mixture cooling and allows super-adiabatic temperature gradients (persistence of hot layers at depth). Thus, an ocean at habitable temperature might be underlain not by a solid seafloor but instead by a hot rock-water mixture.  The true lifetime will be set by double-diffusive convection \citep{Radko2013,FriedsonGonzales2017,Moll2017}, taking account of the exotic behavior of aqueous fluids at $P$~$>$~10~GPa (e.g.,~\citealt{Redmer2011,TianStanley2013}), and could be much less than the  Equation \ref{eqn:tauslurry} upper limit. It is not obvious whether such a system would be more habitable or less habitable than a fluid ocean underlain by a solid crust.

\item Combine eH modeling \citep{Schaefer2016,Wordsworth2018} with pH modeling (this~study).
\vspace{-0.06in}

\item Evaluate the climate feedbacks proposed in Section 6.2. Explore the bestiary of exsolution-driven climate feedbacks suggested by our simple model (Section 4, Figure~\ref{fig:findequilibria}, \citealt{Kite2011}). Determine if any apply to Earth history. %Employ a more sophisticated model to link non-ideal solubility of all plausible atmospheric constituents with atmospheric radiative-convective models. 
\vspace{-0.06in}

\end{itemize}

\section{Conclusion: Long-term stability of habitable surface water on exoplanet waterworlds with no need for carbonate-silicate cycling}

Our model of long-term climate evolution on waterworlds shows long-term stability of habitable surface water can occur without geochemical cycling. Because volcanism is curtailed by seafloor pressure on waterworlds \citep{Kite2009}, the planet is stuck with the ocean mass and ocean cations that it acquires during the first 1\% of its history. Afterwards, ocean-atmosphere exchange sets pH in the ocean and $p_{\mathrm{CO2}}$ in the air.

Our model indicates that the~key controls on the habitability of waterworlds around Sun-like stars are as follows.
\begin{enumerate}[leftmargin=0.8em]

\item Initial water content. As initial water content (and thus, seafloor pressure) is increased relative to that of the Earth, long-term cycling of C between the atmosphere and the convective mantle moves from an active regime to a subdued regime. We studied planets with 1-8~GPa seafloor pressure. Seafloor pressure on such worlds suppresses rock melting, and so  C cycling between the atmosphere and the deep interior is subdued (Figure~\ref{fig:volcaniccutoff}), permitting cycle-independent planetary habitability (Fig \ref{fig:codeflow}).

\item Cation/C ratio. Cations are supplied by crust leaching. [Cation]/C values~$\lesssim$~1 are needed (in our model) for $>$~1~Gyr of habitability, because a cation excess draws C into the ocean from the atmosphere, and worlds with little atmospheric C have short lifetimes  (Figures~\ref{fig:chvsccat} and \ref{fig:mcuninformed}). 

\item Initial atmosphere+ocean C/H. When C/H is large, C sinks are swamped and CO$_2$ accumulates in the atmosphere. This can lead to uninhabitably high surface temperatures (Figures~\ref{fig:bucket}-\ref{fig:mcuninformed}). When C/H is small, the duration of surface liquid water is $<$~1~Gyr. Initial atmosphere+ocean C/H~$\approx$~1\%~by~weight is optimal in our model (corresponding to CO$_2$/H$_2$O $\approx$~0.25\%~by~weight if all C is present as CO$_2$ and all H is present as H$_2$O) (Figures~\ref{fig:lifetimes001}-\ref{fig:mcuninformed}). 
\end{enumerate}

Around a quarter of the parameter combinations that we simulate yield waterworlds with $>$1~Gyr of habitable surface water. We assume that initially frozen surfaces are darkened by meteoritic debris; if they maintain a high albedo, then the fraction of long-lived worlds is cut by a factor~of~$\sim$3. This does not mean that in the real Galaxy, $\gtrsim$10\%~of~waterworlds are habitable for $>$1~Gyr, because many key input parameters are uncertain. Based on these uncertainties, we recommend priority areas for future research (Section 6.6). 

Atmosphere-lithosphere geochemical cycling appears to be necessary for Earth's long-term habitability \citep{Moore2017}. This connection led to the idea that ``[g]eologic activity is crucial for a planet's maintained surface habitability because such habitability depends on the recycling of atmospheric gases like CO$_2$'' \citep{Kaltenegger2017}. However, as we have shown here, such cycling is not needed for multi-Gyr habitability on rocky exoplanets with deep oceans.
 
 %Indeed, if geochemical cycles like those that moderate Earth's climate turn out to be fragile and/or to require fine-tuning of initial conditions, waterworld habitability may be the main pathway for habitability in the Universe.

Our optimistic conclusions for Sun-like stars support optimism for waterworlds that orbit M-dwarfs \citep{Turbet2018}. For Sun-like stars, stellar main-sequence evolution (timescale $\tau_{MS}$ $\sim$10 Gyr) will in the end destroy habitability. But for planets that orbit M-stars, $\tau_{MS}$ $\gg$ 10 Gyr. Because volcanism-requiring (Earth-like) habitability dies with volcanism after $\lesssim$10 Gyr \citep{Kite2009}, most of the habitable volume in the Universe (including the distant future; \citealt{Loeb2016}) is for cycle-independent planetary habitability. For example, the TRAPPIST-1 HZ planets \citep{Gillon2017} are small enough and old enough \citep{BurgasserMamajek2017} that stagnant-lid volcanism should have shut down \citep{Kite2009}. But all seven of the TRAPPIST-1 worlds have densities consistent with seafloor pressures in the 1-8 GPa range studied by this paper \citep{Grimm2018}. Thes.erefore, the mechanisms proposed in this paper offer more hope for habitability in the TRAPPIST-1 system than volcanism-requiring habitability. However, although Earth-radius waterworlds in the HZ of FGK stars can retain their water, it is less clear whether or not this is the case for Earth-radius waterworlds in the HZ of a M star \citep{LugerBarnes2015,RamirezKaltenegger2014,Dong2017,Dong2018}. Although in the absence of volcanism, nutrient supply will ultimately become diffusively limited, habitable surface water on waterworlds can persist for~$\gg$10~Gyr. For these worlds, habitability has no need for geodynamic processes, but only the steady light of the star.

\acknowledgments \noindent We are grateful to Bruce~Fegley for continuing guidance on geochemical thermodynamics. We thank Norm~Sleep for a stimulating and prompt formal review, which improved the manuscript. We~thank Christophe~Cossou, Itay~Halevy, Jim~Kasting, Ravi~Kumar~Kopparapu,  Nataly~Ozak, and Ramses~Ramirez, for unselfish sharing of research output. We thank Dorian~Abbot, Jonathan~Lunine, Nadejda~Marounina, Marc~Neveu, and Ramses~Ramirez for commenting on a draft. We~thank David~Archer, Paul~Asimow, Fred~Ciesla, Eric~Fiegelson, Jim~Fuller, Eric~Gaidos, Marc~Hirschmann, Manasvi~Lingam, Mohit~Melwani~Daswani, Jack~Mustard, Leslie~Rogers, Laura~Schaefer, Hilke~Schlichting, Norm~Sleep, Sarah~Stewart, Cayman~Unterborn, and Steve~Vance, who each provided useful advice. We thank the CHIM-XPT team: Mark~Reed,  Nicolas~Spycher, and Jim~Palandri. We thank Craig~Manning and Tim~Lichtenberg for sharing preprints. We thank the organizers of the University~of~Michigan volatile origins workshop. The Center for Exoplanets and Habitable Worlds is supported by the Pennsylvania State University, the Eberly College of Science, and the Pennsylvania Space Grant Consortium.  Parts of this research were conducted with Advanced CyberInfrastructure computational resources provided by The Institute for CyberScience at The Pennsylvania State University (http://ics.psu.edu), including the CyberLAMP cluster supported by NSF grant \mbox{MRI-1626251}.  The results reported herein benefitted from collaborations and/or information exchange within NASA's Nexus for Exoplanet System Science (NExSS) research coordination network sponsored by NASA's Science Mission Directorate. This work was supported by the U.S. taxpayer, primarily via NASA grant NNX16AB44G.
\vspace{0.15in}

\noindent \emph{Author contributions:} E.S.K. conceived, designed, and carried out the study, and wrote the paper. E.S.K. wrote the volatile tracking code (Section 5.2). E.B.F's contributions were focused on the \emph{N}-body integrations (Section 5.1).

\email{kite@uchicago.edu}

\appendix

%\section{P. Seawater chemistry for astrophysicists.}

\section{A. How seafloor pressure suppresses C~exchange between the convecting mantle and the water ocean.} 
\noindent With reference to Figure~5, %\ref{fig:volcaniccutoff},
four effects allow seafloor pressure to suppress volcanism (and thus C exchange between the convecting mantle and the water ocean) on waterworlds. These four effects (in order of importance) are as follows. (1) The solidus $T$ increases rapidly with $P$, so increasing $P$ truncates the maximum melt fraction, eventually to zero. (2) The upwelling mantle undergoes corner flow (turns sideways), so by reducing the distance between the base of the lithosphere and the maximum $P$ for melting, increasing $P$ reduces the proportion of upwelling mantle that melts in a given time. (3) $T$ just below the lithosphere is regulated to an absolute value that is fairly insensitive to the ocean depth, due to a negative feedback between \mbox{$T$-dependent} mantle viscosity and the viscosity-dependent rate of convective mantle cooling \citep{Schubert2001,KorenagaKarato2008,Stevenson2003}. Therefore, a waterworld mantle is colder for a given $P$ (dashed blue lines in Figure~\ref{fig:volcaniccutoff}) than the mantle of a shallow-ocean planet (dashed gray lines in Figure~\ref{fig:volcaniccutoff}). This further inhibits melting. (4) For the melt fractions corresponding to track A-B-C-D,  melt fraction increases super-linearly as ($T - T_{sol}$)$^{1.5}$/($T_{liq} - T_{sol}$)$^{1.5}$ \citep{Katz2003}. 

Because a planetary mantle undergoing stagnant-lid convection with no volcanism cools less efficiently for a given temperature than a plate-tectonics world with volcanism, a \mbox{stagnant-lid} planet should adjust to a mantle potential temperature that is higher than that of a planet undergoing plate tectonics (Figure 6 in \citealt{Kite2009}). However, models indicate that the temperature rise should be less than the difference between 1350$^\circ$C
 (modern Earth mantle potential temperature) and 1550$^\circ$C (hottest ambient-mantle potential temperature known on Earth, $\sim$3.0 Gya; \citealt{Herzberg2010}).

We use the anhydrous batch-melting solidus of \citet{Katz2003}. Plausible waterworld upper-mantle water contents are similar to that of the Earth's mantle beneath mid-ocean ridges. These rocks have an average of \mbox{0.005-0.02~wt\%}~water \citep{WorkmanHart2005, HirschmannKohlstedt2012}, and this water might depress the solidus by 50K. 

Complete shut-down of eruptions is difficult. This is because anomalously wet parcels of mantle or anomalously hot parcels of mantle (e.g. mantle plumes) will both melt more readily. However, in the waterworld approximation, fitful and low-rate volcanism associated with anomalously wet or anomalously hot mantle cannot buffer waterworld ocean chemistry. 

\section{B. High-pressure ice phases.}
\noindent High-pressure ice appears wherever the H$_2$O adiabat intersects the pure-H$_2$O freezing curve. (We neglect salt effects; \citealt{VanceBrown2013}.) 
To build the fluid-water adiabat, we used

\begin{equation}
\frac{\partial{T_{ad}}}{\partial{P}}  = \frac{ \alpha(T,p) T }{\rho(T_i,p_i) c_{p,f}}
\end{equation}

\noindent where $\rho$ is obtained from Table 1 of \citet{Wiryana1998} and Table 2 of \citet{AbramsonBrown2004}, and $\alpha$ is obtained by differencing these tables. We assumed $c_{p,f}$~$\sim$~3800~J~kg$^{-1}$~K$^{-1}$. The approximation of $c_{p,f}$ as constant is justified by \citet{Myint2017}. We extrapolated the tables above 6~GPa, and for $T$~$<$~80$^\circ$C. Clathrates were neglected. Pure-H$_2$O freezing curves for ices VI and VII are from \citet{IAPWS2011}.
We note that \citet{Alibert2014} did not include the adiabatic heating of liquid water. This effect becomes important for deep oceans (Figure~\ref{fig:H2Oadiabats}).%, the conclusions of \citet{Alibert2014} about the maximum thickness of ice-free ocean are not correct.
%As pointed out by \citet{KirkStevenson1987}, complicated topologies can exist. 

%Once HP ice has formed, we conservatively do not consider the planet to be habitable. Therefore we do not track the exsolution of CO$_2$ gas by the water, which is a negative (stabilizing) feedback on temperature for planets with HP ice.

%
%\begin{figure}
%%\epsscale{1.2}
%\includegraphics[width=0.99\columnwidth]{neutral_planetary_habitability_figure_H2O_VS_PLANET_MASS.pdf}
%\caption{An appendix figure showing the h2o-vs-planet-mass evolution from the N-body runs \textbf{with and without plutonic shielding}} 
%\label{fig:h2ovsplanetmass}
%\end{figure}

%
% \begin{figure}
%%\epsscale{1.2}
%\includegraphics[width=0.99\columnwidth]{neutral_planetary_habitability_figure_FA.pdf}
%\caption{Titration curve showing exhaustion of the Na + Ca buffer when ( [Na]+2*[Ca] / [C] ) $<$1. Circle color corresponds to sea-surface temperature ($^\circ$C). Based on model calculations using CHIM-XPT \citep{Reed1998} at 20 bars.
%} 
%\label{fig:oc2}
%\end{figure}

 \begin{figure}
%\epsscale{1.2}
\includegraphics[width=0.93\columnwidth,clip=true,trim={23mm 10mm 23mm 15mm}]{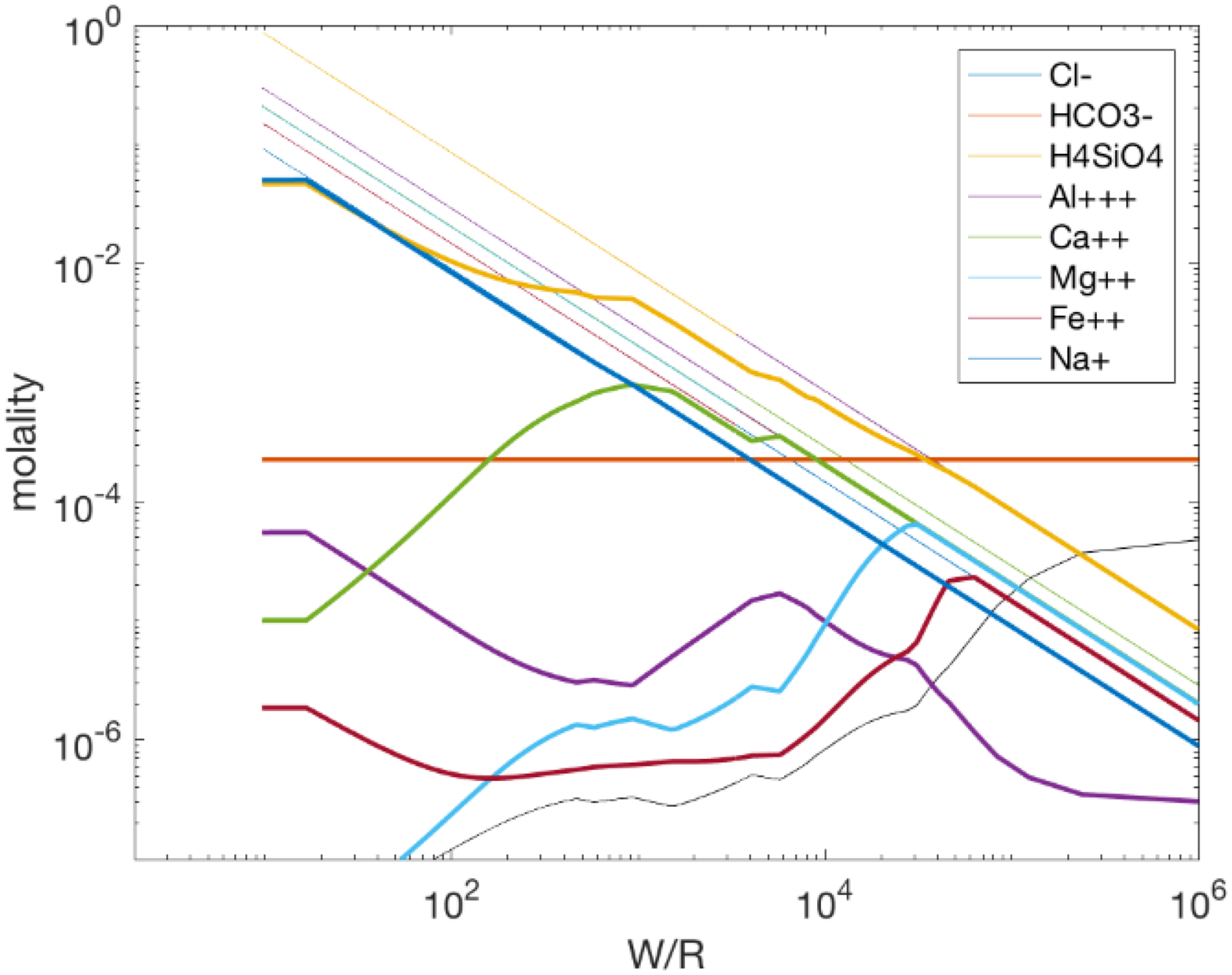}
\includegraphics[width=0.93\columnwidth,clip=true,trim={23mm 10mm 23mm 15mm}]{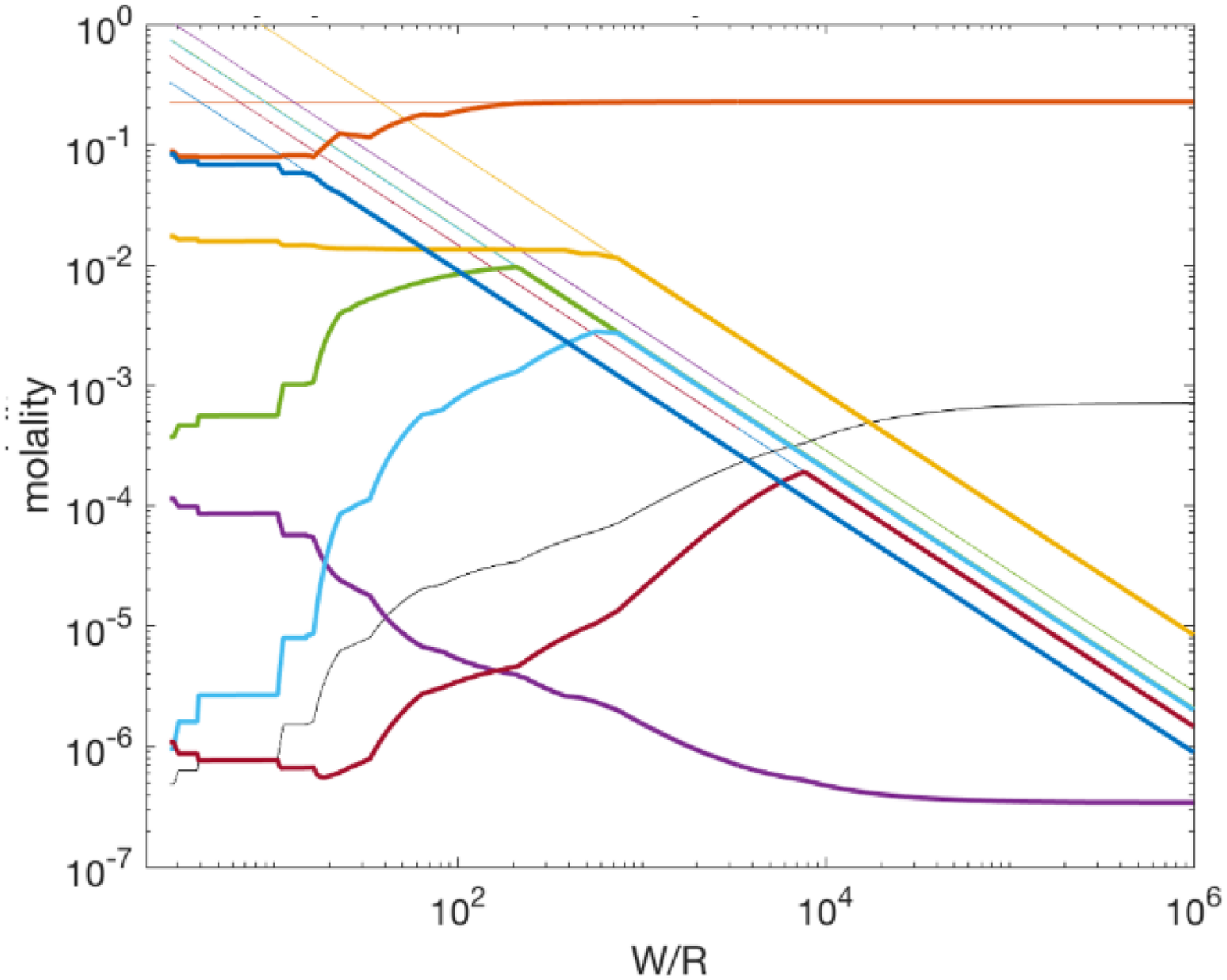}
\caption{Seafloor basalt alteration for (\emph{upper panel}) low [C] concentrations ($P$~=~5000~bar, $T$~=~200$^\circ$C) and (\emph{lower panel}) high [C] concentrations ($P$~=~5000~bar, $T$~=~300$^\circ$C). W/R is water/rock ratio characterizing the reaction. The thick colored lines correspond to concentrations in the fluid. Thin lines of the same color correspond to total mol/kg of rock added to the fluid. Where the total mol/kg of rock added from the fluid deviates from the concentrations in the fluid, secondary minerals have formed. The thin black line corresponds to -log$_{10}$(pH). } 
\label{fig:sfb}
\end{figure}
%
%\begin{figure}
%%\epsscale{1.2}
%%\includegraphics[width=0.99\columnwidth]{neutral_planetary_habitability_figure_LD.pdf}
%\caption{As Figure~\ref{fig:sfb}, but for initial [C] $\sim$0.2 molar. For legend see Figure~\ref{fig:sfb}.} 
%\label{fig:sfbc}
%\end{figure}

% \begin{figure}
%%\epsscale{1.2}
%\includegraphics[width=0.85\columnwidth]{neutral_planetary_habitability_figure_BA.pdf}
%\caption{Calcite solubility products (moles of CaCO$_3$ per kg of H$_2$O) combining the formulae of Table 1 in \citet{DolejsManning2010}, with the water-density tables of \citet{Wiryana1998} and  \citet{AbramsonBrown2004}. The thick black line is the 200$^\circ$C isotherm on an adiabat. Calcite is metastable with respect to aragonite over much of this $T$-$P$ range.% (unfortunately, aragonite solubility data in this $T$-$P$ range are not available). %Calcite can also be unstable with respect to ikaite, but only within 40K of the \mbox{HP-ice} stability for a given pressure: \citep{Shahar2005}
%} 
%\label{fig:caco3solubility}
%\end{figure}

\section{C. Post-magma-ocean timescales.}

%\noindent Here we give our reasons for considering water-rock interaction at $T$ $<$700K \citep{Cannon2017}. 
\noindent Following a giant impact, the characteristic temperature at which water first reacts with the solid crust depends on the ratio between the timescales on which the fluid-envelope cools ($\tau_c$), and the timescale over which the crust forms ($\tau_x$) \citep{Zahnle2007}. If $\tau_c$~$<$~$\tau_x$, crust rocks will first react with hot (supercritical) H$_2$O \citep{Cannon2017}. If $\tau_c$~$>$~$\tau_x$, crust rocks will react with water on an adiabat connected to an sea surface that is in equilibrium with insolation (Figure \ref{fig:H2Oadiabats}). The timescale for cooling $\tau_c$ is 

\begin{equation}
\tau_{c} = \frac{M_{pl} f_W c_{p,f}  (2000\mathrm{K} - 800\mathrm{K}) }{4 \pi R_{pl}^2 (F_{KI} - 0.25 L_* (1 - \alpha ) )}
\end{equation}

\noindent where $c_{p,f}$ is the fluid heat capacity and $F_{KI}$ is the Komabayashi-Ingersoll limit \citep{Ingersoll1969}. This gives $\tau_{c}$~=~10$^5$-10$^6$~yr for the fluid envelopes considered in this paper. By contrast, $\tau_x$, the timescale for the crust to form, is poorly constrained. $\tau_x$ for stagnant-lid mode on waterworlds is the timescale for the initial post-magma-ocean crust to form -- perhaps as long as 10$^8$ yr \citep{Rubin2005,Mezger2013,Moynier2010,Turner2001,Solomatov2015}. Also relevant is the timescale for C delivery by bolide impacts ($\tau_{dyn}$). $\tau_{dyn}$~could~be $>$10$^8$ yr for the Earth, and this may be typical of planets around G-type stars \citep{Lohne2008}. Because $\tau_x$ and $\tau_{dyn}$ are both plausibly much longer than $\tau_c$, but not plausibly shorter than $\tau_c$, we speculate that the characteristic temperature for basalt rock to first encounter C-rich fluid is \mbox{$<$650~K}. Water-rock reactions are well advanced within \mbox{10$^3$-10$^5$~yr} \citep{Hannington2005}, so we expect that even $\tau_x$~$\ll$~1~Myr would allow sufficient time to alter rock.

%, i.e. within the range of Earth hydrothermal vents \citep{Cannon2017}, and within the $T$ range where carbonates are stable \citep{Sleep2001}. 

\section{D. Modeling of seawater-basalt interaction.}

\noindent 
%Calcite precipitation often occurs in our surface-chemistry modeling. 
%
%Our rock-ocean exchange model differs from Earth-centric models (e.g.~\citealt{Charnay2017,Holland1984,RidgwellZeebe2005,Bachan2017}). We do not assume \emph{a priori} that the ocean is saturated with calcite. Calcite-buffering is frequently assumed in ocean chemistry models, because calcite is persistent in the geologic record \citep{Walker1983,WordsworthPierrehumbert2013,HalevyBachan2017}. If we assume that the calcite snow sinks and redissolves swiftly, then the CO$_2$ storage capacity is just the depth-integral of the carbonate-saturation limit, i.e., the ocean is locally carbonate-buffered.

%The set-ups for ocean chemistry are shown below. 
%Hydrothermal $T$ controls whether Ca and Na -- the cations we choose to track -- behave as relatively soluble phases, or alternatively are fixed in secondary minerals \citep{Seyfried1987}.

%As $P$ increases, the ion product of water increases (by~$\sim$1.5 log units going from~0-1~GPa, \citealt{MarshallFranck1981}). 

\noindent To explore seawater-basalt reactions on waterworlds, we used CHIM-XPT v.2.4.6 \citep{Reed1998}. We consider $T$~=~\{298K,~473K,~573K\}. Although temperature changes over geologic time, we assume no ``resetting'' of seafloor alteration. ``Resetting'' appears to be minor for Earth seafloor alteration: seafloor basalt carbonitization occurs near the time of eruption \citep{Coogan2016}. Due to the scarcity of experiments on H$_2$O properties for $>$0.5~GPa, $P$ is fixed at 0.5~GPa. This low $P$ will cause us to understate solubilities. Improved \emph{ab initio} calculations of the properties of H$_2$O \citep{Pan2013} will allow future improvements in waterworld seafloor modeling. We assume thermodynamic equilibrium, with solid solutions suppressed. Basalt composition is from \citet{Gale2013}. A key parameter in CHIM-XPT is the water/rock mass ratio, W/R. The use of W/R by geochemists is easy to misinterpret in an exoplanet context. For geochemists, W/R corresponds to the effective W/R of the reactions. Reactions typically occur in zones that are not hydrologically open to the ocean. Examples are pore spaces and fractures. As a result, the reactions tend to be much more rock-influenced than the ocean as a whole. Therefore, W/R is almost always much less than the (mass~of~the~ocean)/(total~mass~of~altered~rock). For example, Earth hydrothermal vents have W/R of 1-100, even though the mass of pervasively water-altered rock in Earth's oceanic crust is much less than 1\% of the mass of Earth's oceans. Varying W/R is a proxy for the extent of alteration of sub-oceanic crust by water.  W/R is more important than $T$ for setting vent fluid compositions. %The long timescales of warm-fluid circulation through crust on waterworlds favor W/R $\gtrsim$20. %The resulting cations are diluted into the ocean.

Figure \ref{fig:sfb} shows typical results. For W/R~$<$~100 characteristic of hydrothermal reactions on Earth, outlet fluids are dominated by Na$^+$ and by dissolved Si species. However, dissolved Si forms solid particles upon mixing into the colder, lower-pH upper levels of the ocean (as is inferred to occur on Enceladus; \citealt{Hsu2015}). Therefore, we do not consider dissolved Si when calculating the sea-surface chemistry. As carbonates form, they buffer C fluid content to $\sim$0.1~mol/kg for W/R~$\lesssim$~30.  As secondary minerals precipitate, they scrub Fe$^{2+}$, Mg$^{2+}$, and Al$^{3+}$ from the fluid. Outlet fluid composition levels off at [Na] $\approx$ 0.05-0.12 mol/kg for W/R $\approx$ 10 due to Na-mineral formation, but for reasons discussed in Section 6.5, we consider higher [Na] in our models.

\section{E. Details of waterworld evolution calculations.}

Here we provide more detail on the waterworld evolution calculations that are summarized in Section 3.2.

\emph{Greenhouse forcing and climate look-up table.} To track waterworld climate evolution, we start from the OLR($p_{\mathrm{CO2}}$, $T_{surf}$) and albedo($p_{\mathrm{CO2}}$, $T_{surf}$) from the 1D radiative-convective models of \cite{Ramirez2014} and \cite{WordsworthPierrehumbert2013}. Here, OLR is outgoing longwave radiation (W/m$^2$). Using the (safe) assumption that radiative equilibrium is reached so that OLR = insolation~$\times$~(1~-~albedo), we smooth and interpolate to find the $T_{surf}$($p_{\mathrm{CO2}}$,~insolation) in script \texttt{waterworld\_surface\_temperature\_v3.m}. %We extrapolate OLR from $p_{\mathrm{CO2}}$ = 10$^{-4}$ bar down to $p_{\mathrm{CO2}}$~=~3~$\times$ 10$^{-6}$~bar assuming that the OLR decreases linearly with increasing log($p_{\mathrm{CO2}}$) within this range \citep{Pierrehumbert2010}.

\begin{figure}
%\epsscale{1.2}
\includegraphics[width=1.04\columnwidth,clip=true,trim={10mm 10mm 10mm 10mm}]{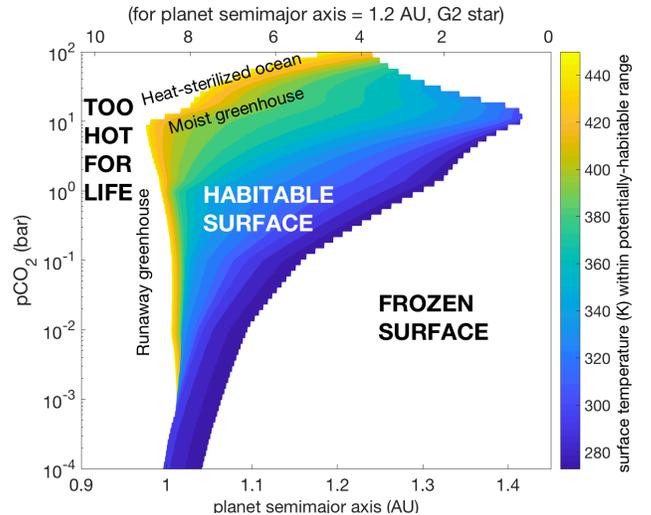} %\\
\caption{As Figure \ref{fig:co2matters}, but using the \citet{Ramirez2014} radiative-convective climate model as a sensitivity test. To show the effect of partial pressure of atmospheric CO$_2$ and insolation $L_*$ on habitability, based on \citet{Ramirez2014}. For G-type stars, insolation increases with age of the star on the main sequence (top axis) and decreases with distance from star (bottom axis). The colored region corresponds to the habitable zone (HZ). The planet fates define a lozenge of habitability that is broadest for \emph{O}(1) bar $p_{\mathrm{CO2}}$. In other words, for a given semimajor axis, the duration of surface habitability is maximized for a ``sweet spot'' of $p_{\mathrm{CO2}}$ in the range 0.2 - 20 bar. The color ramp corresponds to $T_{surf}$(K), and extends from 273K to 450K.}  
\label{fig:ramirezlozenge}
\end{figure}

\begin{figure}
%\epsscale{1.2}
\includegraphics[width=0.9\columnwidth,clip=true,trim={23mm 70mm 30mm 75mm}]{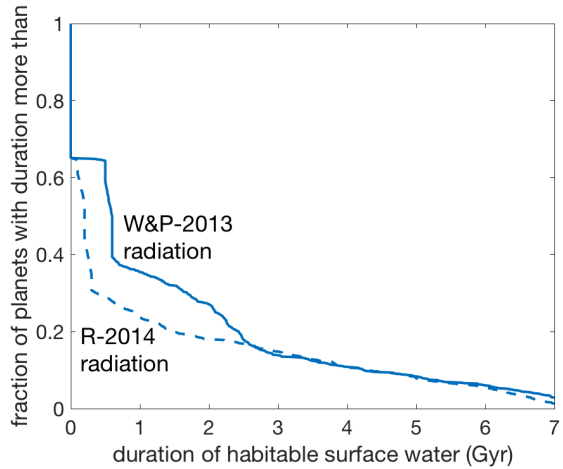}
\caption{Comparison of the duration of habitable surface water obtained using the \citep{Ramirez2014} climate model (dashed line, ``R-2014 radiation''), versus the duration of habitable surface water obtained using the \cite{WordsworthPierrehumbert2013} climate model (solid line, ``W\&P-2013 radiation''). Planet~semimajor~axis~=~1.1 AU, planet~mass~=~1~Earth-mass, planet water mass fraction $f_W$~=~0.01.}  
\label{fig:ramirezvswordsworth}
\end{figure}

To check the sensitivity of our results to the choice of 1D radiative-convective climate model, we re-ran the entire waterworld evolution code substituting the output of \cite{Ramirez2014} in place of the output of \cite{WordsworthPierrehumbert2013}. Among other advantages, the \cite{Ramirez2014} output extends to $p_{\mathrm{CO2}}$ = 100 bar, avoiding the arbitrary cut-off of 40 bar in our main runs. The results are shown in Figure \ref{fig:ramirezlozenge}  and Figure \ref{fig:ramirezvswordsworth}. The main difference between the results is that the duration of habitable surface water for low-$p_{\mathrm{CO2}}$ planets is predicted to be shorter, and there is a greater fraction of such short-lived planets (Figure \ref{fig:ramirezvswordsworth}). However, the result that about 10\%~of~waterworlds have habitable surface water for longer than the age of the Earth does not change (Figure \ref{fig:ramirezvswordsworth}).

Neither the model of \cite{WordsworthPierrehumbert2013} nor that of \cite{Ramirez2014} self-consistently models changes in the effect of clouds on planet albedo as $p_{\mathrm{CO2}}$ increases. Neither model includes ice-albedo feedbacks. We can say little about the effect of planet mass because the gravity correction on the greenhouse effect is important, and not included in this study \citep{Pierrehumbert2010}. On planets bigger than Earth, the high-permeability fractured zone in the uppermost crust is smaller \citep{Sleep2000b}.

%For example, planetary albedo declines with increasing temperature in \citet{WordsworthPierrehumbert2013} because pH2O increases and so the atmospheric absorption of H2O in the visible increasingly offsets the cooling effect of CO2-induced Rayleigh scattering at high $L_*$. This might not be the case in a model that included clouds \citep{Popp2016}.

%We do not calculate water loss explicitly. That is because even the energy-limited water loss rate is too small to make a difference. \cite{WordsworthPierrehumbert2013} further show that at high CO$_2$ mixing ratio, the water loss rate can be lower than the energy-limited water loss rate due to dilution of H$_2$O by CO$_2$ in the upper atmosphere. Thus, the energy-limited water loss rate is conservative and is in any case too slow to make a difference.

\emph{Ocean chemistry look-up grid.} The ocean chemistry look-up grid was constructed using the Gibbs-free-energy-minimization software CHIM-XPT \citep{Reed1998}. We built an ocean chemistry look-up grid in the H-O-C-Ca-Na system. We did this by varying the following component abundances in CHIM-XPT:- Na$^+$ (11 trials); Ca$^{2+}$ (9 trials );  HCO$_3^-$ (12 trials). We also varied
temperature (from 273.25 to 453.15 K; 10 trials).
This gave a total of 11~$\times$~9~$\times$~12~$\times$~10 =~1.2~$\times$~10$^4$ trials. Charge balance was enforced by setting the proton molality, [H$^+$], to [HCO$_3^-$] - 2 $\times$ [Ca$^{2+}$] - [Na$^+$]. Here, the square brackets denote molality. We adjusted the specific values trialled for each of the four parameters, with the final values set non-uniformly in order to ensure good coverage of interesting behavior.

For convenience, we used CHIM-XPT in reaction-transport mode to solve the equilibration problem. Formally, we added a tiny amount of H$_2$O to the system for a single reaction-transport step. CHIM-XPT polices charge balance using a  ``charge balance ion," which we shifted from Na to Ca as appropriate to avoid spurious crashes. We ran CHIM-XPT using the Windows emulator \texttt{wine} and shell scripts with CHIM-XPT settings \texttt{minsolsv}~=~1, \texttt{ipsat}~=~0, and $P$~=~20 bars. From the equilibrated output for each of the 1.2~$\times$~10$^4$ trials , we grepped the abundances of the species H$^+$ ($\leftrightarrow$ pH); HCO$_3^-$; CO$_{\mathrm{aq}}$; CO$_3^{-2}$; CaCO$_{3,\mathrm{aq}}$; Ca(HCO$_3^-$); CO$_{2(g)}$ (where present); calcite; dolomite; siderite; magnesite; and portlandite, to a summary text file.

This text file was ingested by script \texttt{parse\_ocean\_surface\_carbonate\_chemistry.m}. We interpolated to densify the grid to 180 temperatures (1-180$^\circ$C) and 18 [Na$^+$] cases. We then considered 21~cation-content~cases: (1) 0.5 mol/kg Na, minimal Ca; (2) cation-poor (``dilute'') ocean; (3) minimal Na+, high Ca+; (4) a sweep of 18~values of Na$^+$ concentration at minimal Ca concentration. The procedure of using equilibria at a fixed pressure (in this case 20 bar) unavoidably introduces errors at different sea-surface pressures - for example, Ca-bearing minerals are less stable at higher pressure. We also found that there were discrepancies between the saturation pressures interpolated from CHIM-XPT, and the values obtained from \citet{Carroll1991} and  \citet{DuanSun2003}, for the dilute-water case. We chose to use the published tables to get CO$_2$ saturation pressures.
 
 \emph{Ocean-atmosphere equilibriation.}
The ocean chemistry look-up table and the greenhouse forcing and climate look-up table are both passed to the climate evolution script, \texttt{make\_waterworld\_fate\_diagram.m}. For waterworlds, this script takes inputs of planet mass $M_{pl}$, planet initial water mass fraction $f_W$, planet initial atmosphere+ocean carbon content $c_C$, and planet semimajor axis $a$. The output is ocean chemistry, atmospheric partial pressure of carbon dioxide $p_{\mathrm{CO2}}$, and ocean surface temperature $T_{surf}$, as functions of time.

%This script is set up to deal with arbitrary planet mass, although in practice it can only be safely used for the single gravity (that of Earth) used to make the climate look-up table. For each planet mass and planet water content, the script finds planet radius-at-seafloor, seafloor gravity, and seafloor pressure. Planet total carbon content is expressed as kg (CO2-equivalent)/(m$^2$ seafloor/surface). 

A necessary first step is to get the atmosphere-ocean partitioning of carbon. To do this, we use the tables of \cite{DuanSun2003} to get CO$_{2,\mathrm{aq}}$ as a function of $p_{\mathrm{CO2}}$. We then use the CHIM-XPT output to get the pH-dependent ``extra" C (HCO$_3^-$, CO$_3^{2-}$, and solid phases) stored in the ocean for a given CO$_{2,\mathrm{aq}}$. Multiplying the total ocean molality by the ocean column mass (kg/m$^2$) gives the total moles of C in the ocean. Multiplying by 44/1000 gives the CO$_2$-equivalent C column in the ocean, and adding the atmospheric CO$_2$ column mass ($\approx$ $p_{\mathrm{CO2}}$/g) gives the total C column (kg/m$^2$) as a function of $p_{\mathrm{CO2}}$. We then interpolate to find the $p_{\mathrm{CO2}}$ for a given cation content and given total atmosphere+ocean C abundance.

%Specifially, The code also ingests CO$_{2,\mathrm{aq}}$ as a function of $T$ and $P_{total}$ from interpolation/extrapolation of the Carroll et al., 1991, JPCRD, Table 3 and Duan et al., 2003, Chemical Geology, Table 3 results. This gives the CO$_{2,aq}$ molality in fresh water as a function of total atmospheric pressure (pCO2 + pH2O), ignoring the ion effect on CO2 solubility. The pH2O is found using the equations of Duan and subtracted from the total pressure, so .
%
%First, a look up vector is built for each of the 21 cation-abundance cases. For each cation-abundance case, total column carbon is written as the sum of atmospheric carbon and non-atmospheric (aqeous plus suspended-minerals) carbon, where the atmospheric pressure(cations, HCO3) is from \texttt{parse\_ocean\_surface\_carbonate\_chemistry.m}.
% 
%  
%From this, the CO2aqtry2 can be found for each Pco2 of interest. The corresponding non-atmospheric C for that CO2aqtry2 is obtained from the CHIM-XPT output. The total C for each pCO2 can then be found by using the TotalC equation, and the atmospheric fraction of C is computed. Finally, the astrophysical carbon can be interpolated in the resulting look-up table to get the C partitioning and CO2 for each set of assumed \{atm+ocean C, T, fW, planet mass\}.

 \begin{figure}
%\epsscale{1.2}
\includegraphics[width=1.0\columnwidth,trim={15mm 35mm 5mm 35mm}]{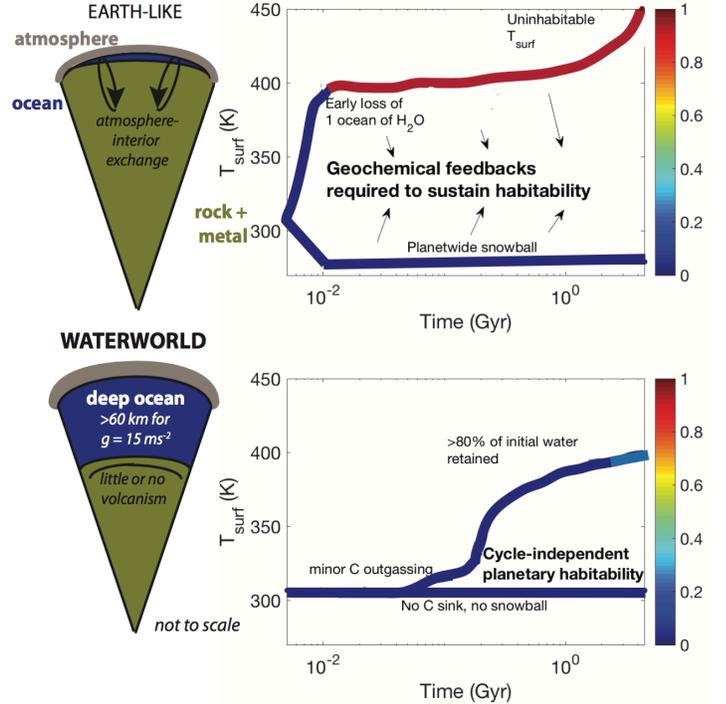}
\caption{How the response of an Earth-mass planet to perturbations of the geologic C~cycle differs depending on ocean depth (in a toy model).
\emph{Upper panel:} a planet with shallow oceans but no geochemical feedbacks. \emph{Lower panel:} a waterworld. The two temperature vs. time tracks for each panel correspond to an increase (upper/hot track in each panel) and a decrease (lower/cold track in each panel) in the CO$_2$ flux from the interior (crust+mantle) into the atmosphere+ocean. Color shows fraction of initial H$_2$O lost to space, assuming XUV-energy-limited escape \citep{Lammer2009} when and only when the stratospheric mixing ratio of water is large \citep{WordsworthPierrehumbert2013}. In the shallow-ocean case (upper panel), the climate rapidly becomes uninhabitable. Therefore, in this toy model, geochemical feedbacks are required to restore and sustain habitability. Specifically, suppose C supply by outgassing exceeds C uptake by weathering. Then, temperature will rise and trigger a moist greenhouse, which will dehydrate the world (hot track). On the other hand, cations made available for weathering by volcanism and tectonics can swiftly draw down atmospheric CO$_2$ to negligible levels, triggering a snowball (cold track).  \emph{Lower panel:} in the deep-ocean case, the planet remains habitable for many Gyr due to the subdued geologic C-cycle, plus ocean dilution. Specifically, volcanism requires a much longer time to trigger the moist greenhouse and the planet stays habitable indefinitely in the moist greenhouse state, which cannot remove the massive initial ocean (hot track). On the other hand, without continents or volcanism, C drawdown quickly becomes cation-limited and diffusive, so the planet cannot enter the snowball state (cold track).}  
\label{fig:toymodels2}
\end{figure}

\section{F. Exsolution-driven climate instabilities.}

Here we provide more discussion of the model-predicted instabilities mentioned in Section 4.1, some of which are shown in Figures~\ref{fig:tsurfevo}~and~\ref{fig:pco2evo}.

Increases in $p_{\mathrm{CO2}}$ of up to a \mbox{factor of 10} accompanied by increases in temperature of up to 100~$^\circ$C can occur in the model within a single~(0.1~Gyr)~timestep. How fast would warming occur? The pace of change is set by deep-ocean thermal inertia and by the ocean mixing time (which is $\sim$10$^3$ yr for Earth). If the ocean mixes slowly (i.e. is stratified), then the timescale for the instability could be as long as the conductive timescale, $>$10$^7$ yr. If instead the ocean mixing time is fast, then as each parcel of seawater is mixed into the near-surface layer with temperature $T_{surf}$, it will exsolve CO$_2$. Therefore, both $p_{\mathrm{CO2}}$ and $T_{surf}$ will rise on the fast ocean mixing timescale. Fast warming of the wave-mixed near-surface layer ($<$0.1~km depth) could stress phototrophs. However, for a $O$(10)~W/m$^2$ change in the surface energy balance, a \mbox{100~km-deep~ocean} will take $>$100~Kyr to warm. Can this rate of change exterminate life? Microbial populations can make major adaptations on 20-yr timescales \citep{Blount2008}. Therefore the rate of change associated with these instabilities is unlikely to risk whole-ocean microbial extinction (in our opinion), provided that both the initial climate and final climate are habitable.

The existence of instabilities implies that surface energy balance for a given solar luminosity and given CO$_2$-equivalent atmosphere+ocean C content can be satisfied by more than one surface temperature. Which temperature is correct?  The low-temperature solution (as long as it exists) is the one followed by our planet evolution tracks. This is because solar luminosity is initially low. Forced by $L_*$, whole-ocean $T$ will start low and increase with time, in the absence of other perturbations. One way to access the warm solutions is an asteroid impact. However the impacts would have to be very large, because of the large thermal inertia of the deep ocean. (A~smaller~impact can raise the temperature of the shallow ocean and exsolve some~CO$_2$, but accessing the stable warm branch requires exsolving CO$_2$ from the bulk of the ocean.) For example, consider one of the largest impact structures on Mars, Hellas. If a~third~of~Hellas' impact energy (total $\sim$4~$\times$~10$^{27}$~J; \citealt{WilliamsGreeley1994}) goes into heating a 100~km~deep ocean on a 1 $M_\earth$ planet, $\Delta T_{surf}$~$\approx$~5 K, which is small. Nevertheless, impacts more energetic than Hellas are possible early in planetary history, and if the warm branch is accessed, then the trace of the impact will be seen in the climate for $O$(Gyr). The time-integrated change in surface energy balance (units J) as the consequence of the impact can easily be 1000$\times$~greater than the impact energy. This is a new mechanism for a metastable impact-driven climate. Metastable impact-triggered climates have previously been proposed using other mechanisms (\citealt{Segura2012,UrataToon2013}, but see also \citealt{Wordsworth2016}). %Our exsolution-driven mechanism requires a deep ocean that stores a lot of CO$_2$. %It is not obvious that this mechanism cannot apply to early Mars. 

\clearpage

\end{document}